\documentclass[acmsmall]{acmart}

\def\BibTeX{{\rm B\kern-.05em{\sc i\kern-.025em b}\kern-.08emT\kern-.1667em\lower.7ex\hbox{E}\kern-.125emX}}

\setcopyright{acmcopyright}

\acmJournal{TIOT}
\usepackage{booktabs} 
\usepackage{subfigure}
\usepackage{multirow}

\renewcommand{\vec}[1]{\mathbf{#1}} 
\newcommand{\mat}[1]{\mathbf{#1}} 
\usepackage{color}

\newcommand{\blue}{}

\newcommand{\revised}{}
\newcommand{\sect}{\textsection}

\DeclareMathOperator*{\argmin}{argmin}

\newtheorem{property}{Property}

\begin{document}

\title[Lightweight Privacy-Preserving Collaborative Learning for IoT]{On Lightweight Privacy-Preserving Collaborative Learning for Internet of Things by Independent Random Projections}

\author{Linshan Jiang}
\affiliation{
  \department{School of Computer Science and Engineering}
  \institution{Nanyang Technological University}
}
\email{linshan001@e.ntu.edu.sg}

\author{Rui Tan}
\affiliation{
  \department{School of Computer Science and Engineering}
  \institution{Nanyang Technological University}
}
\email{tanrui@ntu.edu.sg}

\author{Xin Lou}
\affiliation{
  \department{Advanced Digital Sciences Center}
  \institution{Illinois at Singapore Pte Ltd}
}
\email{lou.xin@adsc-create.edu.sg}

\author{Guosheng Lin}
\affiliation{
  \department{School of Computer Science and Engineering}
  \institution{Nanyang Technological University}
}
\email{gslin@ntu.edu.sg}

\begin{abstract}
  The Internet of Things (IoT) will be a main data generation infrastructure for achieving better system intelligence. This paper considers the design and implementation of a practical privacy-preserving collaborative learning scheme, in which a curious learning coordinator trains a better machine learning model based on the data samples contributed by a number of IoT objects, while the confidentiality of the raw forms of the training data is protected against the coordinator. Existing distributed machine learning and data encryption approaches incur significant computation and communication overhead, rendering them ill-suited for resource-constrained IoT objects. We study an approach that applies independent random projection at each IoT object to obfuscate data and trains a deep neural network at  the coordinator based on the projected data from the IoT objects. This approach introduces light computation overhead to the IoT objects and moves most workload to  the coordinator that can have sufficient computing resources. Although the independent projections performed by the IoT objects address the potential collusion between the curious coordinator and some compromised IoT objects, they significantly increase the complexity of the projected data. In this paper, we leverage the superior learning capability of deep learning in capturing sophisticated patterns to maintain good learning performance. Extensive comparative evaluation shows that this approach outperforms other lightweight approaches that apply additive noisification for differential privacy and/or support vector machines for learning in the applications with light to moderate data pattern complexities.
\end{abstract}



\thanks{A preliminary version of this work appeared in The 4th ACM/IEEE International Conference on Internet of Things Design and Implementation (IoTDI 2019). This research is supported in part by the National Research Foundation (NRF), Singapore and National University of Singapore (NUS) through its National Satellite of Excellence in Trustworthy Computing for Secure Smart Nation Grant (TCSSNG) award no. NSOE-TSS2020-01, in part by the NRF Singapore under its Campus for Research Excellence and Technological Enterprise (CREATE) programme, and in part by the NRF Singapore and the Energy Market Authority (EMA), under its Energy Programme (EP Award NRF2017EWT-EP003-023). Any opinions, findings and conclusions or recommendations expressed in this material are those of the author(s) and do not reflect the views of NRF Singapore, NUS (including its National Satellite of Excellence in Trustworthy Software Systems (NSOE-TSS) office), and EMA}

\maketitle

\section{Introduction}
\label{sec:intro}

The recent research advances of machine learning have led to performance breakthroughs of various tasks such as image classification, speech recognition, and language understanding. The drastically increasing amount of data generated by the Internet of Things (IoT) will further foster machine learning performance and enable new applications in various domains.
In particular,  {\em collaborative learning}, which builds a machine learning model (e.g., a supervised classifier) based on the training data contributed by many {\em participants}, is a desirable and empowering paradigm for smarter IoT systems. By leveraging on the increased volume of training data and coverage of data patterns, collaborative learning will approach the intelligence of a crowd and improve the learning performance beyond that achieved by any single participant alone.
Moreover, a resource-rich learning {\em coordinator} (e.g., a desktop-class edge device or a cloud computing service) allows the execution of advanced, compute-intensive machine learning algorithms to capture deeper structures in the aggregated data, whereas the participants (e.g., IoT objects) are often resource-constrained and insufficient for intensive computation. By contributing training data, the individual participants will {\revised{benefit from}} the improved machine intelligence in return.

However, the data contributed by the participants may contain privacy-sensitive information. {\revised{Various web services (e.g., webmail and social networking) }} generally collect and analyze the user data in the raw forms.
{\revised{In this scheme, users risk their privacy due to both inadvertent or malicious actions by the service provider and due to targeted cyber-attacks by external parties.}} This risk has been evidenced by several recent large-scale user privacy leak incidents \cite{FB2018,Equifax,Lindsey2018}.
Data anonymization can mitigate the concern; but it is inadequate for privacy preservation, because cross correlations among different databases may be used to re-identify data \cite{narayanan2006break}. Moreover, the correlations between different properties of anonymous individuals (e.g., race, income, political views, etc.) can be exploited to identify {\revised{people}} to target for advertisement and advocacy. In the coming era of IoT with many smart objects penetrating into our private space and time, the current raw data collection approach will only raise large privacy concerns and may potentially violate relevant laws such as the recent General Data Protection Regulation in European Union and Personal Data Protection Act in Singapore. {\revised{Therefore, to be successful, IoT-driven collaborative learning applications must preserve privacy.}}




Privacy-preserving collaborative learning (PPCL) has received increasing research recently under the enterprise settings, where the participants are entities with rich computing resources. The existing approaches can be broadly classified into two categories. The first category of approaches \cite{Hamm15,Shokri15,mcmahan2016communication,Phong17,Bonawitz17} follows the distributed machine learning (DML) scheme, such that the participants {\revised{need not}} transmit the training data to the coordinator. Instead, the participants and the coordinator will exchange the parameters of machine learning models. The recently proposed {\em federated learning} \cite{mcmahan2016communication} is a type of DML. In the second category of approaches \cite{graepel2012ml,gilad2016cryptonets,chabanne2017privacy}, each participant applies the homomorphic encryption on the data before being transmitted to the coordinator such that the training and inference computation can be performed on ciphertexts. However, for resource-constrained IoT objects, these DML and data encryption approaches incur significant and even prohibitive {\revised{computation overhead}}.
The DML will require the participants to execute machine learning algorithms to train local models, which is often too compute-intensive for IoT objects.
Moreover, the iterative communication rounds of DML introduce large communication overhead. Currently, the homomorphic encryption algorithms are still too compute-intensive to be realistic for resource-constrained devices. Therefore, these existing approaches are ill-suited or unpractical for the resource-constrained smart objects beneath the IoT edge.

In this paper, we study the design and implementation of a PPCL approach that is lightweight for resource-constrained participants, while {\revised{preserving privacy}} against an honest-but-curious learning coordinator. The coordinator can be a cloud server or a resource-rich edge device, e.g., access points, base stations, network routers, etc.
We propose to apply (1) multiplicative {\em random projection} at the resource-constrained IoT objects to obfuscate the contributed training data and (2) {\em deep learning} at the coordinator to address the much increased complexity of the data patterns due to the random projection. Specifically, each participant uses a private, time-invariant but randomly generated matrix to project each plaintext training data vector and transmits the result to the coordinator. This paper primarily focuses on Gaussian random projection (GRP), because GRP gives several privacy preservation properties of (1) the computational difficulty for the coordinator to reconstruct the plaintext without knowing the Gaussian matrix \cite{liu2006random,rachlin2008secrecy}, and (2) quantifiable plaintext reconstruction error bounds even if the coordinator obtains the Gaussian matrix \cite{liu2006random}. This paper also considers other random projection matrices such as Rademacher and binary matrices. From a system perspective, random projection is computationally lightweight and does not increase the data volume. Thus, random projection is a practical privacy protection method suitable for resource-constrained IoT objects. Regarding random projection's impact on the design of the machine learning algorithms, the projection can be viewed as a process of mapping the original data vectors to
some domain in which the data vectors in different classes are less separable. If the original data vectors are readily separatable (that is, they are features), the inverse or pseudoinverse of the random matrix can be considered as a linear feature extraction matrix. With the deep learning's unsupervised feature learning capability, this inverse matrix can be implicitly captured by the trained deep model.

To achieve robustness of the privacy preservation against the collusion between any single participant and the curious learning coordinator, each participant should generate its own projection matrix independently. However, this presents a challenge on the PPCL system's scalability with respect to the number of participants (denoted by $N$). Specifically, assuming that the training data samples for each class are horizontally distributed among the participants, the number of data patterns for a class will increase from one in the plaintext domain to $N$ in the projection data domain. This increased pattern complexity {\revised{can be}} addressed by the strong learning capability of deep learning. Thus, in the proposed PPCL approach, most of the computational workload is offloaded to the resourceful coordinator at the edge or in the cloud. This is different from the existing DML and homomorphic encryption approaches that introduce significant or prohibitive compute overhead to the smart objects beneath the IoT edge.

To understand the effectiveness of the GRP approach and its scalability with the number of participants, we conduct extensive evaluation to compare GRP with several other lightweight PPCL approaches. The evaluation is based on {\blue four} example applications with data pattern complexity from low to high. They are handwritten digit recognition, spam e-mail detection, {\blue free spoken digital recognition}, and vision-based object classification. The baseline approaches include various combinations between (1) multiplicative GRP versus additive noisification for differential privacy (DP) at the participants, and (2) deep neural networks (DNNs),  including multilayer perceptron (MLP) and convolutional neural network (CNN),
versus support vector machines (SVMs) at the coordinator. The results show that, for the handwritten digit recognition and spam e-mail detection applications with low- and moderate-complexity data patterns, the proposed GRP-DNN approach can support up to hundreds of participants without sacrificing the learning performance much, whereas the GRP-SVM approach may fail to capture the projected data patterns and the performance of the DP-DNN approach is susceptible to additive noisification. The results of this paper suggest that GRP-DNN is a practical PPCL approach for resource-constrained IoT objects observing data with low- or moderate-complexity patterns.  We also compare the learning performance and computation overhead of GRP with the Rademacher and binary random projections.

 We implement GRP-DNN, Crowd-ML \cite{Hamm15} (a federated learning approach based on shallow learning), and CryptoNets \cite{gilad2016cryptonets} (a homomorphic encryption approach) on a testbed of 14 Raspberry Pi nodes. Experiments show that, compared with GRP-DNN, Crowd-ML incurs 350x compute overhead and 3.5x communication overhead to each Raspberry Pi node. Deep federated learning will only incur more compute overhead. CryptoNets incurs 2.6 million times higher compute overhead to the Raspberry Pi node, compared with GRP.

The remainder of this paper is organized as follows. \sect\ref{sec:background} introduces the background and preliminaries. \sect\ref{sec:related} reviews related work. \sect\ref{sec:mov} states the problem and overviews our approach. \sect\ref{sec:evaluation} presents the learning performance evaluation for various lightweight PPCL approaches. \sect\ref{sec:implementation} presents the benchmark results of GRP-DNN, Crowd-ML, and CryptoNets on the testbed.
\sect\ref{sec:conclude} concludes this paper.


\section{Background and Preliminaries}
\label{sec:background}



\subsection{Supervised Collaborative Learning}
\label{subsec:background}

Supervised machine learning has two phases, i.e., the learning phase and the classification phase.
We now formally describe the collaborative learning scheme. The trained classifier, denoted by $h(\vec{x} | \boldsymbol\theta )$, can classify a $d$-dimensional data vector $\vec{x} \in \mathbb{R}^d$ to be one of a finite number of classes represented by a set $\mathcal{C}$, {\blue where $\boldsymbol\theta$ is the classifier parameter and $\mathbb{R}^d$ denotes $d$-dimensional Euclidean space}. The learning process determines the parameter $\boldsymbol\theta$ based on the training data. Let $N$ denote the number of participants of the collaborative learning. Let $\mathcal{D}_i$ denote a set of $M_i$ training data samples generated by the participant $i$, i.e., $\mathcal{D}_i = \{(\vec{x}_{i,j}, y_{i,j}) | j \in \{1,..., M_i\}, y_{i,j} \in \mathcal{C}\}$, where $\vec{x}_{i,j}$ is the training data vector and $y_{i,j}$ is the corresponding class label. For a training data sample consisting of  $(\vec{x}, y)$, denote by $l(h(\vec{x} | \boldsymbol\theta), y)$ the loss function. The collaborative learning solves the following problem to determine the optimal classifier parameter denoted by $\boldsymbol\theta^*$:
\begin{equation}
\boldsymbol\theta^* = \argmin_{\boldsymbol\theta} \sum_{i = 1}^{N} \frac{1}{M_i} \sum_{j=1}^{M_i} l\left( h \left( \vec{x}_{i,j} | \boldsymbol\theta \right), y_{i,j} \right) + \lambda \| \boldsymbol\theta \|^2,
\label{eq:learning}
\end{equation}
where the $\lambda \| \boldsymbol\theta \|^2$ is the regularization term {\blue, $\| \cdot \|$ represents 2-norm,} and $\lambda$ is a parameter affecting the strength of the regularization. With $\boldsymbol\theta^*$, the classification for a test data sample $\vec{x}$ is to compute $h(\vec{x} | \boldsymbol\theta^*)$.

A simple approach is to collect all the plaintext training data to the coordinator and solve Eq.~(\ref{eq:learning}). However, this approach raises the concern of privacy breach, as the raw training data are generally privacy-sensitive. The problem of solving Eq.~(\ref{eq:learning}) without threatening the participants' privacy contained in $\mathcal{D}_i$, $i = 1, \ldots, N$, is called PPCL. Existing approaches to PPCL will be reviewed in \sect\ref{sec:related}.


\subsection{Random Gaussian Projection (GRP) and Other Random Projections}
\label{subsec:random-projection}

This section reviews three random projection approaches: GRP, Rademacher random projection, and binary random projection. Note that this paper primarily focuses on GRP. First, we review two properties of GRP.
Let $\mathbf{R} \in \mathbb{R}^{k \times d}$ represent a random Gaussian matrix, i.e., each element in $\mathbf{R}$ is drawn independently from the normal distribution $\mathcal{N}(0, \sigma^2)$. GRP has the following two properties \cite{liu2006random}:

\begin{property}
  For data vectors $\vec{x}_1$, $\vec{x}_2$ and their projections $\vec{y}_1 = \frac{1}{\sqrt{k} \sigma}\mathbf{R}\vec{x}_1$, $\vec{y}_2 = \frac{1}{\sqrt{k} \sigma} \mathbf{R} \vec{x}_2$, the dot product and Euclidean distance between $\vec{y}_1$ and $\vec{y}_2$ are unbiased estimates of those between $\vec{x}_1$ and $\vec{x}_2$, i.e., $\mathbb{E} \left[ \vec{y}_1^\intercal \vec{y}_2 \right] = \vec{x}_1^\intercal \vec{x}_2$ and $\mathbb{E} \left[ \| \vec{y}_1 - \vec{y}_2 \|_2^2 \right] = \| \vec{x}_1 - \vec{x}_2 \|_2^2$. The estimation error bounds are $\mathrm{Var}[\vec{y}_1^\intercal \vec{y}_2] \le \frac{2}{k}$ and $\mathrm{Var}\left[ \|\vec{y}_1 - \vec{y}_2\|_2^2 \right] \le \frac{32}{k}$.
  \label{property:1}
\end{property}

\begin{property}
  Given a Gaussian matrix instance $\mathbf{R} \in \mathbb{R}^{k \times d}$ where $k < d$ and the projection $\vec{y} = \frac{1}{\sqrt{k}\sigma} \mathbf{R} \vec{x}$, the minimum norm estimate of $\vec{x}$, denoted by $\hat{\vec{x}}$, is an unbiased estimate of $\vec{x}$, i.e., $\mathbb{E}\left[ \hat{\vec{x}} \right] = \vec{x}$. The estimation error for the $i$th element of $\vec{x}$ is $\mathrm{Var}[x_i] = \frac{2}{k}x_i^2 + \frac{1}{k} \sum_{j, j \neq i} x_j^2$.
  \label{property:2}
\end{property}

 Based on Property~\ref{property:1}, the study \cite{liu2006random} shows that a trained SVM classifier can be transferred to classify the projected data. In a recent study \cite{wojcik2018training}, a random projection layer that can be implemented by GRP is added to an MLP for dimension reduction. Such design is also based on Property~\ref{property:1}. However, the studies \cite{liu2006random,wojcik2018training} do not address collaborative learning and privacy. The estimation error given by Property~\ref{property:2} will be used in the later sections of this paper to measure the degree of privacy protection provided by our proposed approach.

Rademacher and binary matrices have also been used for random projections \cite{candes2006compressive,berinde2008combining}. In a Rademacher random matrix, each element is either $\frac{1}{\sqrt{M}}$ or $-\frac{1}{\sqrt{M}}$ with a probability of 0.5, where $M$ is the number of rows in the matrix. In a binary random matrix, each column of the matrix has $S$ ones and $M-S$ zeros, where $S$ is a small integer and $M$ is the number of rows. The position of the $S$ ones are uniformly distributed in a column.

\section{Related Work}
\label{sec:related}

Existing PPCL approaches can be classified into two categories, i.e., distributed machine learning and training data encryption/obfuscation. \sect\ref{subsubsec:distributed-learning} and \sect\ref{subsubsec:data-encryption} review the two categories; \sect\ref{subsubsec:other-related} reviews other related work.

\subsection{Distributed Machine Learning (DML)}
\label{subsubsec:distributed-learning}

DML approaches exploit the computing capability of the participants to solve Eq.~(\ref{eq:learning}) using some variant of stochastic gradient descent (SGD) in a distributed manner. During the learning process, the training data samples are not transmitted. The studies \cite{Hamm15,Shokri15,mcmahan2016communication,mcmahan2018learning} share the similar idea of exchanging gradients and classifier parameters among the participants, which is coordinated by the coordinator. Specifically, in the Crowd-ML approach \cite{Hamm15}, a participant checks out the global classifier parameters $\boldsymbol\theta$ from the coordinator and computes the gradients using its own training data. Then, the participants transmit the gradients to the coordinator that will update $\boldsymbol\theta$.
In \cite{Shokri15}, each participant trains a local deep model using SGD and uploads a selected portion of gradients to the coordinator for combining. Then, each participant downloads a selected portion of the global gradients to update its local deep model.
As the exchanged gradients and classifier parameters may still contain privacy, the approaches \cite{Hamm15,Shokri15} add random noises to the exchanged values for differential privacy \cite{Dwork06}.
In the {\em federated learning} scheme \cite{mcmahan2016communication}, the coordinator periodically pulls the deep models trained by the participants locally based on their training data and returns an average deep model to the participants. In \cite{mcmahan2018learning}, the participant adds random noises to the deep model parameters before being sent to the coordinator for privacy protection in the federated learning process.

However, the above DML approaches have the following limitations. First, the local training introduces computation overhead to the participants. Training a DNN locally may be infeasible for resource-constrained IoT objects.
Second, DML approaches often require many iterations for the learning algorithm to converge, which may incur a high volume of data traffic between each participant and the coordinator. In \sect\ref{sec:implementation}, we will show this by comparing the Crowd-ML \cite{Hamm15} and our proposed approach. Third, as shown recently in \cite{Hitaj17}, generative adversarial networks can generate prototypical training data samples based on the exchanged gradients and model parameters, weakening the privacy preservation claimed in \cite{Shokri15,mcmahan2016communication}. In \cite{Phong17} and \cite{Bonawitz17}, homomorphic encryption and secure aggregation have been applied to enhance the privacy preservation of the approach in \cite{Shokri15} and the federated learning in \cite{mcmahan2016communication}, respectively. With these enhancements, only the encrypted gradients \cite{Phong17} and aggregate model update \cite{Bonawitz17} are revealed to the honest-but-curious coordinator. However, these privacy enhancements further increase the computation overhead of each participant, making it more unsuitable for resource-constrained IoT objects.





\subsection{Training Data Encryption/Obfuscation}
\label{subsubsec:data-encryption}

Different from the DML approaches that transmit classifier's parameters, the approaches in \cite{graepel2012ml,liu2012cloud,shen2018privacy} transmit the encrypted or obfuscated training data to the coordinator to solve Eq.~(\ref{eq:learning}). The approach proposed in this paper also belongs to this category. In the following, we review each of \cite{graepel2012ml,liu2012cloud,shen2018privacy} and then discuss our new design to overcome their shortcomings.

In \cite{graepel2012ml}, homomorphic encryption is integrated with a Linear Means classifier and Fisher's Linear Discriminant classifier. During both the training and classification phases, the participant transmits the homomorphically encrypted data vector to the coordinator. However, homomorphic encryption results in intensive computation and increased volume of data transmissions (cf.~\sect\ref{sec:implementation}). {\blue We now present more details of the high overhead of homomorphic encryption. An integer is often represented by 4 bytes. If a 4-byte integer is homomorphically encrypted using the scheme presented in \cite{graepel2012ml} with default settings, the encrypted cipher text will be 65,536 bytes (specifically, 4,096 coefficients each represented by a 128-bit integer). Thus, the cipher text for a 1MB training data set in plain text will be nearly 16.4GB.  Moreover, the cipher text space is the ring of polynomials modulo a cyclotomic polynomial, with coefficients from a large integer ring (e.g., 128-bit integers). Meanwhile, general arithmetic operations are much more costly than  \revised{the standard arithmetic because a large amount of} polynomial arithmetics related to coefficients introduce the additional overhead of modulo operations on both the coefficients and polynomial \cite{aslett2015review}. }Thus, although the homomorphic encryption approach provides provable confidentiality protection, it is infeasible on many resource-constrained IoT platforms.

To reduce the computation and communication overheads, Liu et al. \cite{liu2012cloud} propose a data obfuscation approach based on random projection. Specifically, the participant $i$ independently generates a Gaussian random matrix $\mathbf{R}_i$ and transmits the obfuscated training dataset $\{(\mathbf{R}_i\vec{x}_{i,j}, y_{i,j}) | j \in \{1,...,M_i\} \}$ to the coordinator.
However, different from Property~\ref{property:1} in \sect\ref{subsec:random-projection} that requires the same projection matrix, the approach \cite{liu2012cloud} uses distinct projection matrices for different participants and thus no longer preserves the Euclidean distance, i.e., $\| \mathbf{R}_{u} \vec{x}_{u,p} - \mathbf{R}_{v} \vec{x}_{v,q} \| \neq \| \vec{x}_{u,p} - \vec{x}_{v,q} \|$. This will result in poor training performance for distance-based classifiers, such as $k$-nearest neighbors and SVM. To address this issue, the study \cite{liu2012cloud} designs a regression phase before the learning phase. Specifically, the coordinator sends a number of {\em public data vectors} $\{\vec{z}_k | k=1, 2, \ldots \}$ to all participants and the participant $i$ returns the projected data $\{ \mathbf{R}_i \vec{z}_k | k=1, 2, \ldots \}$. Based on the original and projected public data vectors, a regress function $f_{uv}(\cdot, \cdot)$ for each participant pair $(u,v)$ is learned such that $f_{uv}(\mathbf{R}_u \vec{x}_{u,p}, \mathbf{R}_v \vec{x}_{v,q}) \simeq \| \vec{x}_{u,p} - \vec{x}_{v,q} \|$. {\blue With the regress function $f_{uv}(\cdot, \cdot)$  that can estimate the distance in the original space based on the projected data vectors, the distance-based SVM and $k$-nearest neighbors ($k$-NN) classifiers can be still trained based on the projected data. Specifically, whenever the training algorithm needs the distance between two original data vectors, the regress function is used to compute the distance based on the projected data vectors. As a result, the distance-based classifiers can be trained in the domain of obfuscated data by using the learned regress functions during the training phase.}

However, the approach \cite{liu2012cloud} has two shortcomings. First, it is only applicable to distance-based classifiers. These conventional classifiers do not scale well with the volume of the training data and the complexity of the data patterns \cite{suykens2003}. It is desirable to support the DNNs that give the state-of-the-art learning performance in a range of applications. Second, obfuscating the public data vectors and returning the results may incur known-plaintext attacks and engender {\revised{a clear privacy concern}}. For instance, a proactively curious coordinator may use a public data vector $\vec{z}_k = [1, 0, 0, \ldots, 0]^\intercal$ to extract the first column of $\mathbf{R}_i$. Other columns of $\mathbf{R}_i$ can be similarly extracted by using specific public data vectors. Even without using these specific public data vectors, in general, the private random projection matrix $\mathbf{R}_i$ can be estimated using regression analysis based on a number of public data vectors and the corresponding projections.




The study \cite{shen2018privacy} also uses random projection to obfuscate the data vector $\vec{x}$ in training and executing a Sparse Representation Classifier. However, all participants use the same random projection matrix, rendering the system vulnerable to the collusion between any single participant and the coordinator.

 Different from \cite{shen2018privacy}, each participant in our approach uses its own private random project matrix, rendering the collusion futile. Different from \cite{liu2012cloud}, our approach uses DNNs and leverages on the deep learning capability to avoid the regression phase that is vulnerable to the known-plaintext attacks. Different from \cite{graepel2012ml} that is too compute-intensive for IoT objects, our approach uses random projection that introduces light computation overhead only.

\subsection{Other Related Work}
\label{subsubsec:other-related}

In CryptoNets \cite{gilad2016cryptonets}, the computation of each neuron in a neural network trained using plaintext data is performed in the domain of homomorphic encryption. During the classification phase, the participant sends the homomorphically encrypted data to the coordinator for classification. The work \cite{chabanne2017privacy} extends \cite{gilad2016cryptonets} to support more hidden layers. However, these studies \cite{gilad2016cryptonets,chabanne2017privacy} address {\em privacy-preserving classification outsourcing} (i.e., offloading the classification computation to a honest-but-curious entity), rather than the collaborative learning addressed in this paper. The training in \cite{gilad2016cryptonets,chabanne2017privacy} is performed based on plaintext data. Moreover, the homomorphic encryption is too compute intensive for resource-constrained IoT devices, which will be shown in \sect\ref{sec:implementation}.




The {\em differentially private machine learning} (DPML) \cite{Abadi16,chaudhuri2009privacy,song2013stochastic} builds a classifier that cannot be used to infer the training data. The training of the classifier is based on plaintext data. For DNNs, DPML can be achieved by perturbing the gradients in each iteration of the SGD with additive noises \cite{Abadi16,song2013stochastic}. DPML and PPCL address different problems: PPCL preserves the privacy of the training data against the honest-but-curious coordinator who builds the classifier, whereas DPML trusts the classifier builder and preserves the privacy of the training data against the curious user of the classifier. Thus, in DPML, the plaintext training dataset is available to the classifier builder; differently, in PPCL, only encrypted or obfuscated training data is made available to the classifier builder (i.e., the learning coordinator).

{\blue Truex et al. \cite{truex2019hybrid} propose an alternative approach that utilizes both the differential privacy and secure multiparty computation (SMC) to balance various trade-offs in federated learning. The proposed federated learning system is a scalable approach that is secure against inference threats and produces models with high accuracy. However, it is not suitable for resource-constrained IoT due to the high computational overhead of SMC.}







\section{Problem Statement and Approach}
\label{sec:mov}

In this section, we state the PPCL problem in \sect\ref{subsec:problem-statement} and present the proposed independent random projection approach in \sect\ref{subsec:approach-overview}. \sect\ref{subsec:example} provides two illustrating examples for insights into understanding the effect of GRP on training DNN-based classifiers. \sect\ref{subsec:alternatives} discusses two other alternative approaches for lightweight PPCL and their limitations.

\subsection{Problem Statement}
\label{subsec:problem-statement}

\begin{figure}
  \centering
  \includegraphics{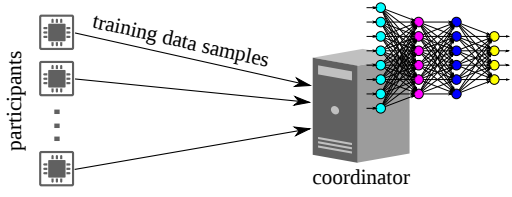}
  \caption{A collaborative learning system.}
  \label{fig:system}
\end{figure}

In this paper, we consider a PPCL system with $N$ resource-constrained {\em participants}  and an honest-but-curious {\em coordinator} with sufficient computation power. {\blue We assume that the data distributed among the participants is homogeneous. Thus, the participants will contribute data in the same format.} Fig.~\ref{fig:system} illustrates the system. During the learning phase, the participants contribute training data samples to build a supervised classifier.
As discussed in \sect\ref{subsec:background}, the training dataset $\mathcal{D}_i$ contributed by the participant $i$ consists of $M_i$ data vectors $\{\vec{x}_{i,j} | j \in \{1,...,M_i\} \}$ and the corresponding class labels $\{y_{i,j} | j \in \{1,...,M_i\} \}$.
As the learning process is often compute-intensive, most of the learning computation should be accomplished by the coordinator. In this paper, we focus on addressing the problem of building an effective supervised classifier while protecting certain privacy contained in the data vectors. We now discuss several aspects of the problem.


The privacy concern regarding the data vectors is primarily {\revised{due to the fact that}} the data vectors may contain information beyond the classification objective in question. For example, consider a PPCL system for training a classifier to recognize human body activity (e.g., sitting, walking, climbing stairs, etc). The recognition is based on various body signals (e.g., motion, heart rate, breath rate, etc) that are captured by wearable sensors. However, the raw body signals can also be used to {\revised{infer the health statuses}} of the participants and even {\revised{pinpoint which people have certain diseases.}}

In this paper, we adopt the following threat and privacy models.




\begin{description}
\item[Threat model:] It consists of the following {three} aspects:
\begin{itemize}
\item {\em Honest-but-curious coordinator:} We assume that the coordinator will honestly coordinate the collaborative learning process, aiming to train the best supervised classifier. Thus, it will neither tamper with any data  collected from or transmitted to the participants.
  However, the coordinator is curious about the participants' private information contained in the training data vectors. {\blue The coordinator may analyze the data received from the participants to infer the participants' privacy. For instance, the coordinator may attempt to reconstruct and manually inspect the original data captured by the participants.}
  \item {\em Potential collusion between participants and coordinator:} We assume that the participants are not trustworthy in that they may collude with the coordinator in finding out {\revised{other participants’ private information contained in}} the data vectors. The colluding participants are also honest, i.e., they will faithfully contribute their training data to improve the supervised classifier. {\blue However, the colluding participants may reveal the details of the adopted privacy-preservation approach to the coordinator. Thus, the design of the PPCL system should maintain the privacy  for a participant {\revised{when any or all of the other}} participants are colluding with the coordinator.
  \item {\em No known input-output attack on non-colluding participants:} We assume that the coordinator cannot launch the known input-output attack on the non-colluding participants due to the following reasons. First, the coordinator cannot access the original data stored at the participants. Second, in our PPCL approach, the communication channel is merely used for uploading obfuscated data samples and their labels. Thus, in our approach, there is no way for the coordinator to obtain the original input of a non-colluding participant. This is different from the approach \cite{liu2012cloud} in which each participant also uses the communication channel to respond to the coordinator's queries by returning obfuscated public data vectors. Without the known input-output attack, it is computationally difficult (practically impossible) for the coordinator to meaningfully estimate the projection matrix and reconstruct the original data vector \cite{rachlin2008secrecy,liu2006random}. Note that, as the participants apply independent Gaussian random projections, the collusion between some participants and the curious coordinator will not enable the known input-output attack on the non-colluding participants.
}
  \end{itemize}
\end{description}

\begin{description}
\item[Privacy model:] {\revised{The raw form of each data vector contains the participant’s private information (e.g., health status) and must be protected from snooping by the curious coordinator.}} The error in estimating the data raw form by the coordinator can be used as a metric to measure the degree of privacy protection. Data form confidentiality is an immediate and basic privacy requirement in many applications.
\end{description}




We now discuss four issues that are related to privacy protection and threat model.

\begin{itemize}
\item {\em Training data anonymization:} We aim to support anonymization of the training data. That is, the coordinator should not expect to know the participant's identity for any received training data sample. Moreover, the coordinator cannot determine whether any two training data samples are from the same participant. To achieve the above anonymity, the training data samples can be transmitted in separate sessions via an anonymous communication network \cite{danezis2008survey}.  Moreover, the transmissions of the data samples from all participants can be interleaved randomly, such that the coordinator cannot associate the data samples from the same participant by their arrival times. Note that the training data anonymization requirement is not mandatory, because the anonymous communication may incur large overhead for some resource-constrained IoT objects. However, the design of our PPCL approach will not leverage the participants' identities to support data anonymization.
\item {\em Label privacy:} The class labels $\{y_{i,j} | j \in \{1,...,M_i\} \}$ may also contain information about the participant. In this paper,  we do not consider label privacy because the participant willingly contributes the labeled data vectors and should have no expectation of privacy regarding labels. In practice, several means can be taken to mitigate the concern of label privacy leak. First, the training data anonymization mitigates the concern during the learning phase.
  Second, during the classification phase, if the participant has sufficient processing capability to perform the classification computation, the coordinator may send the trained model to the participant for local execution. Existing studies have enabled the execution of deep models on personal and low-end devices \cite{yao2017deepiot,huynh2017deepmon}. Low-power inference chips (e.g., Google's Edge TPU \cite{edge-tpu}) will further enhance low-end devices' capabilities in executing classification models. Note that the studies \cite{yao2017deepiot,huynh2017deepmon} and the inference chips are not to support the much more compute-intensive training.
\item {\em Other privacy models:} Differential privacy \cite{Dwork06} aiming at achieving indistinguishability of different data vectors is another widely used quantifiable privacy definition. However, as discussed in \sect\ref{subsec:alternatives} and evaluated in \sect\ref{sec:evaluation}, the additive noisification implementation of differential privacy is ill-suited for PPCL.
\item {\blue Tramer et al.'s work \cite{tramer2016stealing} focuses on the threat of model extraction and reversal to duplicate the functionality of the model. Differently, we focus on the threat from the coordinator on the participants' data privacy. In our problem formulation, the deep model trained by the coordinator is also available to the coordinator. Tramer et al.'s work is applicable to the external threats that aims at extracting the coordinator's model. Thus, their work is out of the scope of this paper.}
\end{itemize}


\subsection{Gaussian Random Projection Approach}
\label{subsec:approach-overview}

Existing DML and homomorphic encryption approaches incur significant computation and communication overhead due to the many computation/communication rounds and data volume swell. In \sect\ref{sec:implementation}, we will provide benchmark results to show this. Thus, these approaches are not promising for resource-constrained participants.
This section describes a GRP-based approach that is computationally lightweight and communication efficient for the participants. The overview of our approach is presented as follows.

At the system initialization, each participant $i$ independently generates a random Gaussian matrix $\mathbf{R}_i \in \mathbb{R}^{k \times d}$, where $d$ is the dimension of the data vector.
During the learning phase, the participant $i$ keeps $\mathbf{R}_i$ secret and uses it to project all the training data vectors. The participant $i$ transmits the projected training dataset $\mathcal{D}_i = \{\mathbf{R}_i\vec{x}_{i,j}, y_{i,j} | j \in \{1,..., M_i\}, y_{i,j} \in \mathcal{C} \}$ to the coordinator. After collecting all projected training datasets $\mathcal{D}_i$, $i = 1, \ldots, N$, the coordinator applies deep learning algorithms to train the classifier $h(\cdot | \boldsymbol\theta^*)$. During the classification phase, the participant $i$ still uses $\mathbf{R}_i$ to project the test data vector $\vec{x}$ and obtains the classification result $h(\mathbf{R}_i\vec{x} | \boldsymbol\theta^*)$. As discussed in \sect\ref{subsec:problem-statement}, the classification computation can be carried out at the participant or the coordinator, depending on whether the participant is capable of executing the trained deep model. In our approach, each participant independently generates its random projection matrix to counteract the collusion between participants and coordinator. Now, we explain the two key components of our approach: GRP and deep learning on projected data.

\subsubsection{Gaussian random projection}
\label{subsubsec:projection}


In this work, we mainly consider Gaussian matrices. Specifically, each element of $\mathbf{R}_i$ is sampled independently from the standard normal distribution \cite{Ailon09}. The rationale of choosing Gaussian matrices will be explained in \sect\ref{subsubsec:10d-example}.
We set the row dimension of $\mathbf{R}_i$ smaller than or equal to its column dimension, i.e., $k \le d$. Thus, the GRP can also compress the data vector. We define the compression ratio as $\rho = d / k$. The understanding regarding the admission of compression into the training data projection is as follows. From the compressive sensing theory \cite{candes2008introduction}, a sparse signal can be represented by a small number of linear projections of the original signal and recovered faithfully. Therefore, in the compressively projected data vector, the feature information still exists, provided that the adopted compression ratio is within an analytic bound \cite{candes2008introduction}.
In \sect\ref{sec:evaluation}, we will evaluate the impact of the compression ratio $\rho$ on the learning performance.


With GRP, if $\mathbf{R}_i$ is kept confidential to the coordinator, it is computationally difficult (practically impossible) for the coordinator to generate a meaningful reconstruction of the original data vector from the projected data vector \cite{liu2006random,rachlin2008secrecy}. Thus, GRP protects the form of the original data. {\blue With sufficient pairs of input and output vectors, the coordinator can train a well-designed deep neural network (e.g., the decoder of an autoencoder) to reconstruct the raw forms of original data vectors. However, as discussed in \sect\ref{subsec:problem-statement}, the coordinator cannot launch the known input-output attack in our considered context. }
In the worst case where the coordinator obtains $\mathbf{R}_i$, the estimation error given by Property~\ref{property:2} in \sect\ref{subsec:random-projection} can be used as a measure of privacy protection.
Random projection has been used as a lightweight approach to protect data form confidentiality in various contexts \cite{li2013compressed,tan2017joint,wang2013privacy,xue2017kryptein}.

\subsubsection{Deep learning on projected data}
\label{subsubsec:dl}

Feature extraction is a critical step of supervised learning. With the traditional {\em shallow learning}, the classification system designer needs to handcraft the feature. {\blue As an example, in the study \cite{liu2012cloud}, the system trains a regress function to recover the Euclidean distance between any two projected samples as the feature. However, the training of the regress function creates a privacy vulnerability as discussed in \sect\ref{subsubsec:data-encryption}. Our approach uses deep learning to avoid involving feature engineering that can potentially introduce privacy vulnerabilities.} The emerging deep learning method \cite{lecun2015deep} automates the design of feature extraction by {\em unsupervised feature learning}, which is often based on a neural network consisting of a large number of parameters. Thus, the deep model is often a tandem of the feature extraction stage and the classification stage. For example, a convolutional neural network (CNN) for image classification consists of convolutional layers and dense layers, which are often considered performing the feature extraction and classification, respectively.

Our approach {\blue utilizes} the unsupervised feature learning capability of deep learning to address the data distortion introduced by the GRP. We now illustrate this using a simple example system, in which there is only one participant and the projection matrix $\mathbf{R}$ is a square invertible matrix. Moreover, we make the following two assumptions to simplify our discussion. First, we assume that a linear transform $\boldsymbol\Psi \in \mathbb{R}^{f \times d}$ gives effective features of the data vectors, where $f$ is the feature dimension. That is, $\vec{f} = \boldsymbol\Psi \vec{x}$ is an effective representation of the data vector $\vec{x}$ for classification. Second, we assume that $\boldsymbol\Psi$ can be learned in the form of a neural network by the unsupervised feature learning. Now, we discuss the impact of the random projection on the unsupervised feature learning. After the projection, the data vector becomes $\mathbf{R} \vec{x}$.
Moreover, the linear transform $\boldsymbol\Psi \mathbf{R}^{-1}$ will be an effective feature extraction method, since $\vec{f} = \left( \boldsymbol\Psi \mathbf{R}^{-1} \right) \left( \mathbf{R} \vec{x} \right)$. It is reasonable to expect that the unsupervised feature learning can also build a neural network to capture the linear transform $\boldsymbol\Psi \mathbf{R}^{-1}$, similar to the unsupervised feature learning to capture the $\boldsymbol\Psi$ based on the plaintext training data $\vec{x}$. {\blue When the projection matrix is non-invertible, we may consider its {\em pseudoinverse} denoted by $\mathbf{R}^{+}$ \cite{ben2003generalized}. As the Gaussian random projection matrix is most likely of full rank \cite{paige1982lsqr}, the linear transform $\boldsymbol\Psi \mathbf{R}^{+}$ can be regarded as an effective feature extraction. Similarly, it is reasonable to assume that the unsupervised feature learning can capture the linear transform $\boldsymbol\Psi \mathbf{R}^{+}$ by a neural network.  }As a result, the deep model trained using the projected data can still classify future projected data vectors. In \sect\ref{subsec:example}, we will use a numerical example to illustrate this.

The above discussion based on linear features provides a basis for us to understand how the unsupervised feature learning helps address the distortion caused by the GRP. In practice, effective feature extractions are generally non-linear mappings. Neural network-based deep learning has shown strong capability in capturing sophisticated features beyond the above ideal linear features. In this paper, based on multiple datasets, we investigate the effectiveness of deep learning to address the distortion caused by the GRP.



As discussed earlier, each participant independently generates a Gaussian matrix to counteract the potential collusion between participants and the coordinator. However, this introduces a challenge to deep learning, because the pattern for a class of projected data vectors from $N$ participants will be a composite of $N$ different patterns. Thus, intuitively, a deeper neural network and a larger volume of training data will be needed to well capture the data patterns with increased complexity due to the participants' independence in generating their projection matrices. {\blue The participants' independence can also cause the following possible situation leads to classification errors: $\mathbf{R}_u \vec{x}_u = \mathbf{R}_v \vec{x}_v$, where $\vec{x}_u$ and $\vec{x}_v$ are respectively generated by participants $u$ and $v$ and belong to different classes. However, the probability of the above situation is low, especially when the data vectors are of high dimension. Instead, the overlaps between the distributions of any two classes' projected data vectors should receive attention. Fortunately, advanced machine learning algorithms such as SVM and deep learning can learn the mapping from the space of the input data in which the classes overlap to a different space possibly with higher dimensions in which the classes are separated. This issue will be discussed in detail with examples in \sect\ref{subsubsec:toysoverlap}. Nevertheless, }the more complex data patterns due to the independent projection matrix generation do cause a challenge. In this paper, we conduct extensive experiments to assess how well deep learning can scale with the number of participants, compared with the traditional learning approaches.




\subsection{Illustrating Examples}
\label{subsec:example}
In this section, we present a number of examples to illustrate the intuitions discussed in \sect\ref{subsec:approach-overview}.

\subsubsection{A 2-dimensional example}
\label{subsec:toys}

\begin{figure*}
  \centering
  \subfigure[Original data]
  {
    \includegraphics[width=0.2\textwidth]{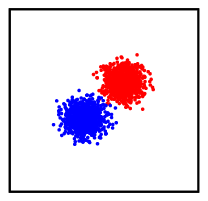}
   \label{fig:toy1-original}
  }
  \subfigure[Participant 1]
  {
    \includegraphics[width=0.2\textwidth]{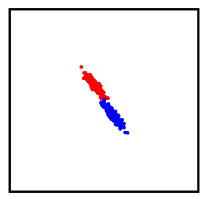}
    \label{fig:toy1-p1}
 }
  \subfigure[Participant 2]
 {
    \includegraphics[width=0.2\textwidth]{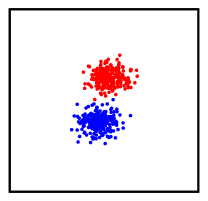}
    \label{fig:toy1-p2}
  }
  \subfigure[Participant 3]
  {
    \includegraphics[width=0.2\textwidth]{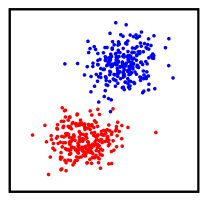}
   \label{fig:toy1-p3}
 }
  \subfigure[Participant 4]
  {
    \includegraphics[width=0.2\textwidth]{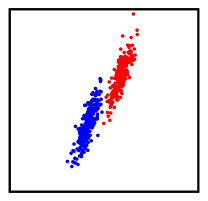}
    \label{fig:toy1-p4}
  }
  \subfigure[Coordinator]
  {
    \includegraphics[width=0.2\textwidth]{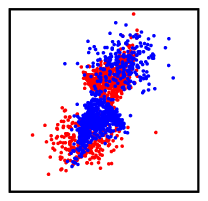}
    \label{fig:toy1-projection1}
  }
  \subfigure[Coordinator]
  {
    \includegraphics[width=0.2\textwidth]{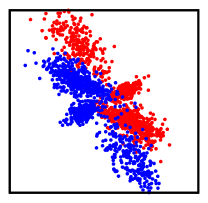}
    \label{fig:toy1-projection2}
  }
  \subfigure[Coordinator]
  {
    \includegraphics[width=0.2\textwidth]{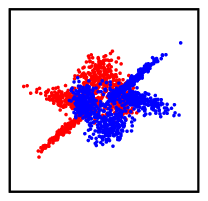}
    \label{fig:toy1-projection3}
  }
  \caption{Two-dimensional example. Original data vectors and projected data vectors (red: class 0; blue: class 1). The ranges for the $x$ and $y$ axes are $[-10,10]$.}
  \label{fig:random_projection}
\end{figure*}

We consider a PPCL system with four participants (i.e., $N=4$) to build a two-class classifier. The original data vectors in the two classes follow two 2-dimensional Gaussian distributions with means of $[-2,-2]^\intercal$ and $[2,2]^\intercal$, and the same covariance matrix of $[1, 0; 0, 1]$. Fig.~\ref{fig:toy1-original} shows the plaintext data vectors generated by the four participants. From the figure, the plaintext data vectors of the two classes can be easily separated using a simple hyperplane. Each participant independently generates a Gaussian random matrix. Figs.~\ref{fig:toy1-p1}-\ref{fig:toy1-p4} show the projected data vectors of each participant. We can see that the patterns of the projected data vectors are different across the participants. Fig.~\ref{fig:toy1-projection1} shows the mixed projected data vectors received from all participants. Compared with Fig.~\ref{fig:toy1-original}, the pattern of the mixed projected data from all participants is highly complex. Moreover, no simple hyperplane can well divide the two classes.

We also generate two other sets of the random projection matrices for all participants. Figs.~\ref{fig:toy1-projection2} and \ref{fig:toy1-projection3} show the mixes of all participants' projected data vectors with the two sets of random projection matrices, respectively. Similarly, the pattern of the mixed projected data from all participants is highly complex.

\begin{figure}
  \centering
  \includegraphics[width=0.36\textwidth]{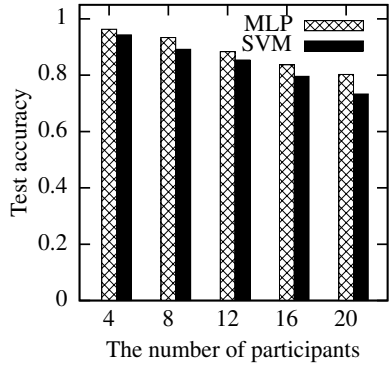}
  \caption{Test accuracy based on projected data vs. the number of participants.}
  \label{fig:toydata1}
\end{figure}

We construct a classifier based on an MLP with two hidden layers of 30 and 40 rectified linear units (ReLUs), respectively. The input layer admits a 2-dimensional data vector, whereas the output layer consists of two ReLUs. The final classification result is generated using a softmax function based on the output layer's ReLU values. Moreover, we construct an SVM classifier as a baseline approach. We use LIBSVM \cite{libSVM} to implement the classifier. The SVM classifier uses {\blue the} radial basis function (RBF) kernel with two configurable parameters $C$ and $\lambda$. During the training phase, we apply grid search to determine the optimal settings for $C$ and $\lambda$.

First, we use disjoint subsets of the original data shown in Fig.~\ref{fig:toy1-original} to train and test the MLP and SVM classifiers. Both classifiers can achieve 99\% test accuracy. This shows that the MLP and the SVM are properly designed for the 2-dimensional data vectors.

Then, we use disjoint subsets of the randomly projected data shown in Fig.~\ref{fig:toy1-projection1} to train and test the MLP and SVM classifiers. Moreover, we also increase the number of participants in the PPCL system. Fig.~\ref{fig:toydata1} shows the test accuracy versus the number of participants. We can see that the MLP classifier always outperforms the SVM classifier. Moreover, the test accuracy decreases with the number of participants. This is because, with more participants, the pattern of the projected data becomes more complex, introducing challenges to both MLP and SVM. The mean test accuracy difference between MLP and SVM increases from 2\% to 7\%, when the number of participants increases from 4 to 20. This result is also consistent with the understanding that deep learning is more effective in capturing complex patterns than traditional learning.
{\blue
\subsubsection{Impact of inter-class overlaps on learning performance}
\label{subsubsec:toysoverlap}
 After the participants apply independent GRPs, the consolidated training samples at the coordinator may have inter-class overlaps. We conduct a set of numerical experiments based on the previous 2-class 2-dimensional example system to investigate the impact of the inter-class overlaps on the learning performance.}

\begin{figure*}
	\centering
	\subfigure[{\blue CDFs of Euclidean distance between any two data vectors respectively from the two classes.}]
	{
		\includegraphics[width=0.28\textwidth]{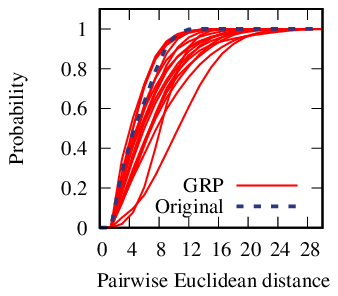}
		\label{fig:expected_distancecdf}	
	}
    \hspace*{1em}
    \subfigure[{\blue Projected data vectors in the most overlapped case.}]
    {
	\includegraphics[width=0.28\textwidth]{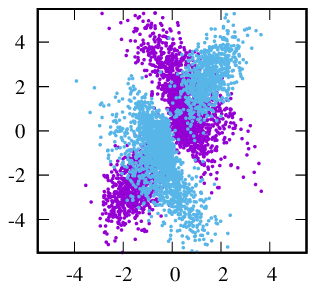}
	\label{fig:datasamplesthe}	
    }
    \hspace*{1em}
    \subfigure[{\blue Overlap rates of 20 cases.}]{
     \includegraphics[width=0.28\textwidth]{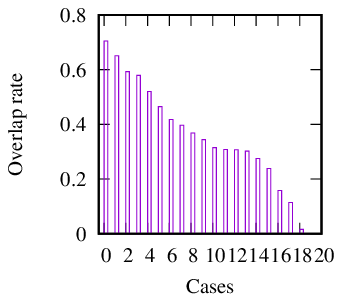}
     \label{fig:overlap}  
     }
     \caption{{\blue Impact of inter-class overlaps.}}
\end{figure*}

{\blue For each set of the random projection matrices, we compute the cumulative distribution function (CDF) of the Euclidean distance between any two projected data vectors respectively from the two classes. The solid curves in Fig.~\ref{fig:expected_distancecdf} are the CDFs, each corresponding to one set of the random projection matrices among the 20 sets. The dashed curve shows the CDF of the Euclidean distance between any two original data vectors respectively from the two classes. We can see that the solid curves are in general below the dashed curve, which suggests that the GRPs likely disperse the two classes in terms of inter-sample Euclidean distance.}

{\blue Fig.~\ref{fig:datasamplesthe} shows the consolidated data vectors after GRPs corresponding to the highest solid CDF curve shown in Fig.~\ref{fig:expected_distancecdf}. Among the 20 cases, the two classes in the case shown in Fig.~\ref{fig:datasamplesthe} are most overlapped. We quantify the inter-class overlap using a metric called \textit{overlap rate}. It is defined as the ratio of overlapped data vectors to all data vectors. A data vector is overlapped if there are $k$ data vectors of different classes within a distance of $r$  from the considered data vector. In this set of experiments, we set $k = 3 $, $r = 0.01$. Note that as the data vectors shown in Fig.~\ref{fig:datasamplesthe} are distributed in a $10 \times 10$ area, the distance threshold $r=0.01$ is a stringent requirement on the proximity of data vectors in defining overlap. Fig.~\ref{fig:overlap} shows the ordered overlap rates of the projected data in the 20 cases. The case shown in Fig.~\ref{fig:datasamplesthe} has the largest overlap rate, i.e., 0.705. For this most overlapped case, SVM and MLP achieve test accuracies of $87.24\%$ and $91.08\%$, respectively, which are still satisfactory. SVM projects the overlapped distributions of the classes to a space with a higher dimension, such that the higher-dimension data distributions of different classes can be separated by linear planes. Compared with SVM, MLP can better handle the overlaps among the data distributions of different classes. The above results show that, although different classes may have overlapped areas in the projected data domain, advanced machine learning algorithms such as SVM and MLP may still be able to differentiate the two classes.

}



\begin{figure}
  \centering
  \includegraphics[width=0.36\textwidth]{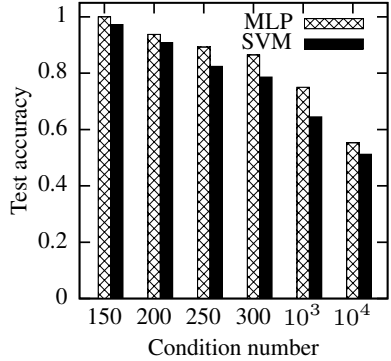}
  \caption{Test accuracy based on projected data vs. the condition number.}
  \label{fig:toydata2}
\end{figure}

\subsubsection{A 10-dimensional example}
\label{subsubsec:10d-example}

Now, we use another example system to understand the effect of deep learning's unsupervised feature learning capability in addressing the data distortion caused by the random projection. This example is a PPCL system with only one participant (i.e., $N=1$).
The original data vectors in two classes follow two 10-dimensional Gaussian distributions, with the $[-2,-2,\ldots,-2]^\intercal$ and $[2,2,\ldots,2]^\intercal$ as the respective mean vectors, and the 10-dimensional identity matrix as their identical covariance matrix.

In our discussions in \sect\ref{subsubsec:dl}, we assume that the projection matrix $\mathbf{R}$ is invertible and the unsupervised feature learning tend to capture $\boldsymbol\Psi \mathbf{R}^{-1}$. As learning algorithms are based on numerical computation on the training data, an ill-conditioned matrix $\mathbf{R}$ will impede efficient fitting of $\boldsymbol\Psi \mathbf{R}^{-1}$. We verify this intuition by assessing the learning performance of the single-participant PPCL system using different $\mathbf{R}$ matrices with varying condition numbers.
Specifically, by following a method described in \cite{bierlaire1991iterative}, the participant generates a random square matrix $\mathbf{R}$ that has a certain condition number value.
The condition number is defined as $\|\mathbf{R}\|_F \| \mathbf{R}^+ \|_F$ \cite{paige1982lsqr}, where $\mathbf{R}^+$ denotes the pseudoinverse of $\mathbf{R}$ and $\|\cdot\|_F$ represents the Frobenius norm.
Fig.~\ref{fig:toydata2} shows the test accuracy of the MLP and SVM classifiers trained using data projected by $\mathbf{R}$ versus the condition number of $\mathbf{R}$. Note that a larger condition number means that the matrix is more ill-conditioned. We can see that the test accuracy decreases with the condition number, consistent with the intuition.

\begin{figure}
  \centering
  \includegraphics[width=0.85\textwidth]{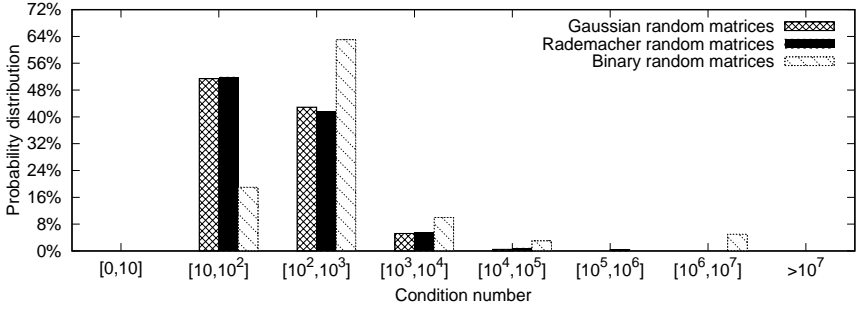}
  \caption{The distributions of the condition number of Gaussian, Rademacher, binary random matrices with dimension $28 \times 28$. }
  \label{fig:con1}
\end{figure}
\subsubsection{Condition numbers of various projection matrices}
\label{subsec:conv}
\sect\ref{subsubsec:10d-example} shows that the condition number affects the impact of the random projection on the learning performance. In this section, we compare the condition numbers of Gaussian, Rademacher, and binary random matrices. The comparison will help understand the superior learning performance of the GRP-based approach. In this section, the Gaussian, Rademacher, and binary random matrices have an identical dimension of $28 \times 28$. For each type of random matrix, we generate 1,000 instances and investigate the distribution of their condition numbers. Fig.~\ref{fig:con1} shows the distributions of the condition numbers for the three types of random matrices. We can see that the condition number distributions of Gaussian and Rademacher matrices are similar, while Rademacher's distribution has a longer tail. Specifically, the probability that a Rademacher matrix's condition number is within $[10^4, 10^5]$ is $0.8\%$. In contrast, the corresponding probability of Gaussian matrix's condition number within same range is $0.5\%$. In addition, a binary random matrix can be extremely ill-conditioned. For instance, as shown in Fig.~\ref{fig:con1}, the condition number of a binary random matrix can be up to $10^7$.
The study \cite{chen2005condition} has analyzed the distribution of the condition numbers of Gaussian random matrices. The results show that a Gaussian random matrix is well-conditioned with a high probability. For instance, it is shown in \cite{chen2005condition} that for a $10 \times 5$ Gaussian random matrix, the probability that its condition number is larger than 100 is less than $6 \times 10^{-7}$.
From the above discussions, Gaussian random matrices are preferred based on their condition numbers. However, Gaussian random projection has higher computation overhead than binary and Rademacher random projections. With a binary matrix defined in \sect\ref{subsec:random-projection}, the projection can be implemented using $SN-M$ addition operations. With a Rademacher matrix defined in \sect\ref{subsec:random-projection}, the projection can be implemented with $M(N-1)$ addition operations and just one multiplication operation. In contrast, GRP needs $M(N-1)$ additions and $MN^2$ multiplications. Thus, there is a trade-off between the condition of the chosen random matrix type and the associated computation overhead  that will be borne by the collaborative learning participants.






\subsection{Alternative Approaches and Limitations}
\label{subsec:alternatives}

This section discusses two alternative approaches to PPCL and their limitations. These two alternatives will be used as the baseline approaches in our comparative performance evaluation in \sect\ref{sec:evaluation}.

\subsubsection{Non-collaborative learning}
\label{subsubsec:nc-learn}

If the data anonymity requirement  is not enforced, the coordinator can train a separate deep model based on the projected data vectors contributed by each participant. This alternative approach can address the challenge of the complex mixed patterns due to different random projection matrices adopted by different participants as illustrated in \sect\ref{subsec:example}. However, it loses the advantages of collaborative learning, i.e., the increased data volume and pattern coverage. From our evaluation in \sect\ref{sec:evaluation}, compared with our proposed approach, despite that this non-collaborative learning approach additionally uses the participant identity information, it yields inferior average accuracy.

\subsubsection{Differential privacy}
\label{subsubsec:dp}

Differential privacy (DP) \cite{Dwork06} is a rigorous information-theoretic approach to prevent leak of individual records by statistical queries on a database of these records. The $\epsilon$-DP \cite{Dwork06} is formally defined as follows:
\begin{definition}
	A randomized algorithm $\mathcal{A}: \mathbb{D}\rightarrow \mathbb{R}^t$ gives $\epsilon$-DP if for all adjacent datasets $D_1 \in \mathbb{D}$ and $D_2 \in \mathbb{D}$ differing on at most one element, and all $S\subseteq Range(\mathcal{A})$, $\Pr(\mathcal{A}(D_1) \in S)\leq \exp(\epsilon)\cdot \Pr(\mathcal{A}(D_2) \in S)$.
\end{definition}

 The $\epsilon$, a positive real number, is a measure of privacy loss, i.e., a smaller $\epsilon$ implies better privacy. When $\epsilon$ is very small, $\Pr(\mathcal{A}(D_1) \in S) \simeq \Pr(\mathcal{A}(D_2) \in S)$ for all $S \subseteq Range(\mathcal{A})$, which means that the query results $\mathcal{A}(D_1)$ and $\mathcal{A}(D_2)$ are almost indistinguishable based on any ``test criterion'' of $S$.
The indistinguishability between the query results $\mathcal{A}(D_1)$ and $\mathcal{A}(D_2)$ decreases with $\epsilon$. The study \cite{Dwork061} develops the {\em Laplace mechanism} of adding Laplacian noises to implement $\epsilon$-DP. Specifically, for all function $\mathcal{F}: \mathcal{D} \rightarrow \mathbb{R}^t$, the randomized algorithm $\mathcal{A}(D) = \mathcal{F}(D) + [n_1, n_2, \ldots, n_t]^\intercal$ gives $\epsilon$-DP, where each $n_i$ is drawn independently from a Laplace distribution $\mathrm{Lap}(S(\mathcal{F})/\epsilon)$ and $S(\mathcal{F})$ denotes the global sensitivity of $\mathcal{F}$. Note that $\mathrm{Lap}(\lambda)$ denotes a zero-mean Laplace distribution with a probability density function of $f(x|\lambda)=\frac{1}{2\lambda}e^{\frac{|x|}{\lambda}}$; the global sensitivity is
\begin{equation*}
S(\mathcal{F})=\max_{\forall D' \in \mathbb{D}, \forall D'' \in \mathbb{D}}||\mathcal{F}(D')-\mathcal{F}(D'')||_1.
\end{equation*}

Essentially, $\epsilon$-DP gives quantifiable indistinguishability of the query results based on different datasets. The $\epsilon$-DP framework has been applied in various privacy preservation problems in machine learning. As discussed in \sect\ref{subsubsec:distributed-learning}, the DML approaches to PPCL \cite{Hamm15,Shokri15} add random noises to the parameters exchanged between the participants and the coordinator to achieve $\epsilon$-DP. The original parameters can be viewed as deterministic query results of the training data. Adding random noises to the parameters ensures certain levels of indistinguishability between the noise-added parameters based on different training datasets.
The achieved $\epsilon$-DP mitigates the privacy concern that the curious coordinator may use the received parameters to infer the existence of particular data vectors in the training dataset.
However, these DML approaches \cite{Hamm15,Shokri15}
incur significant overhead to resource-constrained participants.
For PPCL based on resource-constrained participants, an approach to achieving $\epsilon$-DP is to add a Laplacian noise vector to the original data vector $\vec{x}$ and then transmit the noise-added data vector to the coordinator for building the classifier. By doing so, certain levels of indistinguishability between the noise-added data vectors based on different original data vectors are achieved.

{\blue The recently proposed local differential privacy (LDP) \cite{bebensee2019local} is an $\epsilon$-DP realization different from the Laplace mechanism. It allows statistical computation while protecting each individual user's privacy. As LDP does not require the global sensitivity, it does not depend on the trust in a central authority, which presents practical advantages. However, as shown in \cite{erlingsson2014rappor}, LDP needs greater noise levels than the Laplace mechanism and thus reduces the utility of data. Google has implemented LDP in the RAPPOR project \cite{erlingsson2014rappor}. We apply RAPPOR to achieve LDP in this paper.}

Additive noisification and multiplicative GRP preserve different forms of privacy. Compared with protecting indistinguishability under the DP framework, we believe that protecting the confidentiality of the raw data form, which can be achieved by GRP, is a more immediate and basic privacy requirement in many applications.
The additive noisification, though achieving $\epsilon$-DP, falls short of protecting the confidentiality of the raw data form. Specifically, under the $\epsilon$-DP framework based on zero-mean Laplacian noises, a noise-added data vector can be considered an unbiased estimate of the original data vector with an estimation variance related to $\epsilon$. Thus, the coordinator always has a meaningful (i.e., unbiased) estimate of the raw data.
According to Property~\ref{property:2} in \sect\ref{subsec:random-projection}, this only happens to the GRP approach in the worst (and unrealistic) case that the projection matrix is revealed to the coordinator; other than the worst case, the coordinator cannot have a meaningful estimate of the raw data form. In the image classification case studies in \sect\ref{sec:evaluation}, we will show that when $\epsilon$ is small (i.e., good DP), the contents of the noise-added images can still be interpreted. In contrast, the projected images cannot be interpreted visually at all.


Applying $\epsilon$-DP to PPCL with resource-constrained participants also introduces the following two challenges:

\begin{itemize}
\item {\em Non-trivial computation overhead:} From the DP theory, an independent random noise vector should be generated and added to every data vector $\vec{x}$. However, random number generation is often a costly operation due to the use of various mathematical functions.
The continuous generation of Laplacian noises will incur non-trivial computation overhead for the resource-constrained participants. Differently, in our approach, the random projection matrix generation is a one-off overhead. The projection to compute $\mathbf{R}\vec{x}$ is a lightweight operation consisting of multiplications and additions only.  Our previous work \cite{tan2017joint} has implemented the projection operation on an MSP430-based platform. Moreover, the projection can be sped up if a parallel computing chip (e.g., Google's Edge TPU \cite{edge-tpu}) is available. {\blue In the RAPPOR implementation of LDP, randomized response \cite{warner1965randomized} needs to generate random numbers continuously. Note that continuous random number generation presents substantial overhead to resource-constrained platforms [54].}



\item {\em Learning performance degradation:} As discussed in \sect\ref{subsubsec:dl}, the projection matrix can be implicitly learned by the deep learning algorithms. Differently, the additive Laplacian noises to ensure $\epsilon$-DP can be considered neither a pattern nor an embedding that can be learned by learning algorithms. Thus, the Laplacian noises will only negatively affect the learning performance. {\blue Similarly, the random response mechanism of LDP cannot be considered as a pattern that can be learned.} Our evaluation in \sect\ref{sec:evaluation} shows that {\blue both the Laplace mechanism and RAPPOR significantly degrade the learning performance}.
\end{itemize}

From the above discussions and the evaluation results in \sect\ref{sec:evaluation}, adding Laplacian noises to the training data for $\epsilon$-DP is not a promising approach to PPCL with resource-constrained participants.



\section{Performance Evaluation}
\label{sec:evaluation}

 In this section, we extensively compare the accuracy achieved by various approaches. The computation and communication overhead of these approaches will be profiled in \sect\ref{sec:implementation} based on their implementations on a testbed. {\revised{The source code of the evaluation can be found from \cite{codes}.}}


\subsection{Evaluation Methodology and Datasets}

We conduct extensive evaluation to compare several approaches:
\begin{itemize}
\item {\bf GRP-DNN:} This is the main proposed approach consisting of GRP at the participants and collaborative learning based on a DNN at the coordinator. The design or choice of the DNN model will be application specific. The DNN models and training algorithms are implemented based on PyTorch \cite{pytorch}.
\item {\bf RRP-DNN:} This approach replaces the GRP in GRP-DNN with Rademacher random projection (RRP). The DNN models and training algorithms are same as GRP-DNN.
\item {\bf BRP-DNN:} This approach replaces the GRP in GRP-DNN with binary random projection (BRP). The DNN models and training algorithms are same as GRP-DNN.
\item {\bf GRP-SVM:} This baseline approach applies GRP at the participants and trains an SVM-based classifier at the coordinator. The SVM-based classifier is implemented using LIBSVM \cite{libSVM}. The classifier uses RBF kernel with two configurable parameters $C$ and $\lambda$. During the training phase, we apply grid search to determine the best settings for $C$ and $\lambda$. This grid search is often lengthy in time (e.g., several days).
\item {\bf GRP-NCL:} This is the non-collaborative learning (NCL) baseline approach described in \sect\ref{subsubsec:nc-learn}. It runs GRP at the participants and trains a separate DNN for each participant at the coordinator. Compared with other approaches, this approach additionally requires the identity of the participant for each training sample.
\item {\bf $\epsilon$-DP-DNN:} As described in \sect\ref{subsubsec:dp}, this approach implements $\epsilon$-DP by adding Laplacian noise vectors to the data vectors and performs collaborative deep learning based on a DNN at the coordinator. {\blue Note that this implementation corresponds to the case where $\mathcal{F}(D)$ defined in Definition~4.1 returns $D$ itself. This case is more related to our privacy objective of protecting the raw form of the original data vector. If the DP noises are added to a certain statistics as usually performed in DP applications, the relationship between the additive perturbation and the objective of protecting the raw data form is weakened. As a result, the DP approach and our GRP approach become less comparable. Thus, our DP implementation adds noises to the individual records.}
\item {\bf $\epsilon$-DP-SVM:} This approach implements $\epsilon$-DP by adding Laplacian noise vectors to the data vectors and performs collaborative learning based on SVM at the coordinator.{\blue
\item {\bf $\epsilon$-LDP-DNN:} This approach implements $\epsilon$-LDP using RAPPOR \cite{warner1965randomized} and performs collaborative deep learning based on a DNN at the coordinator.}
\item {\bf CNN, SVM, MLP, ResNet-152:} These are the plain learning approaches based on the CNN, SVM, MLP, and ResNet-152 models, respectively. They do not protect any privacy.
\end{itemize}

\begin{figure}
  \subfigure[Original images]
  {
    \includegraphics[width=.044\textwidth]{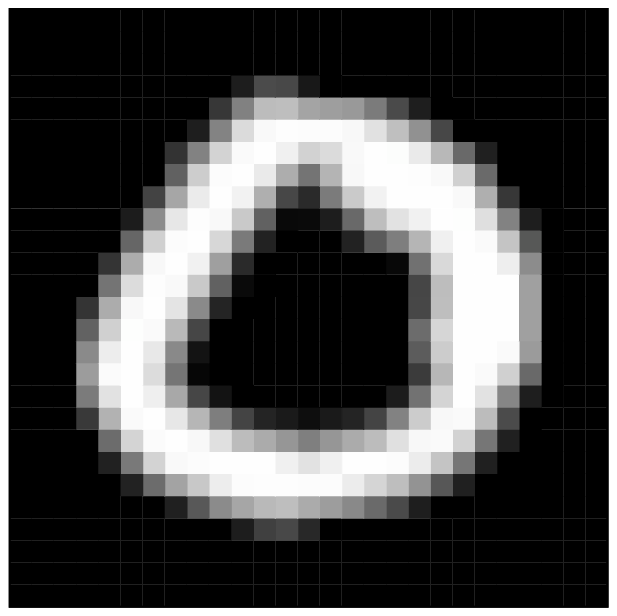}
    \includegraphics[width=.044\textwidth]{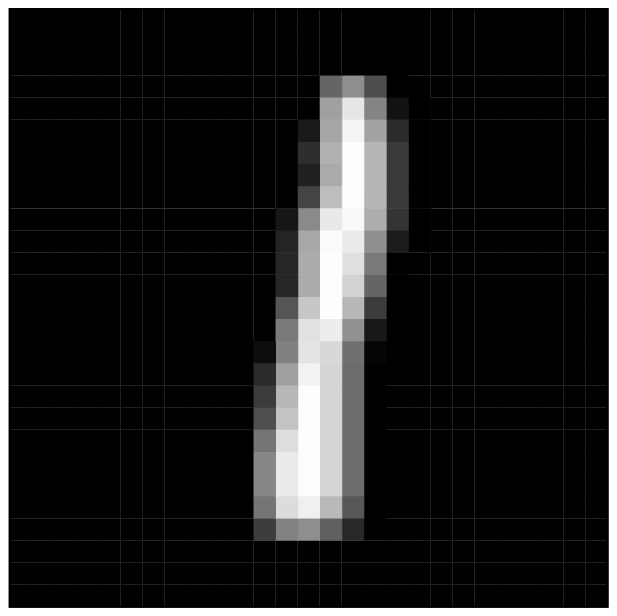}
    \includegraphics[width=.044\textwidth]{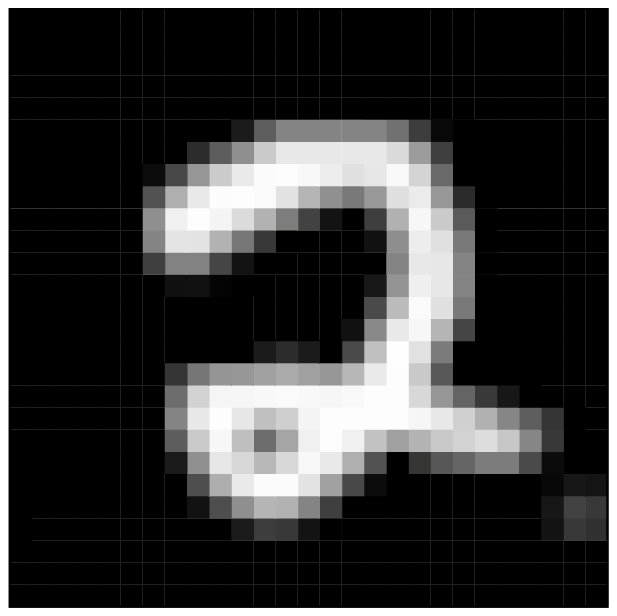}
    \includegraphics[width=.044\textwidth]{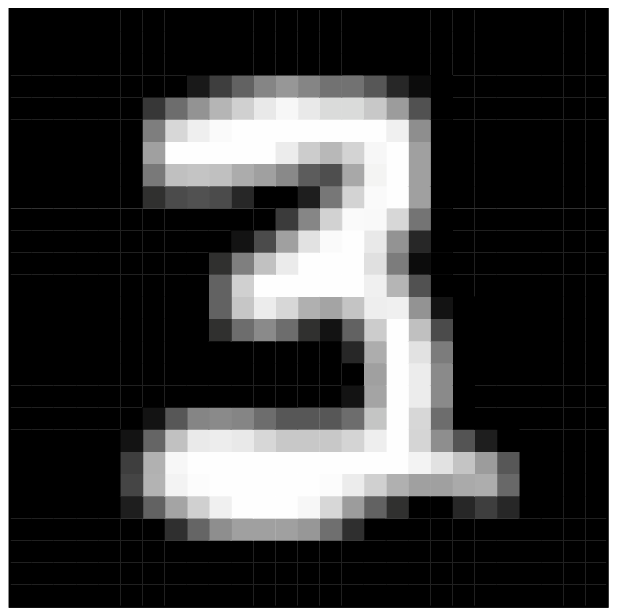}
    \includegraphics[width=.044\textwidth]{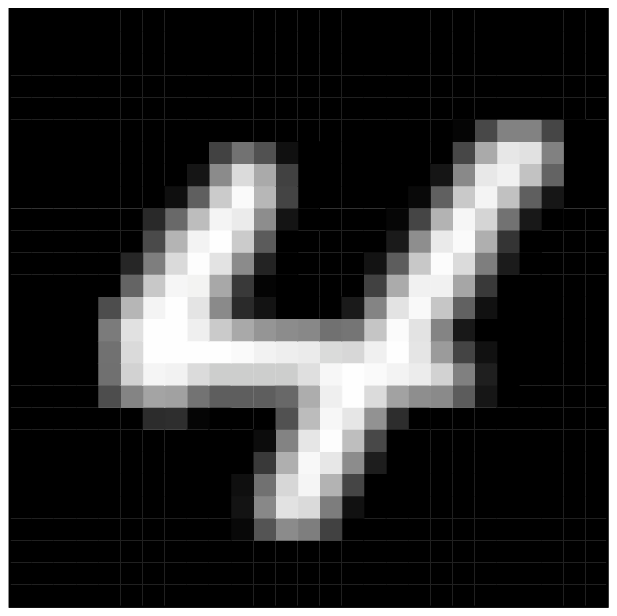}
    \includegraphics[width=.044\textwidth]{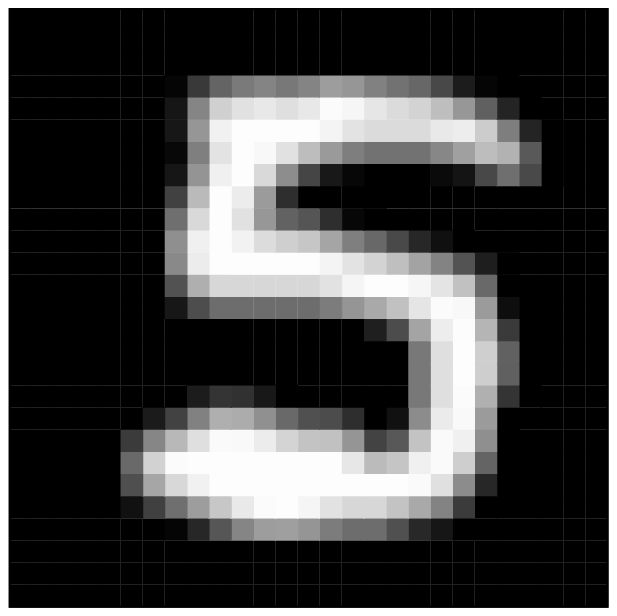}
    \includegraphics[width=.044\textwidth]{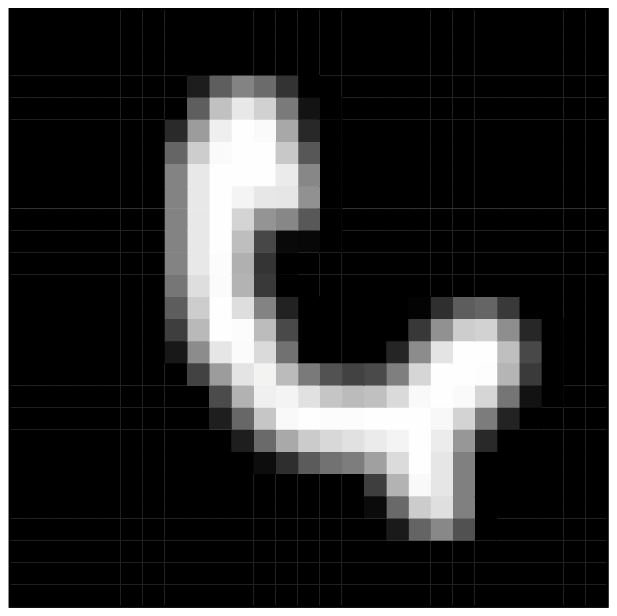}
    \includegraphics[width=.044\textwidth]{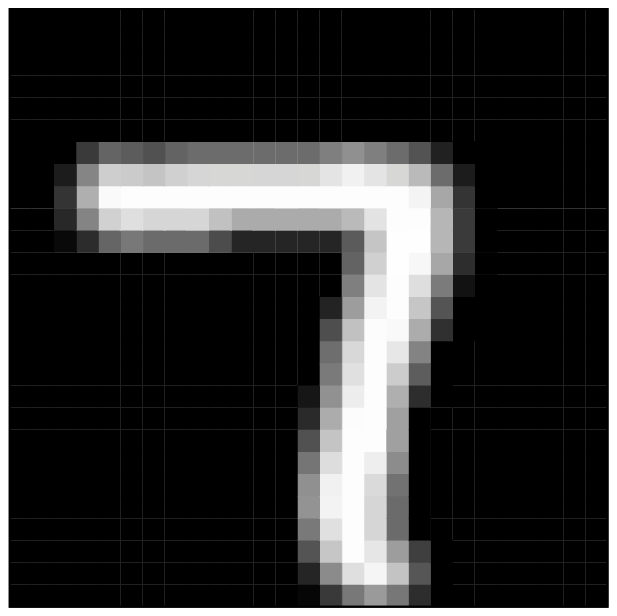}
    \includegraphics[width=.044\textwidth]{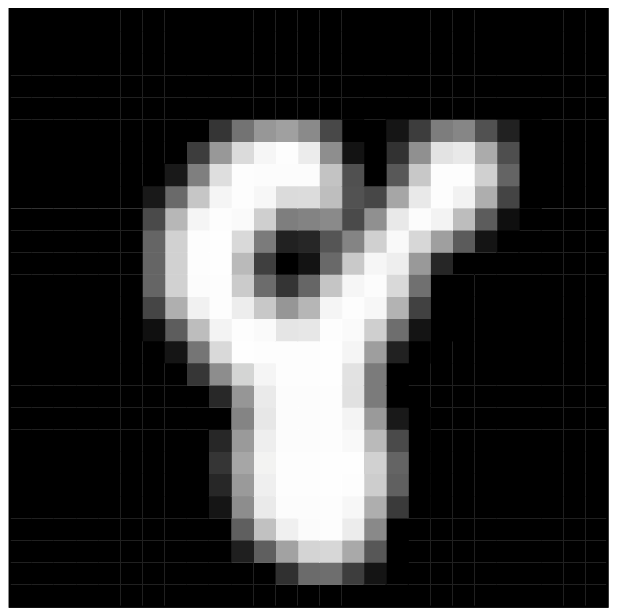}
    \includegraphics[width=.044\textwidth]{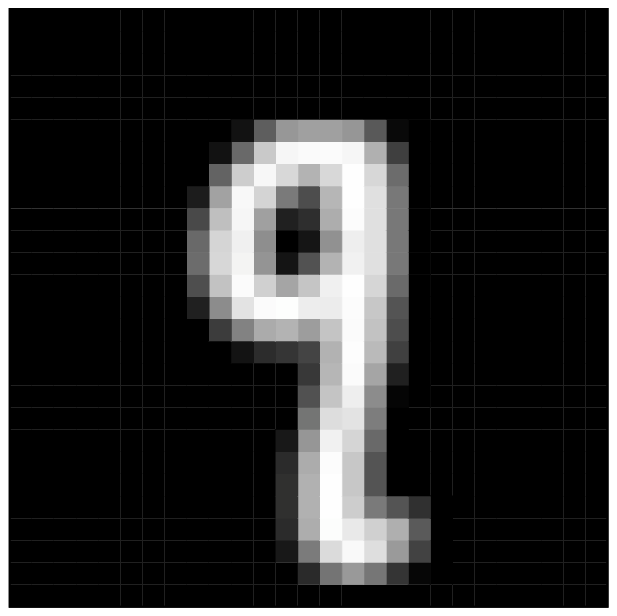}
    \label{fig:mnist}
  }
  \subfigure[Projected images in GRP-DNN]
  {
    \includegraphics[width=.044\textwidth]{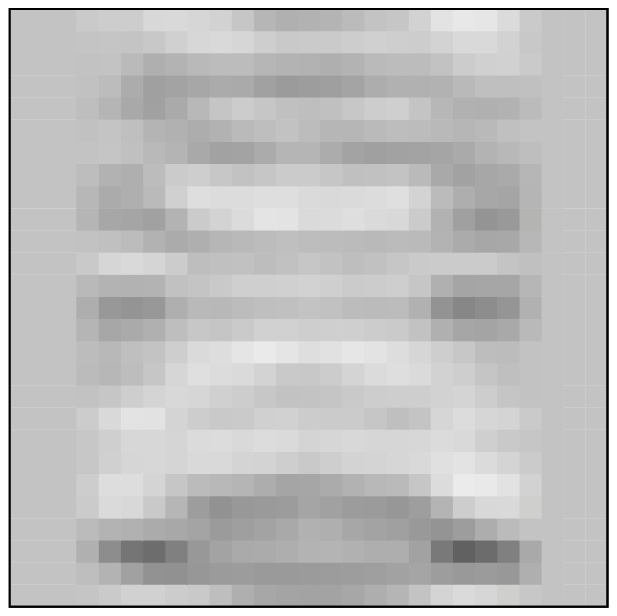}
    \includegraphics[width=.044\textwidth]{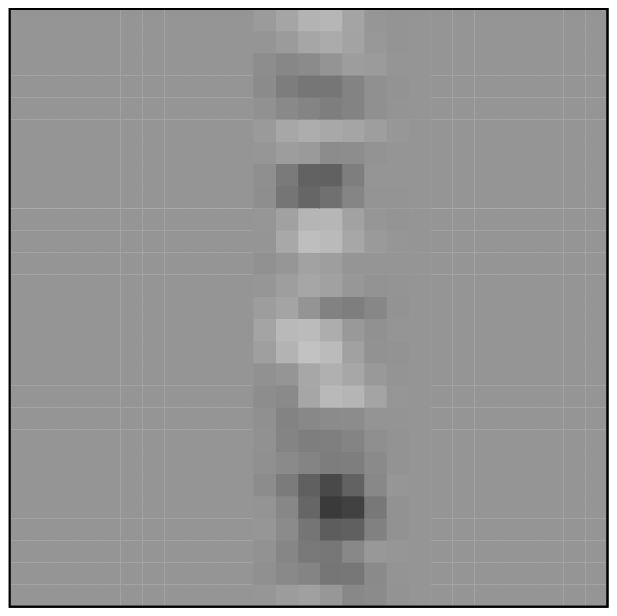}
    \includegraphics[width=.044\textwidth]{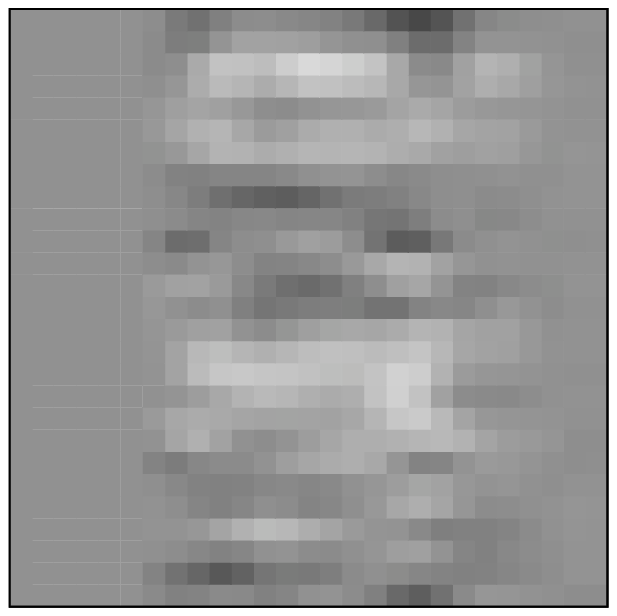}
    \includegraphics[width=.044\textwidth]{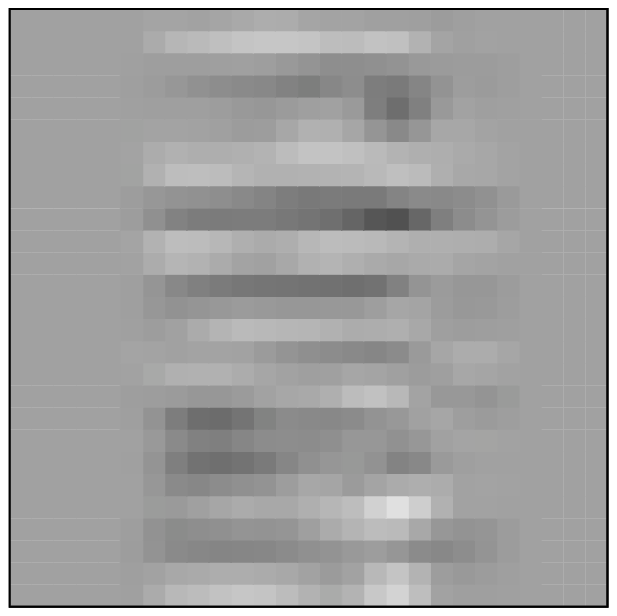}
    \includegraphics[width=.044\textwidth]{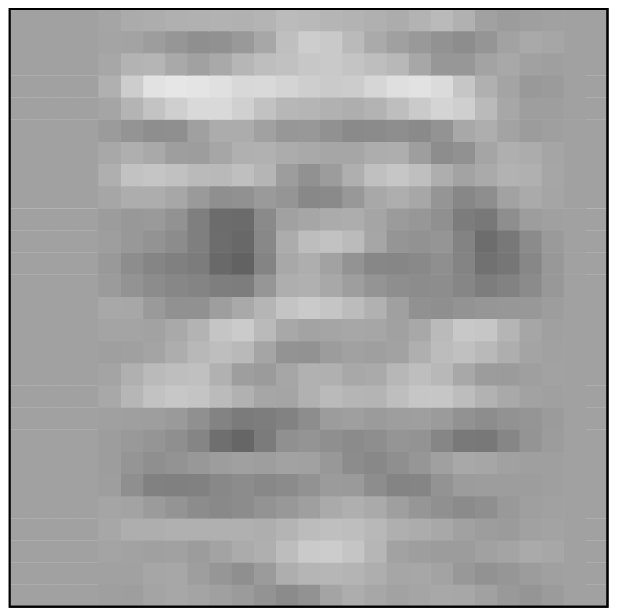}
    \includegraphics[width=.044\textwidth]{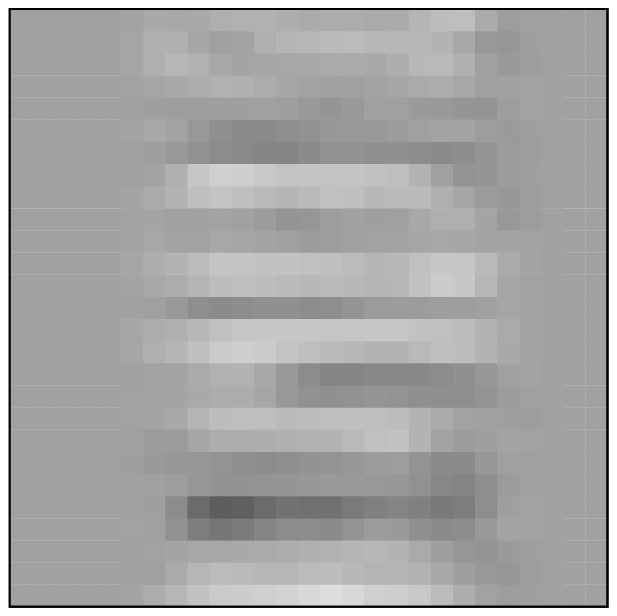}
    \includegraphics[width=.044\textwidth]{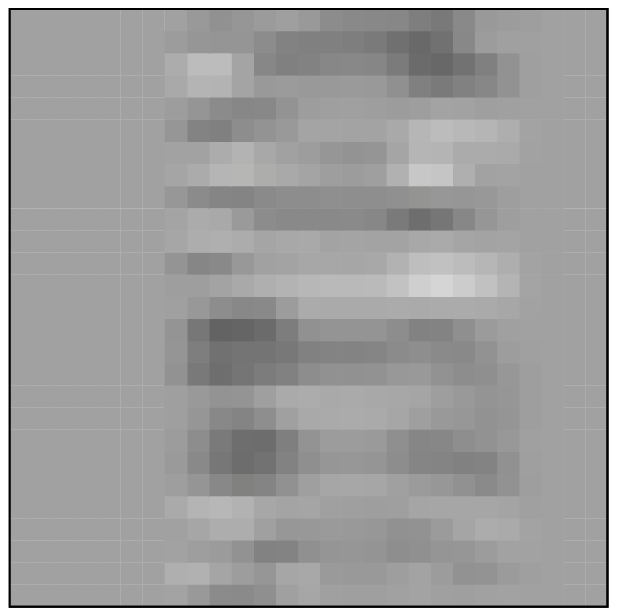}
    \includegraphics[width=.044\textwidth]{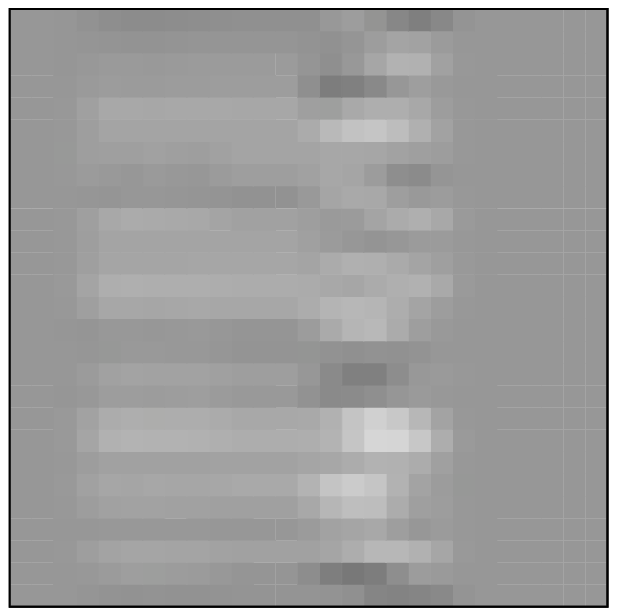}
    \includegraphics[width=.044\textwidth]{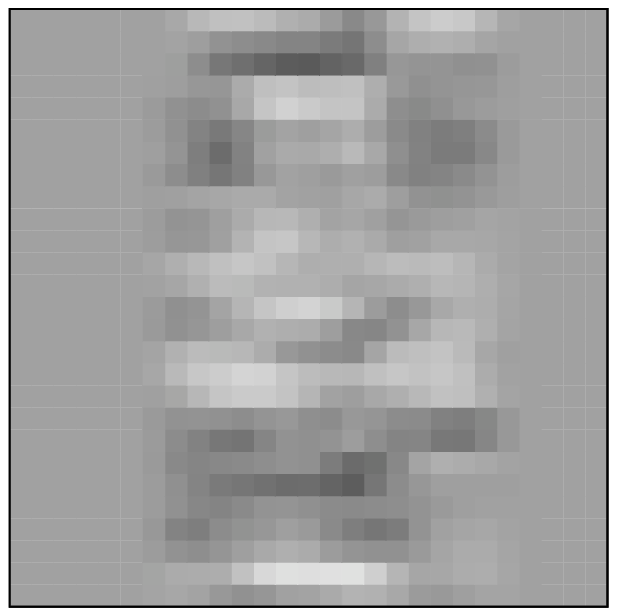}
    \includegraphics[width=.044\textwidth]{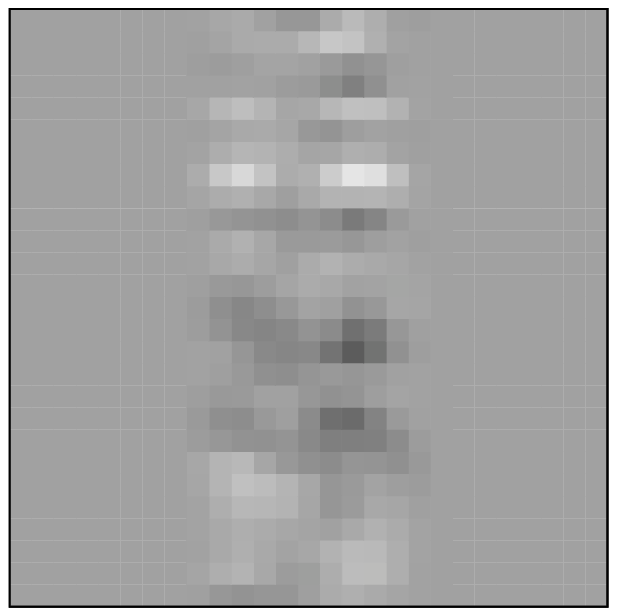}
    \label{fig:mnist-rp}
  }
  \subfigure[Noise-added images in $\epsilon$-DP-DNN ($\epsilon=50$)]
  {
    \includegraphics[width=.044\textwidth]{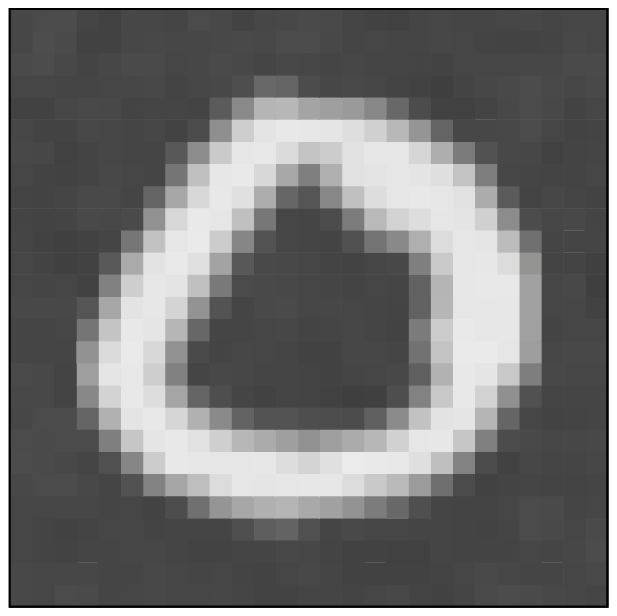}
    \includegraphics[width=.044\textwidth]{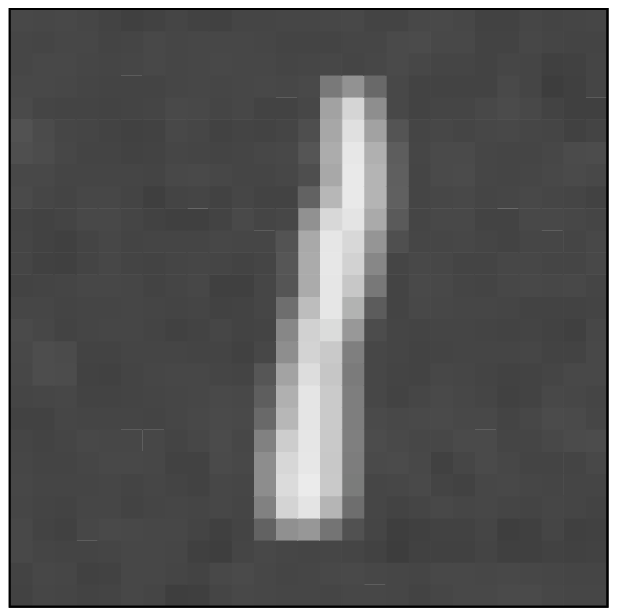}
    \includegraphics[width=.044\textwidth]{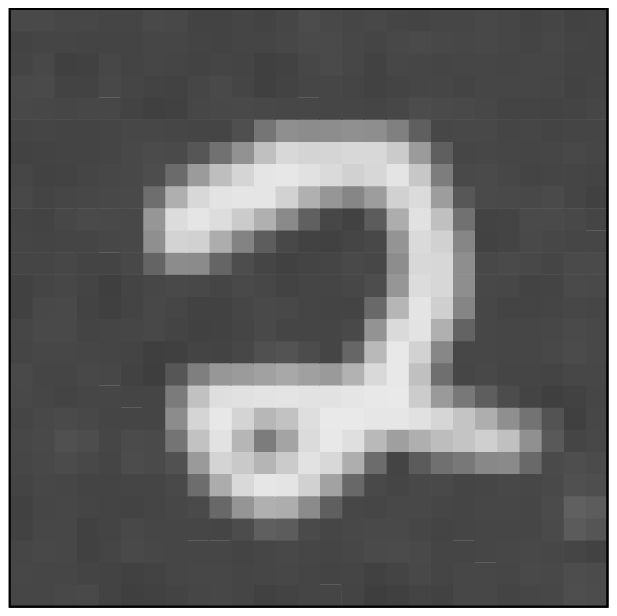}
    \includegraphics[width=.044\textwidth]{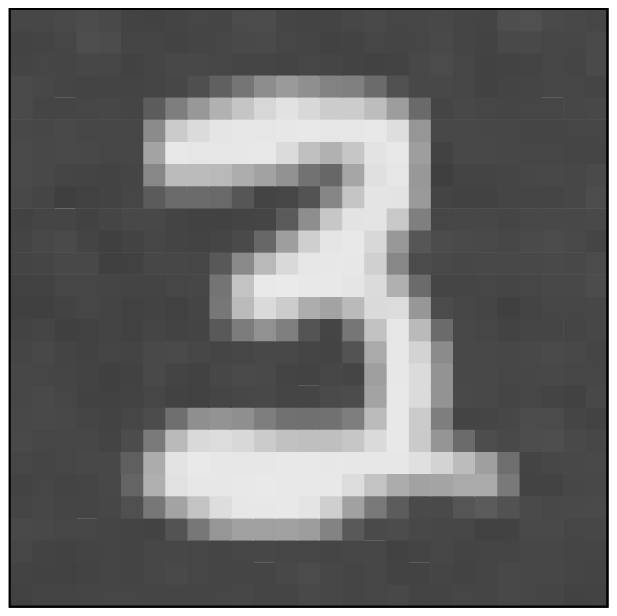}
    \includegraphics[width=.044\textwidth]{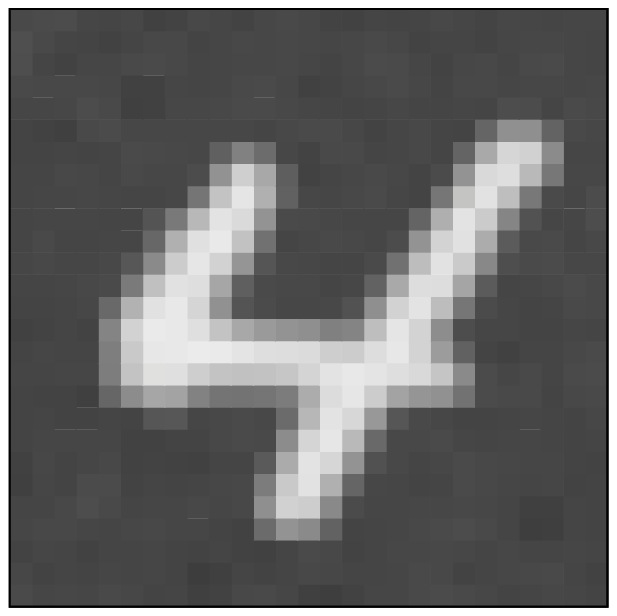}
    \includegraphics[width=.044\textwidth]{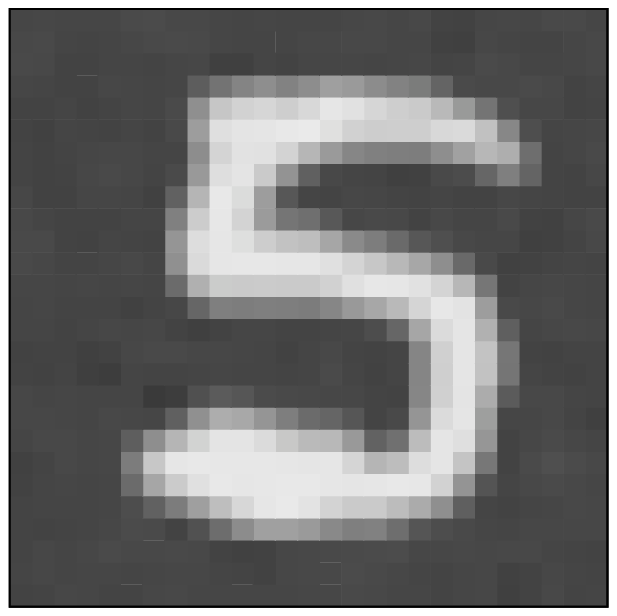}
    \includegraphics[width=.044\textwidth]{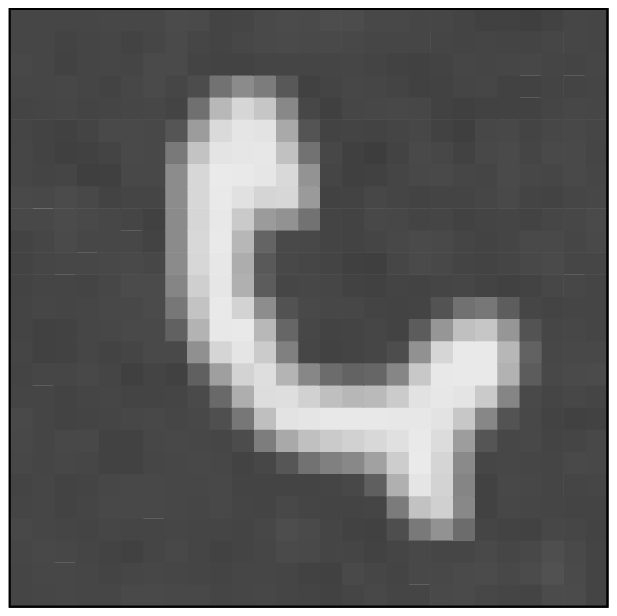}
    \includegraphics[width=.044\textwidth]{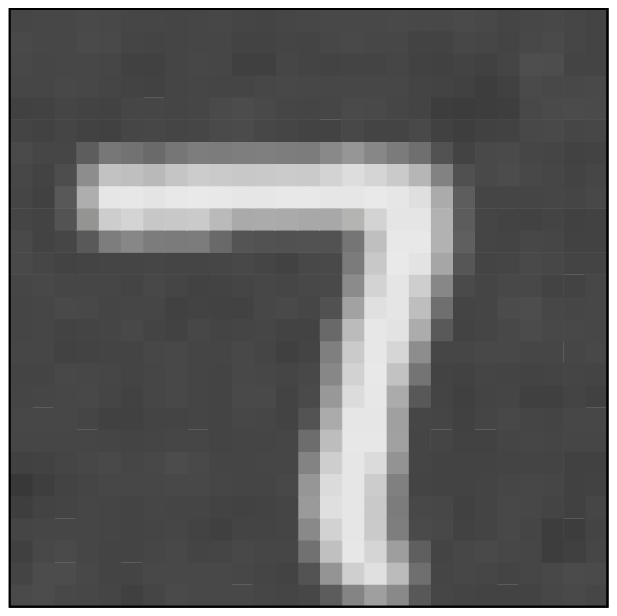}
    \includegraphics[width=.044\textwidth]{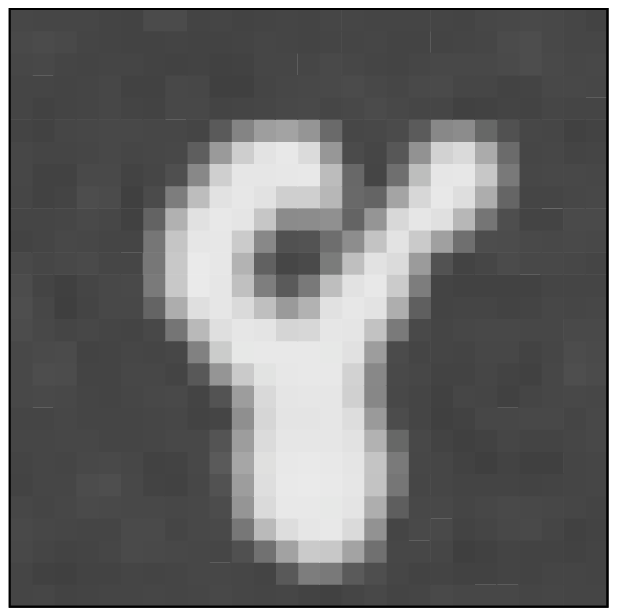}
    \includegraphics[width=.044\textwidth]{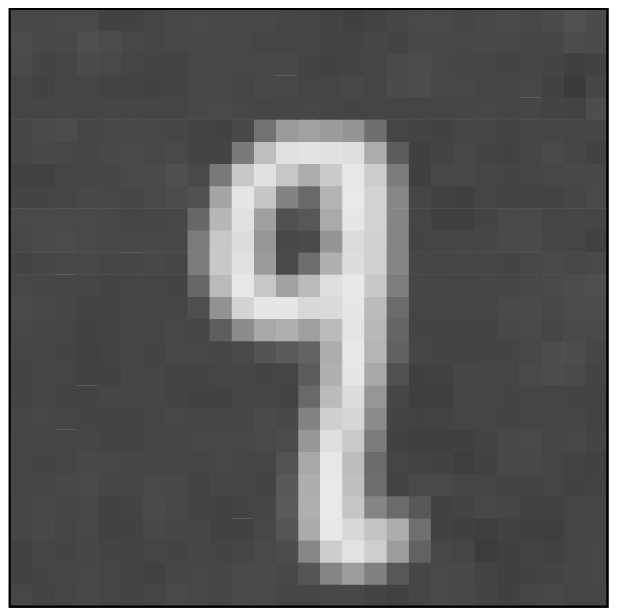}
    \label{fig:mnist-dp50}
  }
  \subfigure[Noise-added images in $\epsilon$-DP-DNN ($\epsilon=10$)]
  {
    \includegraphics[width=.044\textwidth]{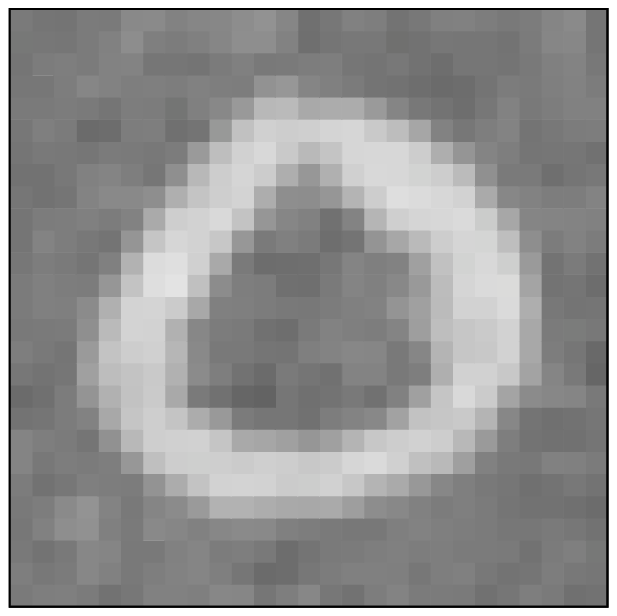}
    \includegraphics[width=.044\textwidth]{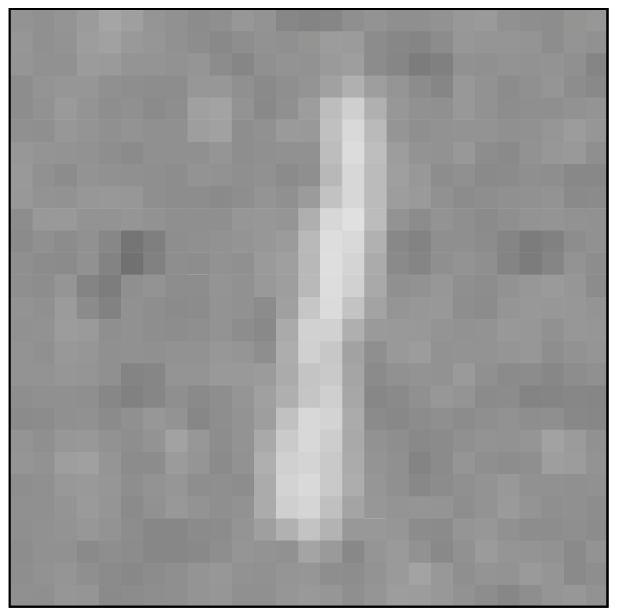}
    \includegraphics[width=.044\textwidth]{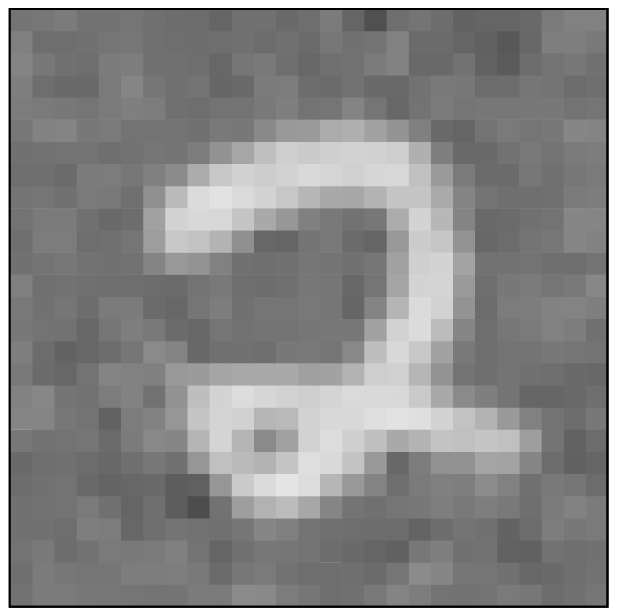}
    \includegraphics[width=.044\textwidth]{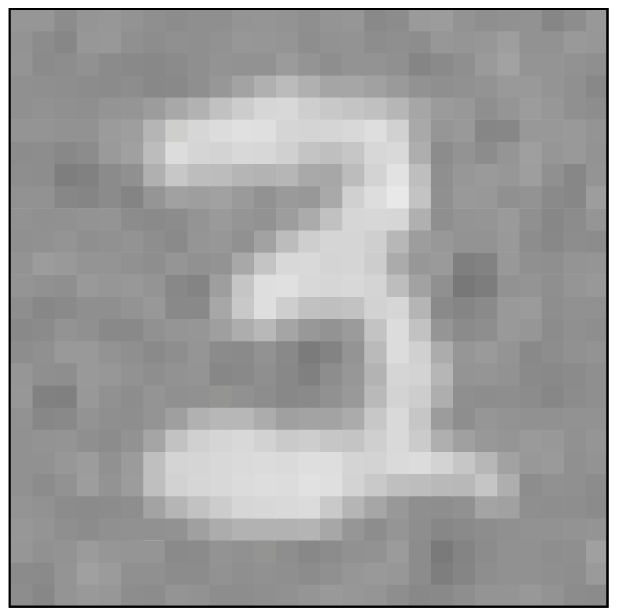}
    \includegraphics[width=.044\textwidth]{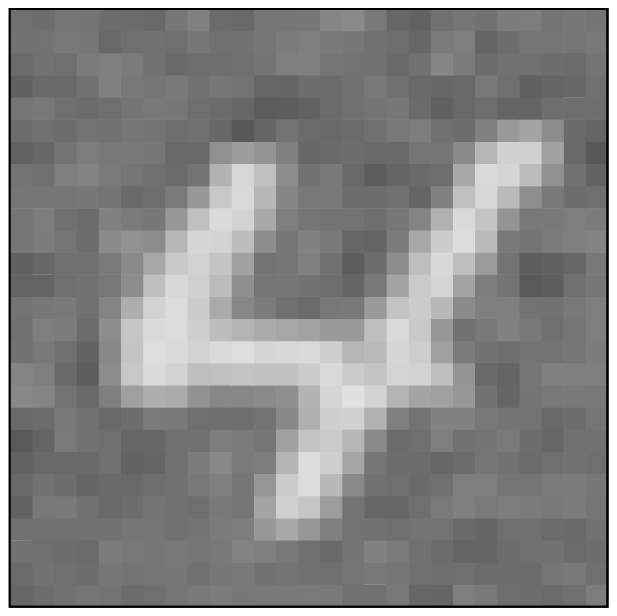}
    \includegraphics[width=.044\textwidth]{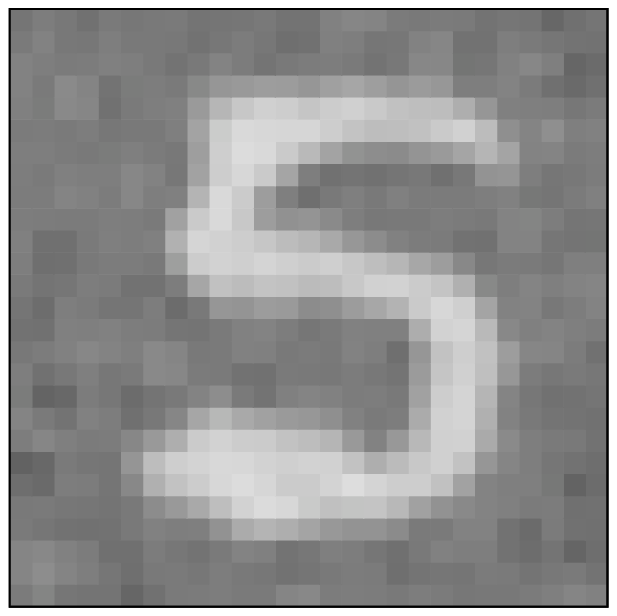}
    \includegraphics[width=.044\textwidth]{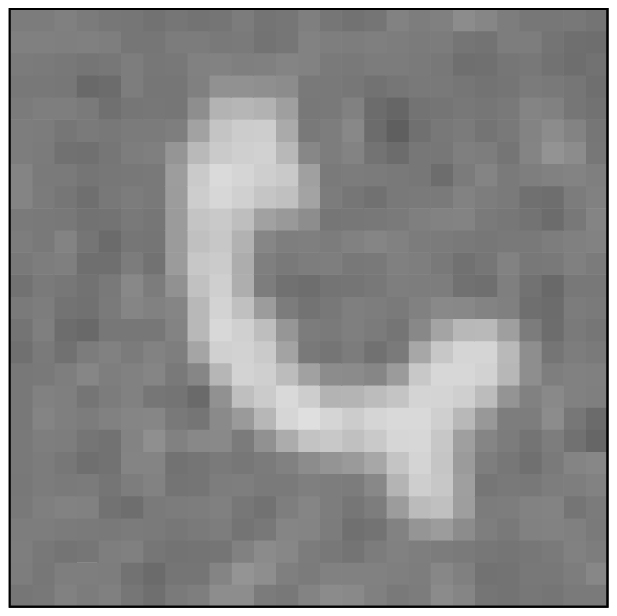}
    \includegraphics[width=.044\textwidth]{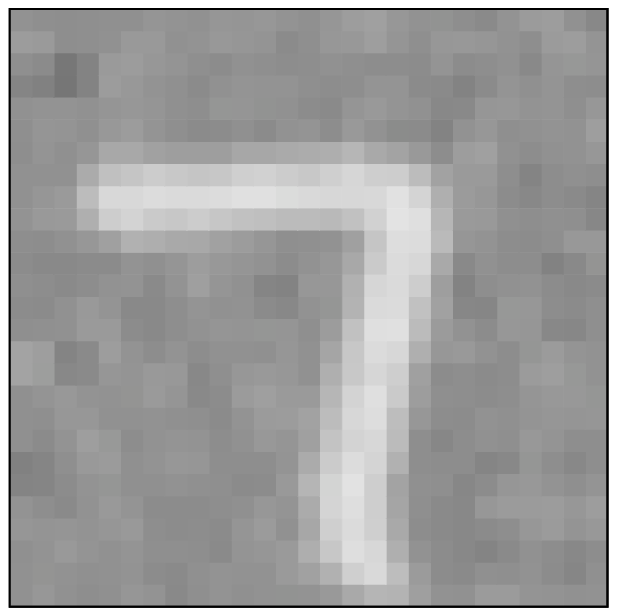}
    \includegraphics[width=.044\textwidth]{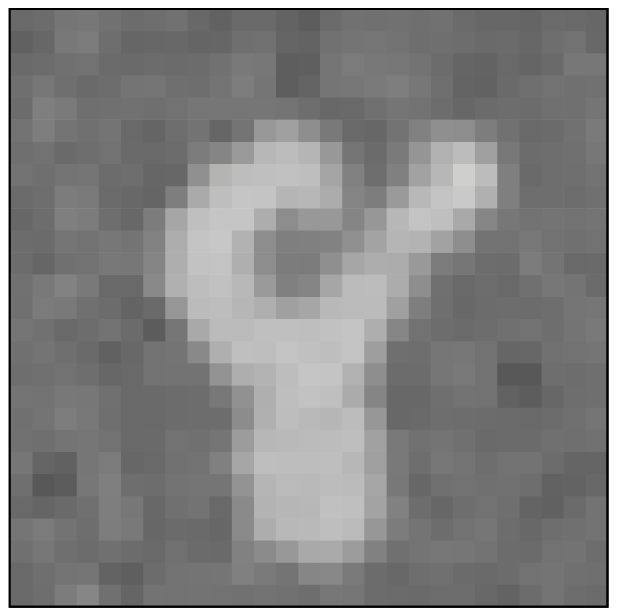}
    \includegraphics[width=.044\textwidth]{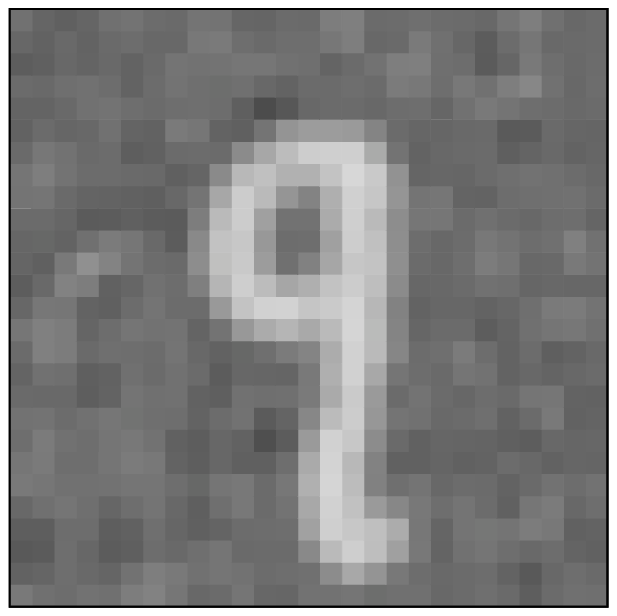}
    \label{fig:mnist-dp10}
  }
  \caption{Example images from MNIST dataset.}
  \label{fig:mnist-examples}
\end{figure}

{\revised{The performance evaluation is performed based on four datasets, i.e., MNIST \cite{mnist}, spambase \cite{spambase}, FSD \cite{sound-mnist}, and CIFAR-10 \cite{krizhevsky2009learning}.
\begin{itemize}
\item {\bf MNIST:} The MNIST dataset consists of 60,000 training samples and 10,000 testing samples. Each sample is a $28 \times 28$ grayscale image showing a single, handwritten digit. Fig.~\ref{fig:mnist} shows an instance of each digit.
\item {\bf Spambase:} The spambase dataset consists of 4,601 samples. Each sample consists of (i) a 57-dimensional feature vector that is extracted from an e-mail message and (ii) a class label indicating whether the e-mail message is an unsolicited commercial e-mail. The details of the feature vector can be found in \cite{spambase}. As the data volume of this spambase dataset is limited, we apply data augmentation to the spambase by adding zero-mean Gaussian noises, resulting in 40,000 training samples and 400 testing samples.
\item {\bf FSD:}  {\blue  The free spoken digit (FSD) dataset consists of 2,000 WAV recordings of spoken digits from 0 to 9 in English. We randomly split the data into 80$\%$ for training, 10$\%$ for validation, and 10$\%$ for testing. We extract the mel-frequency cepstral coefficients (MFCC) \cite{logan2000mel} as the features to represent a	segment of audio signal. MFCC can well represent the pertinent aspects of the short-term speech spectrum. As the recordings are of different lengths, we apply constant padding to unify the number of MFCC feature vectors for each recording. As a result, the extracted MFCC feature vectors over time for each recording form a $20 \times 45$ matrix.}
\item {\bf CIFAR-10:} The CIFAR-10 dataset consists of 60,000 $ 32 \times 32 $ RGB color images in ten classes, in which 50,000 images are for training and 10,000 images are for testing. The 10 classes are airplanes, cars, birds, cats, deers, dogs, frogs, horses, ships, and trucks. Each class has 6,000 images. Fig.~\ref{fig:cifar10} shows an instance of each class. 
\end{itemize}
We choose these {\blue four} datasets because the small sizes of the data vectors are commensurate with the limited computing and communication capabilities of IoT end devices.
}  }

 Training a spam detector based on user-contributed samples (e.g., e-mails) may cause privacy concerns. Thus, our proposed approach {\revised{is quite appropriate}}. The choice of the vision-based character recognition and object classification tasks with the MNIST and CIFAR-10 datasets allows us to leverage on the learning capabilities of the latest deep models that are often designed for image classification. Moreover, by using images as the data vectors, the effect of the distortion caused by noise adding or random projection can be visualized for intuitive understanding.
The CIFAR-10 images have varying backgrounds and object appearances, i.e., complex patterns. Thus, the vision-based object recognition task using CIFAR-10 is more challenging.  Although the character and object recognition tasks are not privacy-sensitive, the results based on MNIST and CIFAR-10 will provide understanding on other image classification-based privacy-sensitive applications, such as collaboratively training a mood classifier using the photos in the album of the users' smartphones. {\blue The choice of the FSD dataset is to diversify the application scenarios in evaluating our approach. Recently, voice recognition has been integrated into various smart systems such as smartphones and voice assistants found in households and cars. In many scenarios, voice recordings are privacy sensitive. {\revised{Our approach matches the privacy expectations for PPCL applied to voice recognition}}. In summary, our evaluation datasets cover image, text, and voice modalities, and represent important IoT applications.


For a PPCL system with $N$ participants, by default we divide both the training and testing samples into $N$ disjoint sets evenly. Each set is assigned to a participant. Note that in \sect\ref{subsubsec:hor}, we will evaluate the impact of the horizontal distribution of the data on the learning performance, where the training and testing samples are not evenly distributed among the participants. Under GRP-DNN, GRP-SVM,GRP-NCL, RRP-DNN, and BRP-DNN, each participant independently generates its random matrix and uses the matrix to project its plaintext data vectors. {\revised{The coordinator trains the deep models and SVM }} based on the projected or noise-added training data vectors from the participants. The trained deep models and SVM are used to classify the projected or noise-added testing data vectors to measure the test accuracy as the evaluation results.



\subsection{Evaluation Results with MNIST Dataset}
\label{subsec:MNIST}



\begin{figure}
\begin{tabular}{cc|c|c|c|c|c|c|c|c|c|c}
\cline{3-6} \cline{8-11}
\begin{minipage}{0.05\textwidth}
\centering
\includegraphics[width=\textwidth]{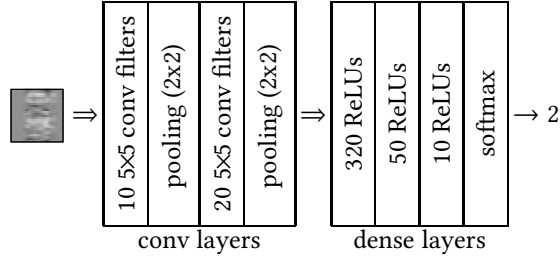}
\end{minipage}
& $\!\!\!\!\Rightarrow \!\!$
& \rotatebox[origin=c]{90}{$\;$ 10 $5\!\!\times \!\!5$ conv filters $\;$} & \rotatebox[origin=c]{90}{pooling (2x2)} & \rotatebox[origin=c]{90}{20 $5\!\! \times \!\! 5$ conv filters} & \rotatebox[origin=c]{90}{pooling (2x2)} & $\!\!\Rightarrow \!\!$ & \rotatebox[origin=c]{90}{320 ReLUs} & \rotatebox[origin=c]{90}{50 ReLUs} & \rotatebox[origin=c]{90}{10 ReLUs} & \rotatebox[origin=c]{90}{softmax} & $\!\!\rightarrow 2$\\
\cline{3-6} \cline{8-11}
\multicolumn{2}{c}{} & \multicolumn{4}{c}{conv layers} & \multicolumn{1}{c}{} & \multicolumn{4}{c}{dense layers} & \multicolumn{1}{c}{} \\
\end{tabular}
\caption{CNN with a projected MNIST image as input.}
\label{fig:CNN-MNIST}
\end{figure}

We design a CNN that is used in the GRP-DNN, GRP-NCL, and $\epsilon$-DP-DNN approaches. The CNN consists of two convolutional layers and three dense layers of ReLUs.
We apply max pooling after each convolutional layer to reduce the dimension of data after convolution. The max pooling controls overfitting effectively and improves the CNN's robustness to small spatial distortions in the input image. The last dense layer has ten ReLUs corresponding to the ten classes of MNIST.
A softmax function is used to make the classification decision based on the outputs of the last dense layer.
Fig.~\ref{fig:CNN-MNIST} illustrates the design of the CNN.
Note that, without random projection, the CNN and the SVM with grid search for kernel parameters achieves test accuracy of 98.7\% and 98.52\%. This shows that the CNN and SVM {\revised{capture the patterns of MNIST well}}.

\begin{figure}
  \centering
  \includegraphics[width=0.85\textwidth]{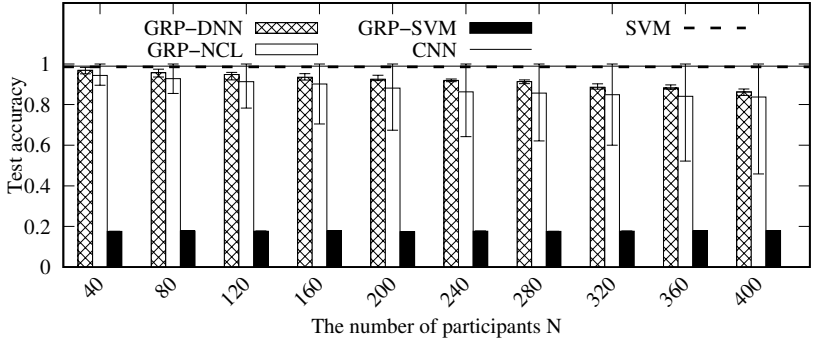}
  \caption{Impact of the number of participants (MNIST). The error bars represent min and max.}
  \label{fig:mnist-N}
\end{figure}



\subsubsection{Impact of $N$ on learning performance}
 We evaluate the impact of the number of participants $N$ on the learning performance of GRP-DNN, GRP-NCL, and GRP-SVM. {\blue
  We randomly split the training data and testing data equally into $N$ parts and assign to $N$ participants. The amount of data with a participant decreases with the increase of $N$ since the total amount of data is fixed.} Fig.~\ref{fig:mnist-N} shows the results.  The two horizontal lines in Fig.~\ref{fig:mnist-N} represent the test accuracy of the plain CNN and SVM without any privacy protection. The two lines overlap. When $N$ increases from 40 to 400, the {\blue mean} test accuracy of GRP-DNN decreases from 96.87\% to 86.18\%. If $N$ is no greater than 280, GRP-DNN maintains a test accuracy greater than 90\%. The drop of accuracy with increased $N$ is consistent with the understanding that distinct random projection matrices increase the pattern complexity of the aggregated data. However, for MNIST data with light pattern complexities, the GRP-DNN approach can support up to 280 IoT objects for a satisfactory classification accuracy of 90\%.
Under the GRP-NCL approach, the deep models corresponding to the participants have different test accuracy values. The histogram and error bars in Fig.~\ref{fig:mnist-N} represent the average, minimum, and maximum of the test accuracy values across all trained deep models. Under each setting of $N$, the maximum test accuracy is 100\%. However, the average test accuracy is consistently lower than that of GRP-DNN. This shows that the GRP-NCL that needs to compromise data anonymity yields inferior average learning performance compared with GRP-DNN. This result shows the advantage of collaborative learning. Lastly, the GRP-SVM approach gives poor test accuracy around 17.5\% because no efficient RBF kernels can be found to create proper hyperplanes for classification. {\revised{This observation suggests}} that DNNs are {\revised{more efficient in coping with}} the distortions caused by projections.

\begin{figure}
  \centering
  \includegraphics{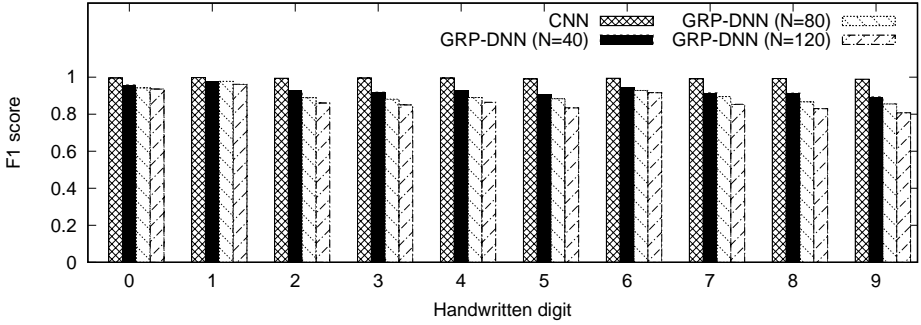}
  \caption{The F1 scores of different handwritten digits in the MNIST dataset under the CNN and the GRP-DNN approaches (MNIST).}
  \label{fig:F-score}
\end{figure}
\subsubsection{Classification accuracy of different classes}

 We also evaluate the F1 scores of different classes (i.e., different handwritten digits) under the GRP-DNN and the plain CNN approaches. The F1 score of a particular class characterizes the classification accuracy for the class. Thus, from the F1 score distribution among all classes, we can assess whether the classifier is biased for certain classes. Fig.~\ref{fig:F-score} shows the results. We can see that the F1 score distributions of the GRP-DNN with 40, 80 and 120 participants are similar with the F1 score distribution of the plain CNN. Thus, the DNN trained with the projected data is not biased towards certain classes.

\begin{figure}
  \centering
  \includegraphics{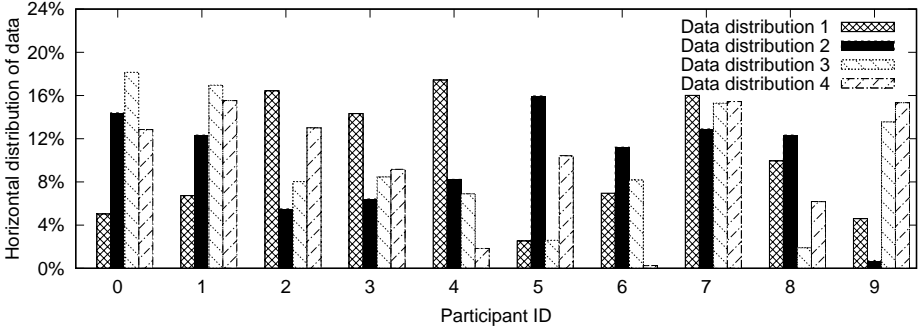}
  \caption{Four horizontal distributions of the training data among 10 participants. The test accuracies for the four distributions are $96.17\%$, $96.33\%$, $96.24\%$, $96.32\%$, respectively (MNIST).}
  \label{fig:unblanced}
\end{figure}

\subsubsection{Impact of the horizontal distribution of data}
\label{subsubsec:hor}

 In practice, different participants may have different amounts of training data. In this set of experiments, we evaluate the impact of the horizontal distribution of the training data on the learning performance. Fig.~\ref{fig:unblanced} shows four different horizontal distributions of the training data among 10 participants. During the collaborative learning phase, the participants contribute different amounts of training data. During the classification phase, the horizontal distribution of the testing data is same as that of the learning phase. The corresponding test accuracies of the four horizontal distributions are $96.17\%$, $96.33\%$, $96.24\%$, and $96.32\%$, respectively. From the results, we can see that the horizontal distribution of the data has little impact on the collaborative learning performance.

\begin{figure}
  \centering
  \includegraphics[width=0.44\textwidth]{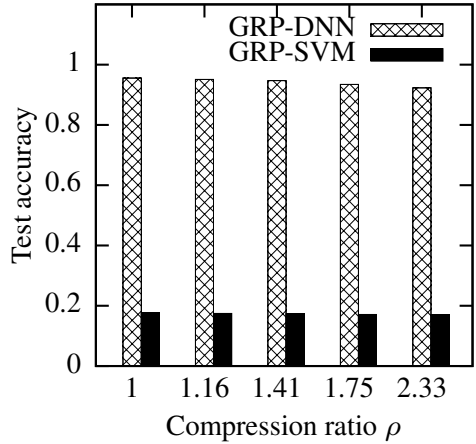}
  \caption{Impact of data compression on learning performance (MNIST, $N=100$).}
  \label{fig:mnistdata1}
\end{figure}

\subsubsection{Impact of data compression}
 We evaluate the impact of GRP's data compression on the learning performance. Fig.~\ref{fig:mnistdata1} shows the results when $N=100$. When the compression ratio increases from 1 (i.e., no compression) to 2.33 (i.e., 43\% of data volume is retained), the test accuracy of GRP-DNN decreases from 95.52\% to 92.85\% only. From our discussion in \sect\ref{subsubsec:projection}, the good tolerance of GRP-DNN against data compression is due to the high sparsity of the MNIST images.
In contrast, the GRP-SVM approach performs poorly under all compression ratio settings.

\subsubsection{Various random projection approaches}

\begin{figure}
  \centering
  \includegraphics{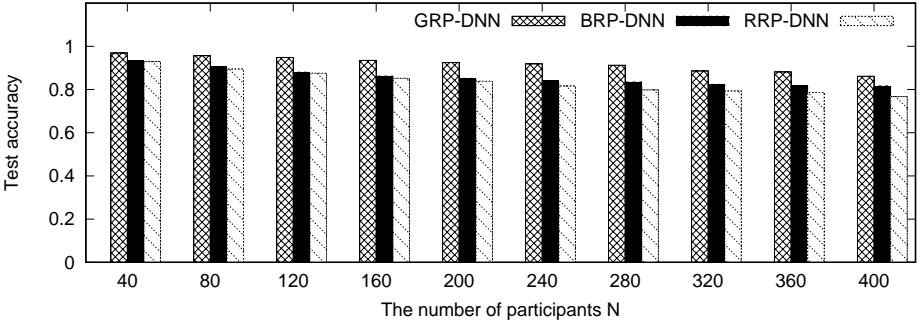}
  \caption{The test accuracy of GRP-DNN, BRP-DNN, RRP-DNN when $N$ varies (MNIST).}
  \label{fig:mixN}
\end{figure}

\begin{figure}
  \centering
  \includegraphics{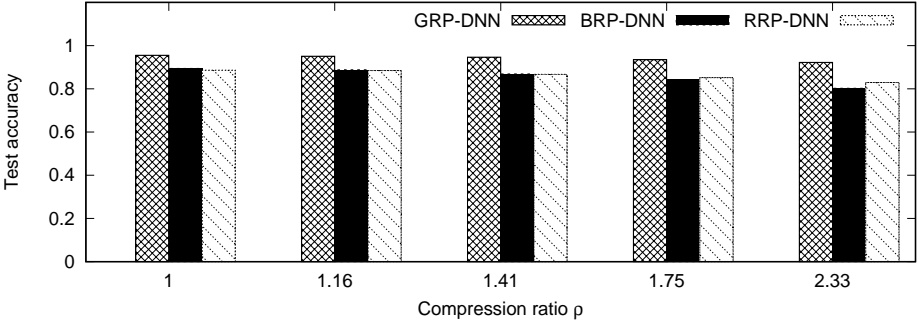}
  \caption{The test accuracy of GRP-DNN, BRP-DNN, RRP-DNN when the compression ratio varies (MNIST, $N=100$).}
  \label{fig:mixC}
\end{figure}

 This set of experiments compare the performance of collaborative learning from the data obfuscated using GRP, RRP, and BRP. Fig.~\ref{fig:mixN} shows the test accuracy of GRP-DNN, RRP-DNN, and BRP-DNN when the number of participants $N$ varies. For all three projection approaches, when $N$ increases from 40 to 400, the test accuracy drops. The GRP-DNN approach gives higher test accuracy than the other two approaches. Recall that \sect\ref{subsubsec:10d-example} has shown the better condition of Gaussian random matrices compared with Rademacher and binary random matrices. The results here are consistent with the understanding that better condition numbers will lead to better learning performance. We also compare the learning performance of GRP-DNN, RRP-DNN, BRP-DNN when the compression ratio $\rho$ varies. The number of participants is 100. Fig.~\ref{fig:mixC} shows the results. When the compression ratio increases, the test accuracy of all the three projection approaches decreases. From Fig.~\ref{fig:mixC}, in terms of test accuracy, GRP-DNN outperforms RRP-DNN and BRP-DNN.
\subsubsection{Impact of DP noises}

\begin{figure*}
	\centering
	\subfigure[{\blue Impact of privacy loss of DP on learning performance (MNIST).}]
	{
		\includegraphics{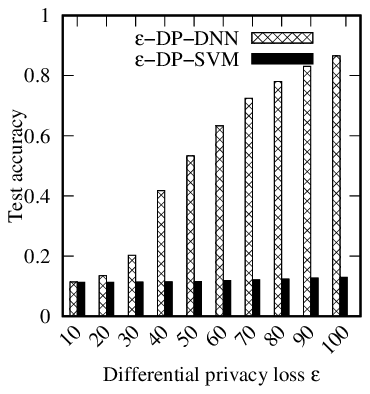}
		\label{fig:mnist-dp-acc}	
	}
    \hspace{1em}
	\subfigure[{\blue Impact of privacy loss of LDP on learning performance (MNIST).}]
	{
		\includegraphics{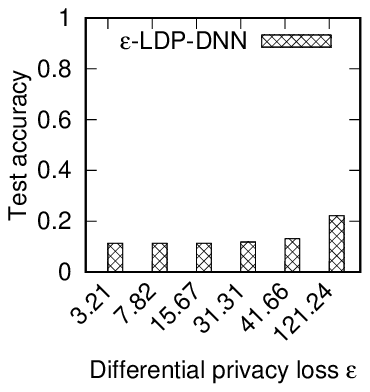}
		\label{fig:mnist-ldp-acc}
	}
	\caption{Impact of privacy loss of DP and LDP on learning performance (MNIST).}
\end{figure*}

In this set of experiments, we evaluate the impact of adding Laplacian noises to implement $\epsilon$-DP {\blue and RAPPOR to implement LDP} on  the learning performance. Fig.~\ref{fig:mnist-dp-acc} shows the test accuracy of $\epsilon$-DP-DNN versus the privacy loss level $\epsilon$. {\blue Under the considered $\epsilon$-DP-DNN or $\epsilon$-DP-SVM approaches, an $\epsilon$ setting smaller than 1 (which is the usual $\epsilon$ setting range \cite{Dwork06}) will lead to large noise levels such that the learning performance is very poor. To achieve the learning performance comparable to that of our GRP approach, we relax the range for $\epsilon$.} When $\epsilon=100$ (small Laplacian noises and large differential privacy loss), the $\epsilon$-DP-DNN achieves a test accuracy of 86.6\%, lower than those achieved by GRP-DNN when $N$ is up to 400. When $\epsilon=10$, the performance of $\epsilon$-DP-DNN drops to 11.4\%, close to the performance of random guessing. For comparison, we visualize the projected and noise-added images with two $\epsilon$ settings in Fig.~\ref{fig:mnist-examples}. From Fig.~\ref{fig:mnist-rp}, we cannot visually interpret the projected images. However, from Figs.~\ref{fig:mnist-dp50} and \ref{fig:mnist-dp10}, the noise-added images are easily interpreted when $\epsilon$ is down to 10. Note that in our evaluation, we use the same CNN model as shown in Fig.~\ref{fig:CNN-MNIST} for the GRP-DNN, GRP-NCL, and $\epsilon$-DP-DNN approaches. We do not spend special efforts to improve the CNN design in favor of any approach; we only make sure the CNN fed with the original MNIST images achieves satisfactory performance. The poor performance of $\epsilon$-DP-DNN is consistent with the understanding that the performance of deep learning can be susceptible to small perturbations to the data vectors \cite{zheng2016improving}. There are also systematic approaches to generating adversary examples with small differences from the original samples \cite{goodfellow15,bose2018adversarial}. The adversary examples will be wrongly classified by the deep models. Special care is needed in the deep model design to improve robustness against human-indiscernible perturbations \cite{zheng2016improving}. Significant noises, which are required to achieve good DP protection, are still open challenges to deep learning. Thus, under the $\epsilon$-DP framework, it is challenging to achieve a desirable trade-off between the privacy protection strength and learning performance.

We discussed in \sect\ref{subsubsec:dp} that the additive noisification for $\epsilon$-DP is ineffective in achieving a good trade-off between learning performance and protecting the confidentiality of the raw forms of the training data. Now, we compare the results of GRP-DNN ($N=1$, $k = d-1$) and $\epsilon$-DP-DNN. We consider the worst case for GRP-DNN, i.e., the projection matrix $\mat{R}$ is revealed to the curious coordinator. From Property~\ref{property:2} in \sect\ref{subsec:random-projection}, the minimum norm estimate of the original data vector by the coordinator will have a per-element variance of about 410 for any MNIST image. Under this setting, GRP-DNN  achieves a test accuracy of 94.82\%. To achieve the same per-element variance of 410, the $\epsilon$ value adopted by the $\epsilon$-DP-DNN should be 18.89. Under this $\epsilon$ setting, the test accuracy of $\epsilon$-DP-DNN is only 12.86\%.

Fig.~\ref{fig:mnist-dp-acc} also shows the test accuracy of the $\epsilon$-DP-SVM approach. It performs poorly when $\epsilon \le 100$. {\revised{This approach achieves good test accuracy only when}} the added noises are very small under the settings of $\epsilon=400$ and $\epsilon=500$.

{\blue We adopt {the BASIC RAPPOR} \cite{erlingsson2014rappor} scenario for $\epsilon$-LDP-DNN on MNIST dataset. BASIC means that each string can be deterministically mapped to a single bit in the bit array. By arranging the pixels of an MNIST sample into a 8-bit array, we adjust the parameter $f,p,q$ in BASIC RAPPOR to achieve the required privacy loss $\epsilon$. Fig.~\ref{fig:mnist-ldp-acc} shows the test accuracy  of $\epsilon$-LDP-DNN versus the privacy loss level $\epsilon$. When $\epsilon = 3.21$, the $\epsilon$-LDP-DNN only achieves a test accuracy of $11.35\%$, which is just slightly higher than that of random guessing  (i.e., 10\%). When $\epsilon = 121.74$, the $\epsilon$-LDP-DNN achieves a test accuracy of 22.21\%, much lower than that achieved by $\epsilon$-DP-DNN when $\epsilon = 100$. This result is consistent with the observation in \cite{cormode2018privacy} that LDP requires larger noise levels than the Laplace mechanism.   }


\subsection{Evaluation Results with Spambase Dataset}
\label{subsec:spambase}

\begin{figure}
  \includegraphics[width=0.85\textwidth]{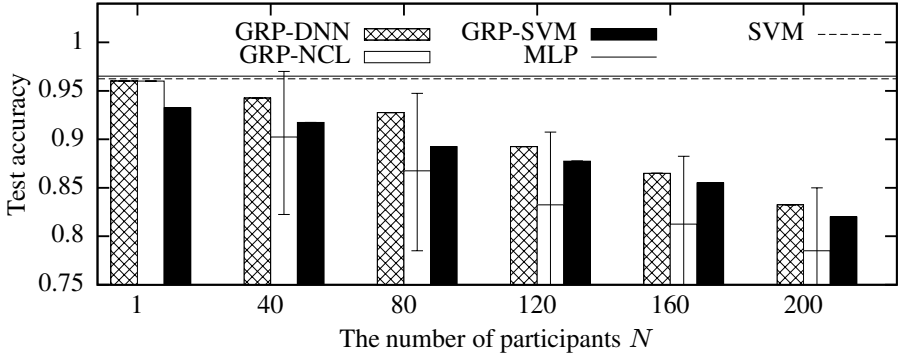}
  \caption{Impact of the number of participants (spambase). The error bars represent min and max.}
  \label{fig:spam-n}
\end{figure}

\begin{figure*}
  \centering
  \subfigure[Original images]
  {
    \begin{minipage}{\textwidth}
      \includegraphics[width=0.08\textwidth]{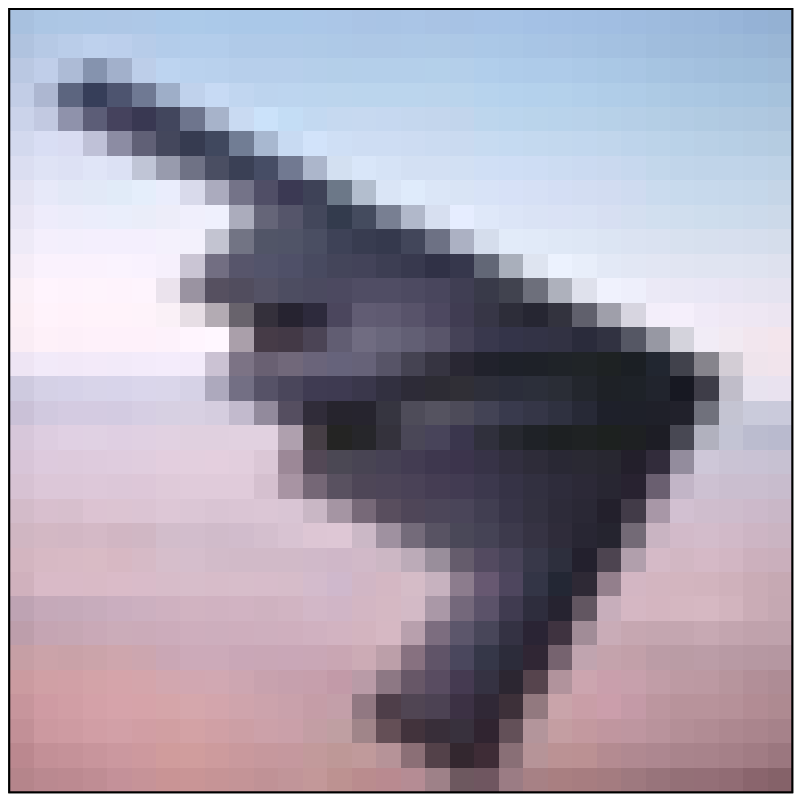}
      \hspace{0.005\textwidth}
      \includegraphics[width=0.08\textwidth]{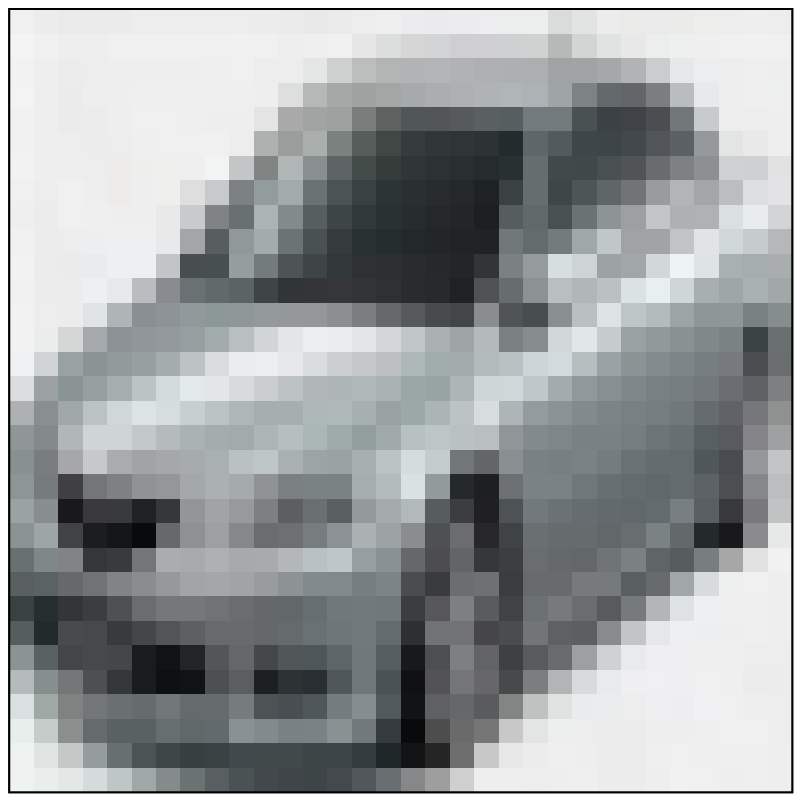}
      \hspace{0.005\textwidth}
      \includegraphics[width=0.08\textwidth]{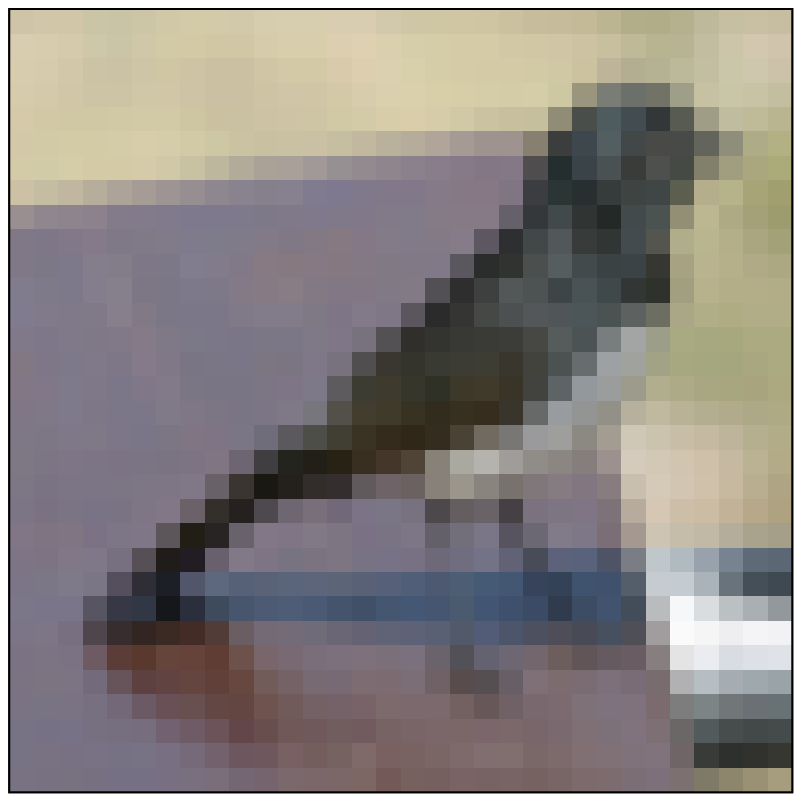}
      \hspace{0.005\textwidth}
      \includegraphics[width=0.08\textwidth]{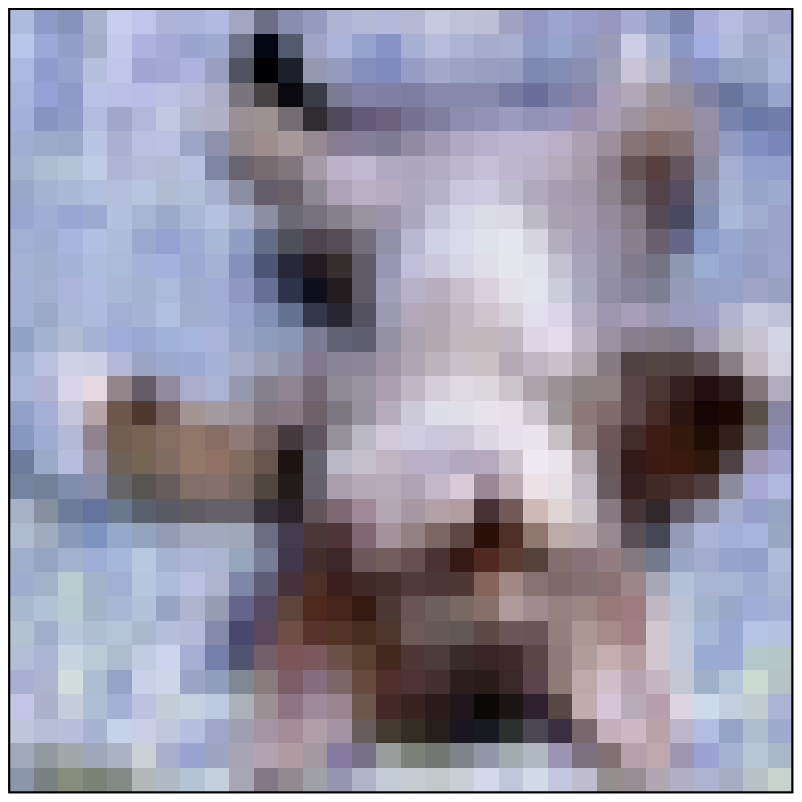}
      \hspace{0.005\textwidth}
      \includegraphics[width=0.08\textwidth]{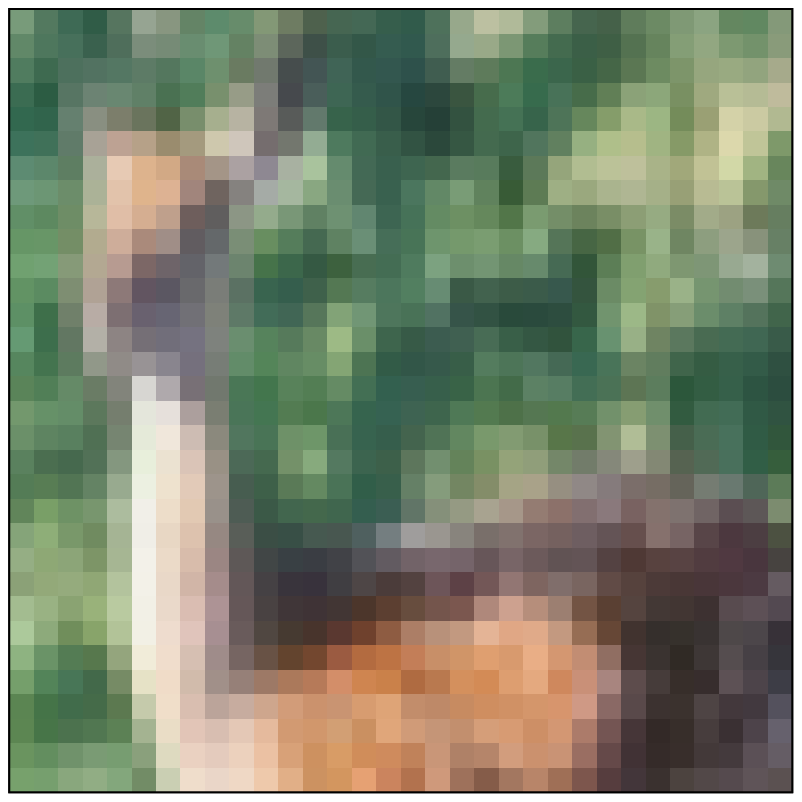}
      \hspace{0.005\textwidth}
      \includegraphics[width=0.08\textwidth]{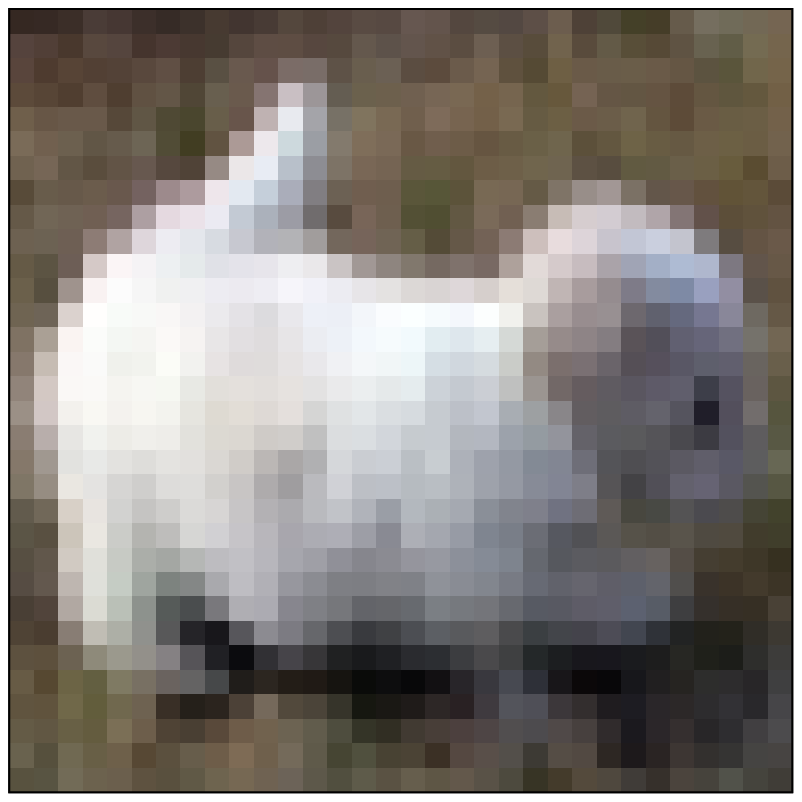}
      \hspace{0.005\textwidth}
      \includegraphics[width=0.08\textwidth]{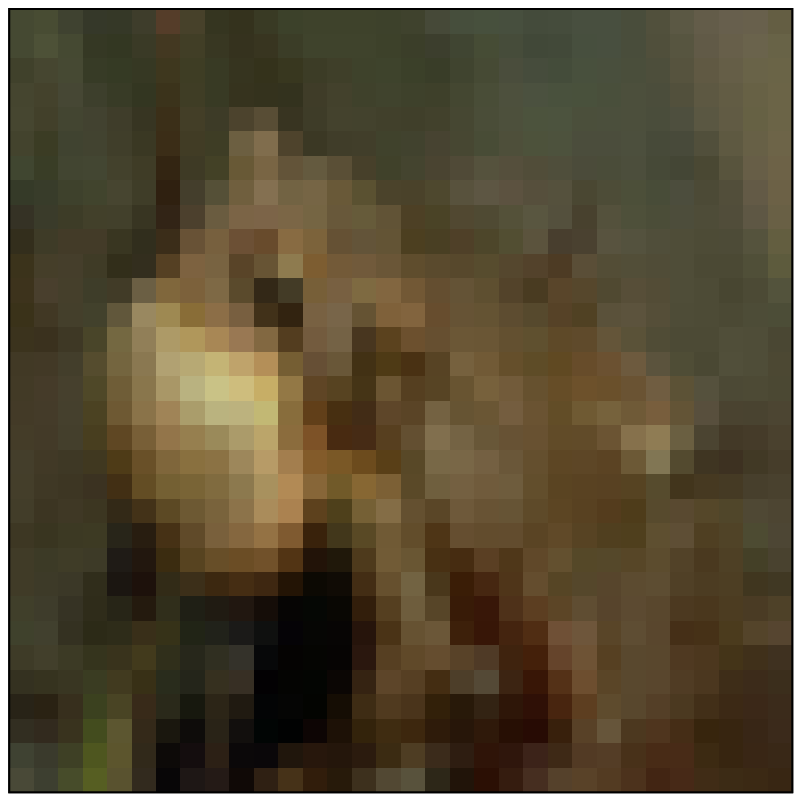}
      \hspace{0.005\textwidth}
      \includegraphics[width=0.08\textwidth]{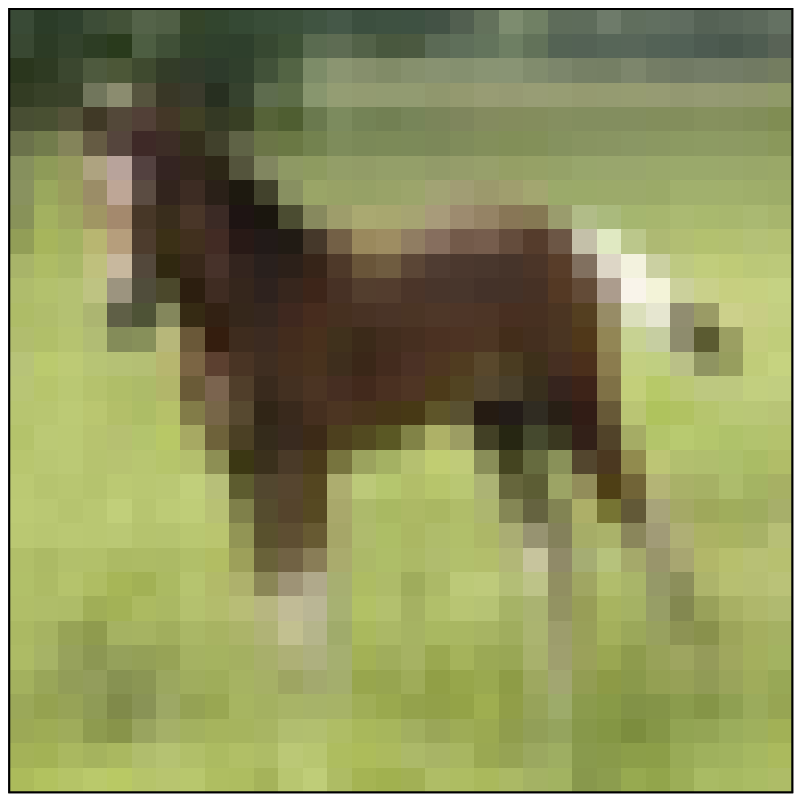}
      \hspace{0.005\textwidth}
      \includegraphics[width=0.08\textwidth]{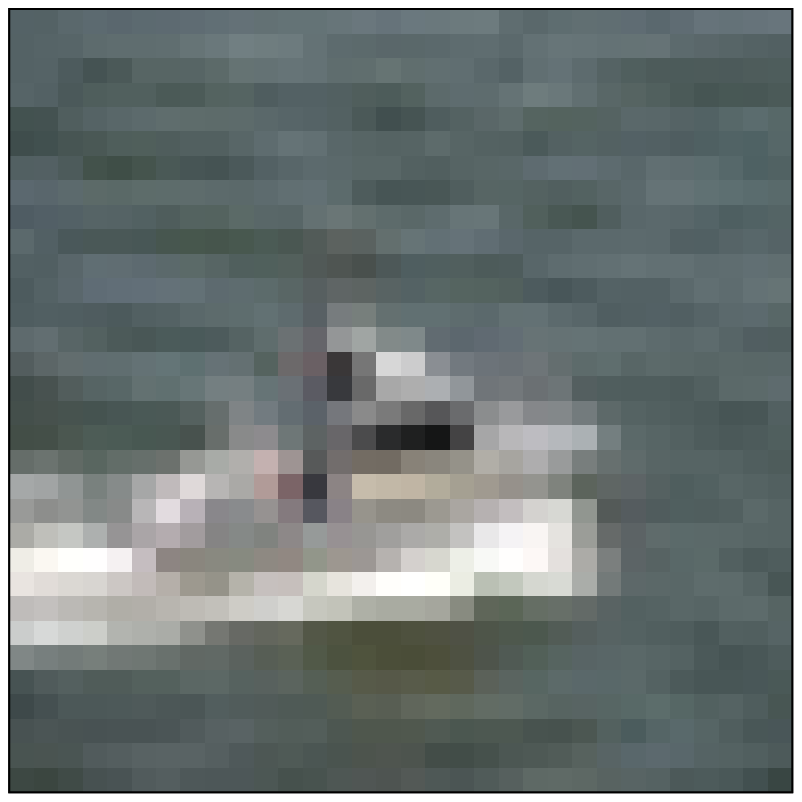}
      \hspace{0.005\textwidth}
      \includegraphics[width=0.08\textwidth]{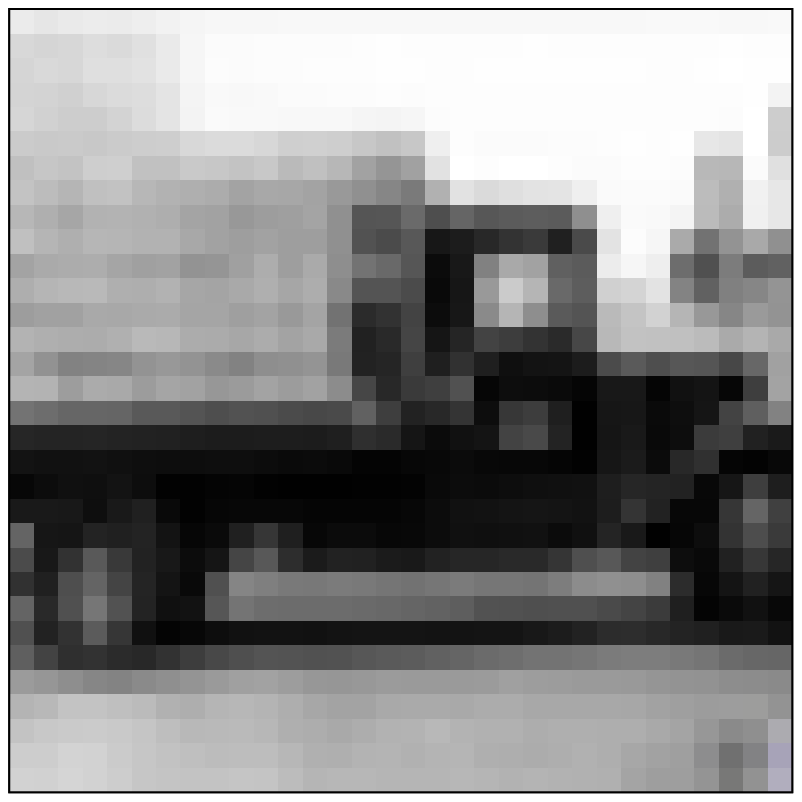}
    \end{minipage}
    \label{fig:cifar10}
  }
  \subfigure[Projected images in GRP-DNN]
  {
    \begin{minipage}{\textwidth}
      \includegraphics[width=0.08\textwidth]{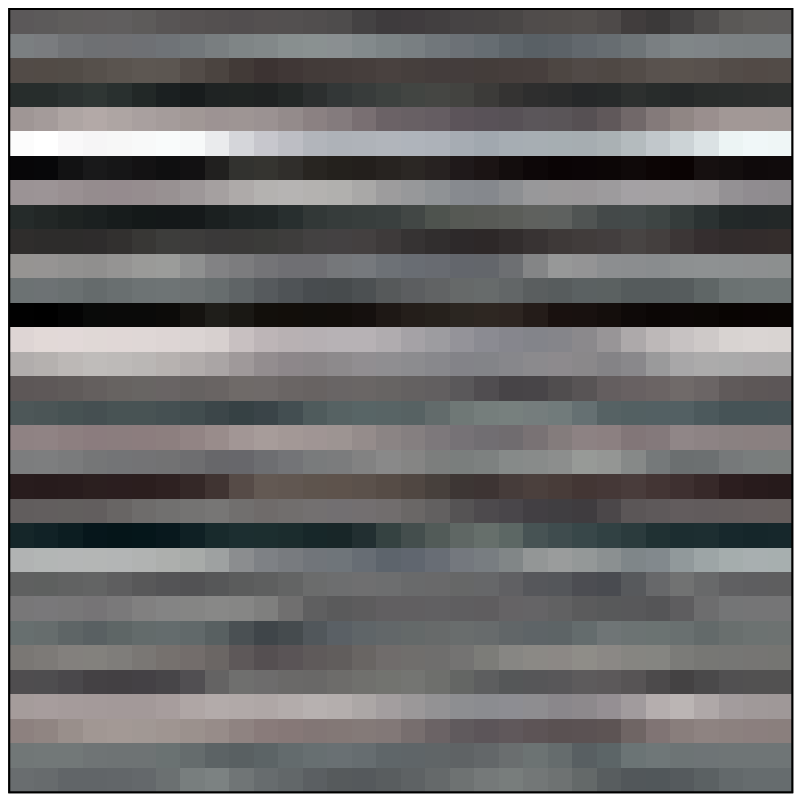}
      \hspace{0.005\textwidth}
      \includegraphics[width=0.08\textwidth]{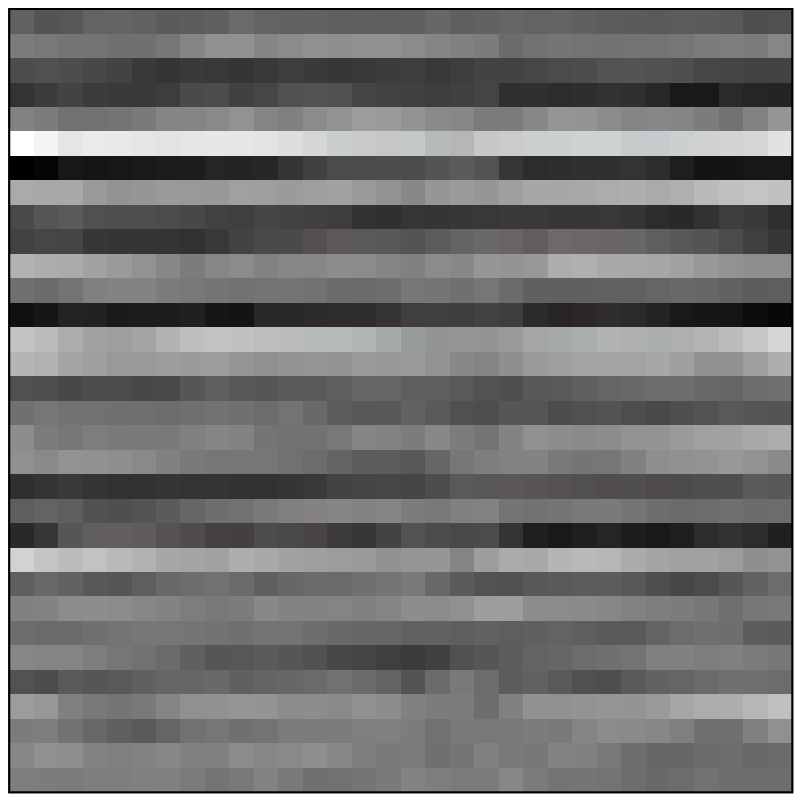}
      \hspace{0.005\textwidth}
      \includegraphics[width=0.08\textwidth]{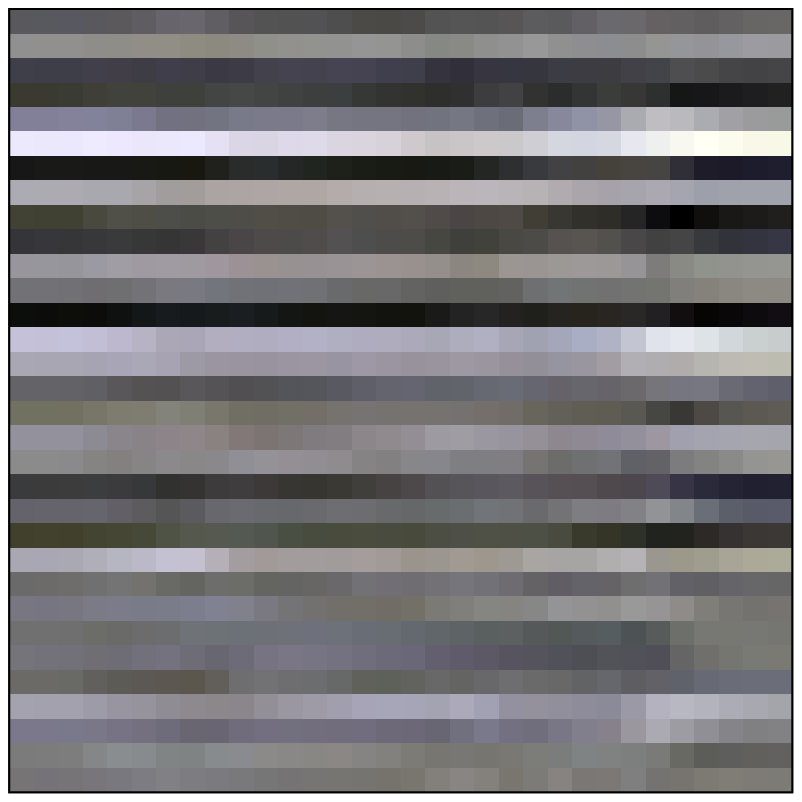}
      \hspace{0.005\textwidth}
      \includegraphics[width=0.08\textwidth]{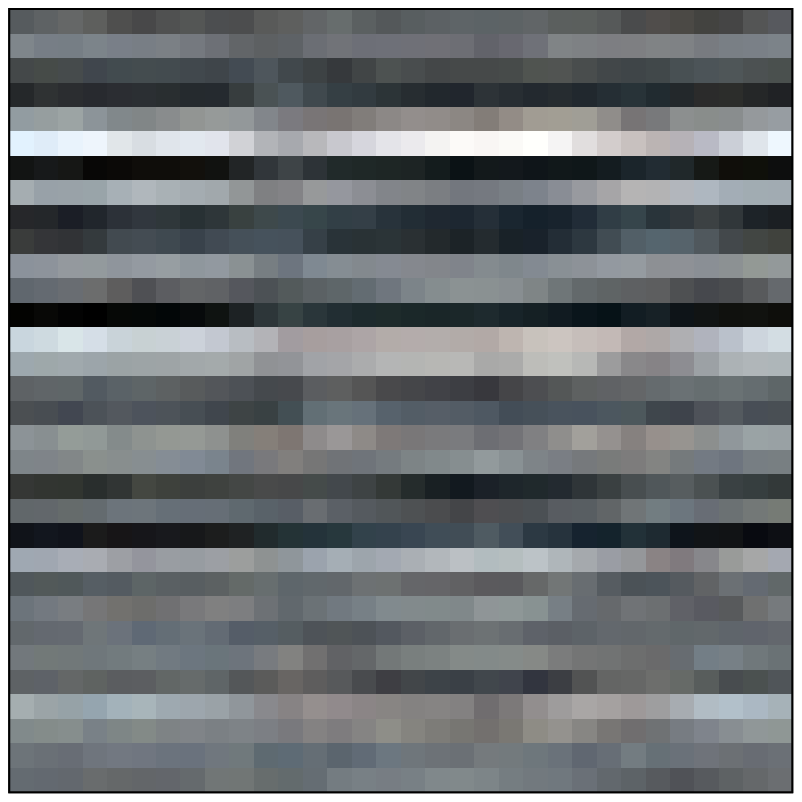}
      \hspace{0.005\textwidth}
      \includegraphics[width=0.08\textwidth]{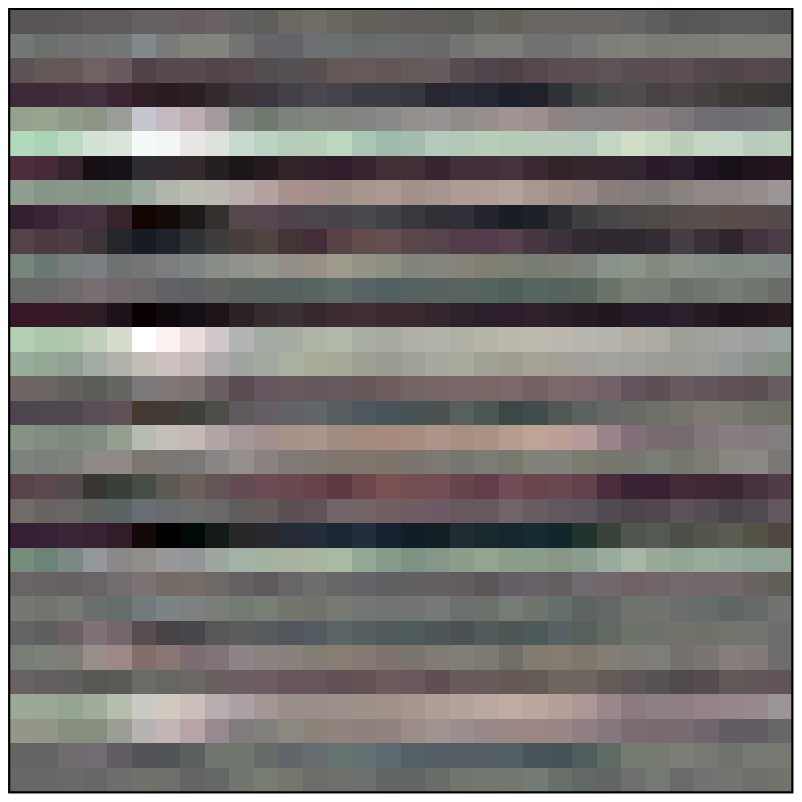}
      \hspace{0.005\textwidth}
      \includegraphics[width=0.08\textwidth]{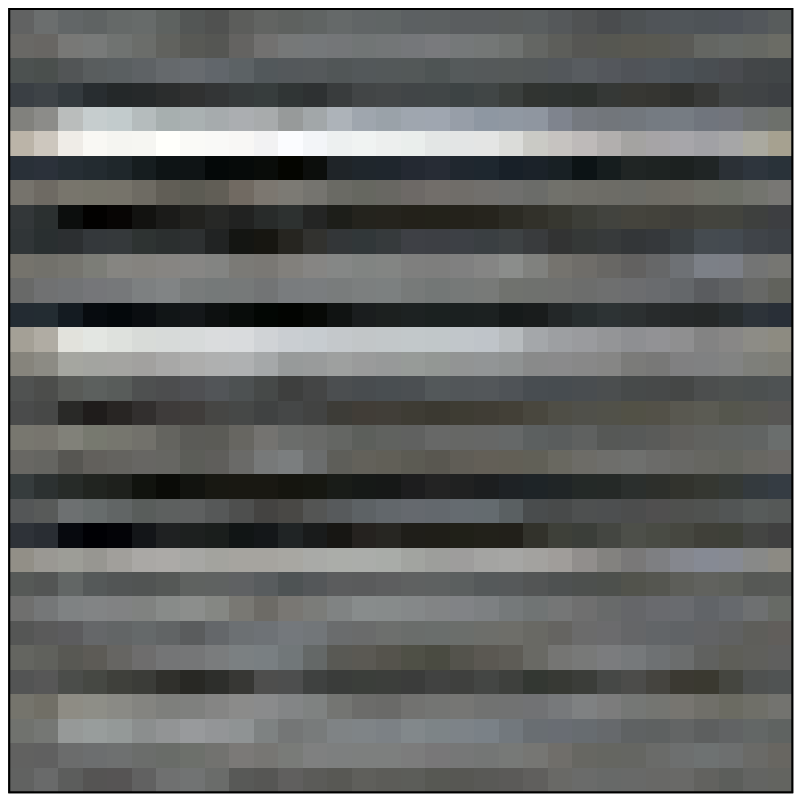}
      \hspace{0.005\textwidth}
      \includegraphics[width=0.08\textwidth]{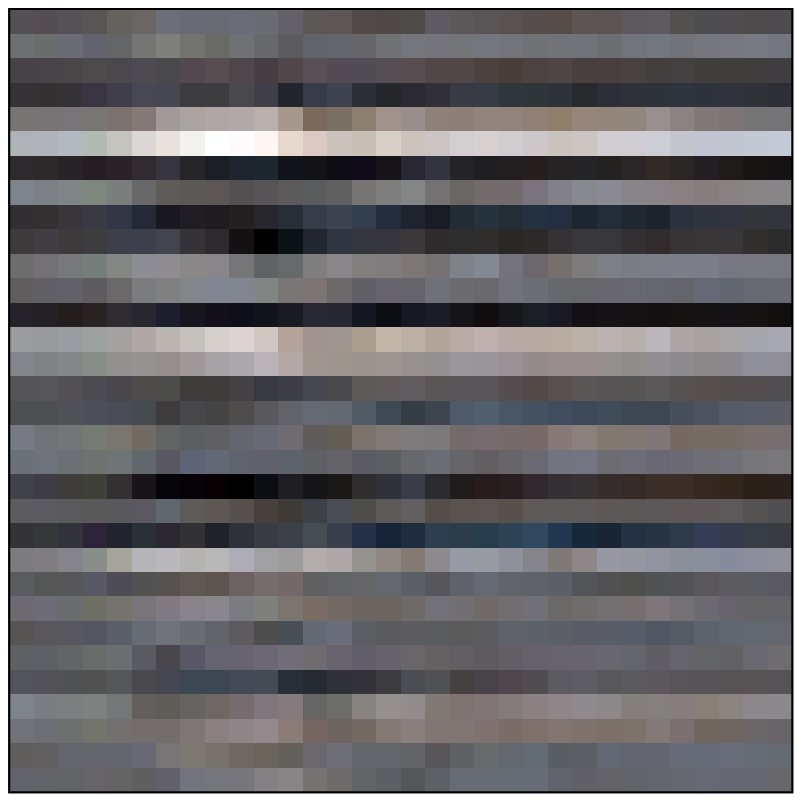}
      \hspace{0.005\textwidth}
      \includegraphics[width=0.08\textwidth]{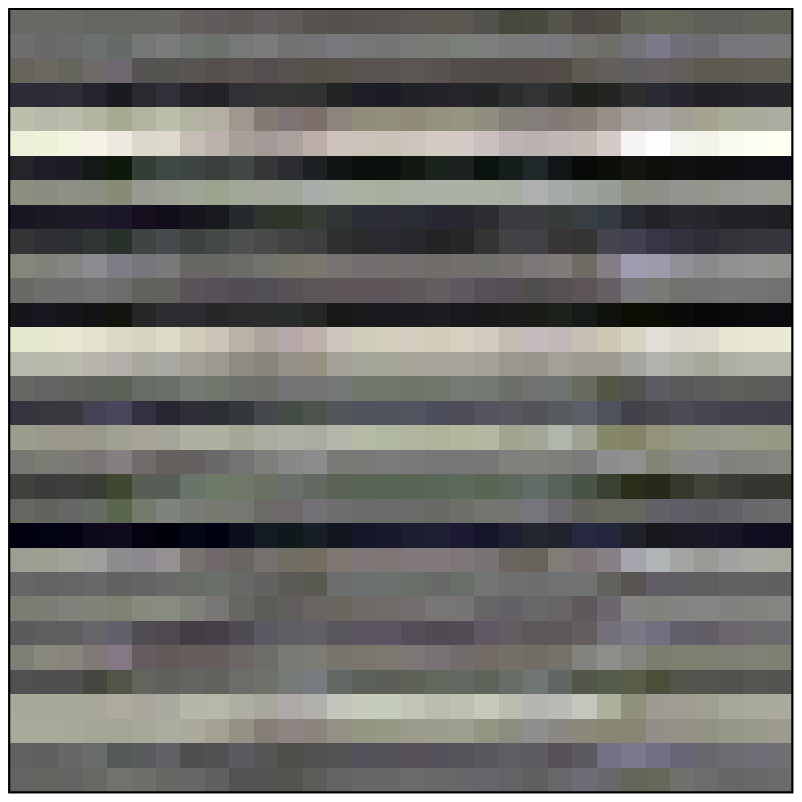}
      \hspace{0.005\textwidth}
      \includegraphics[width=0.08\textwidth]{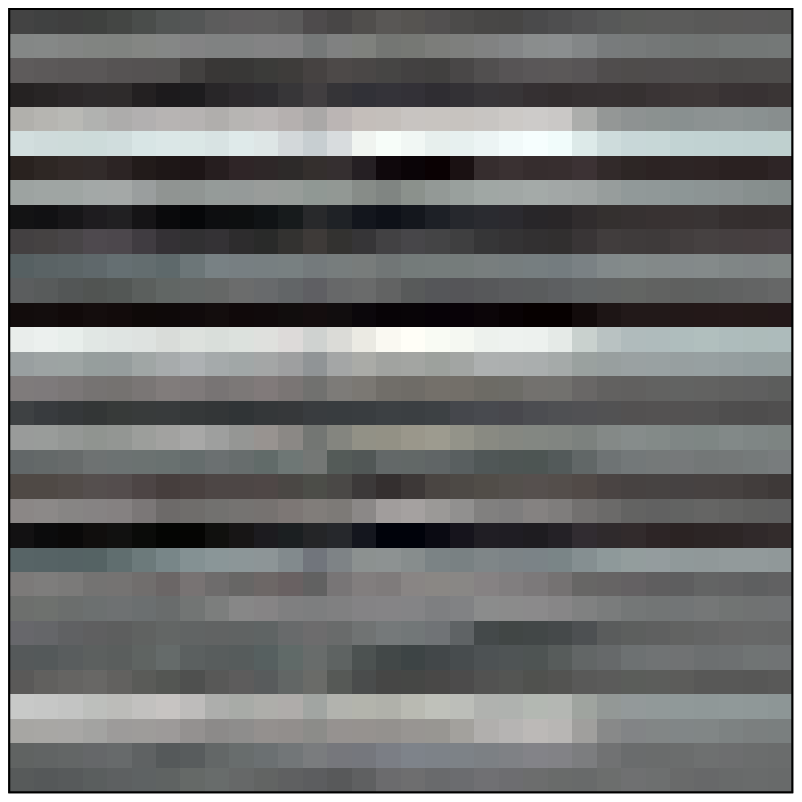}
      \hspace{0.005\textwidth}
      \includegraphics[width=0.08\textwidth]{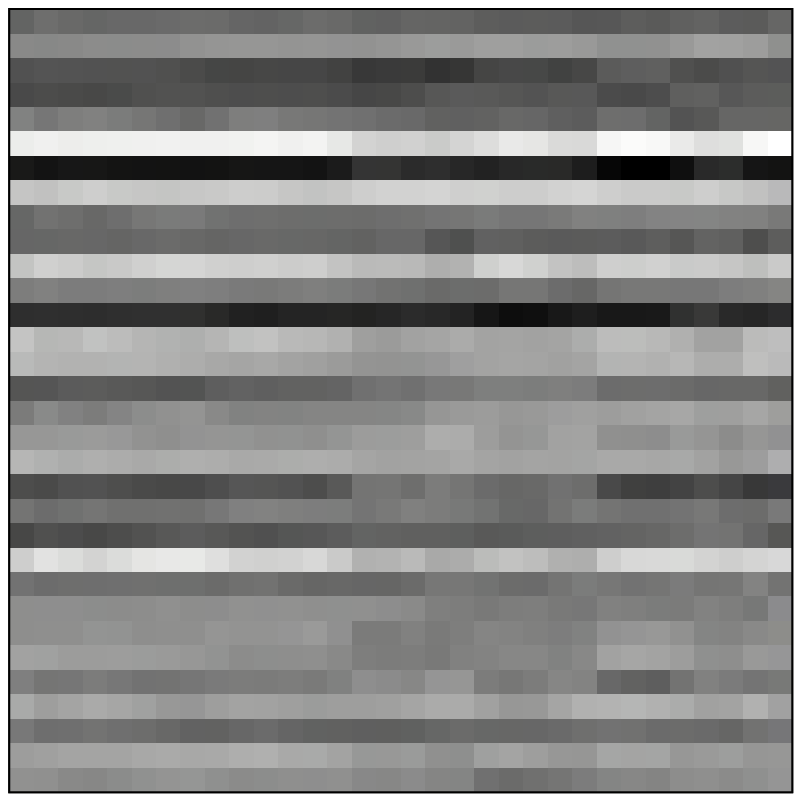}
    \end{minipage}
    \label{fig:cifar-rp}
  }
  \subfigure[Noise-added images in $\epsilon$-DP-DNN ($\epsilon=100$)]
  {
    \begin{minipage}{\textwidth}
      \includegraphics[width=0.08\textwidth]{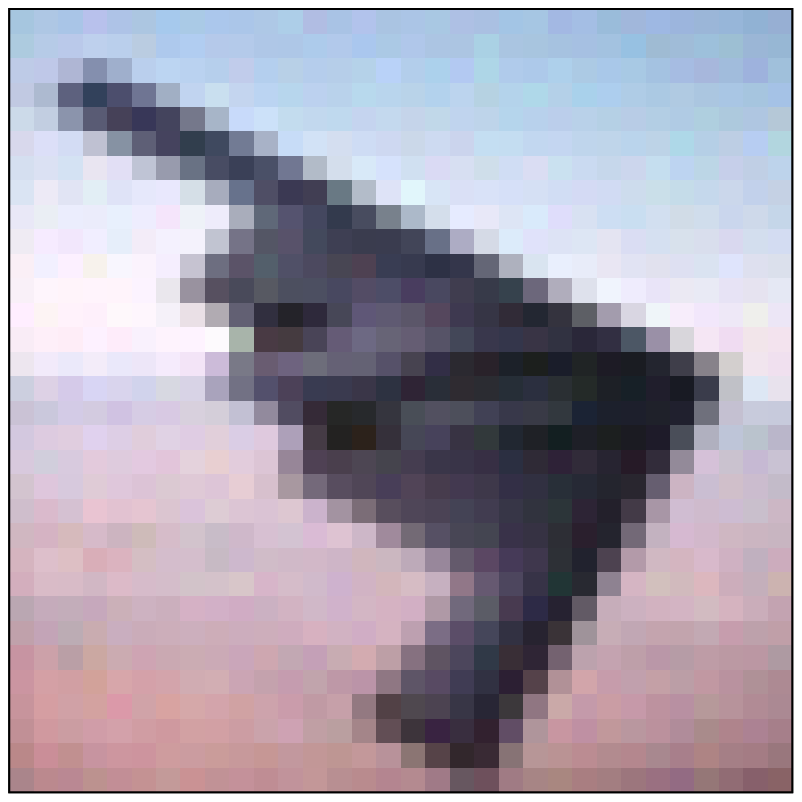}
      \hspace{0.005\textwidth}
      \includegraphics[width=0.08\textwidth]{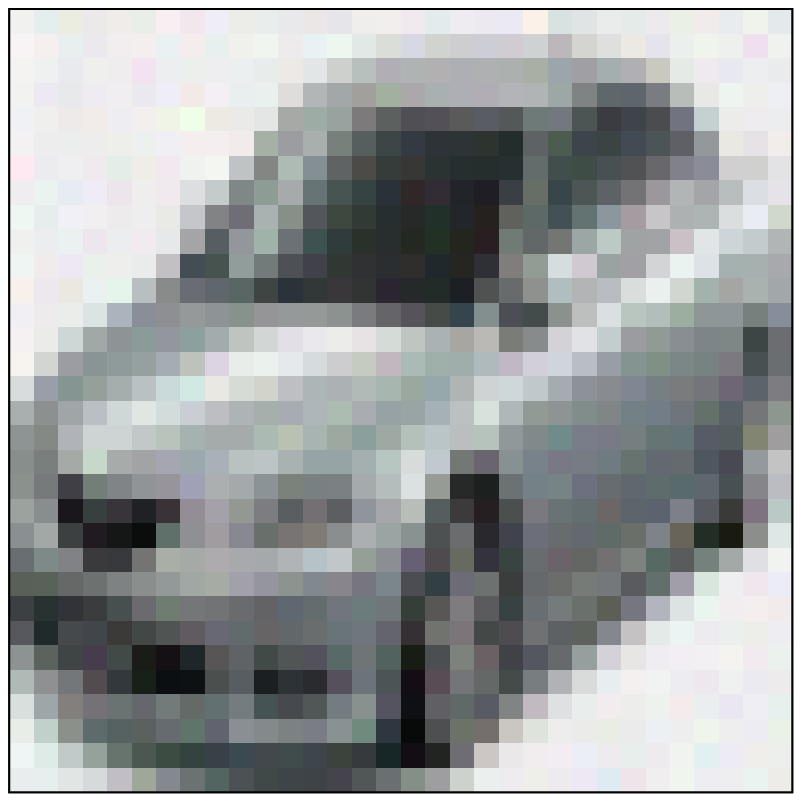}
      \hspace{0.005\textwidth}
      \includegraphics[width=0.08\textwidth]{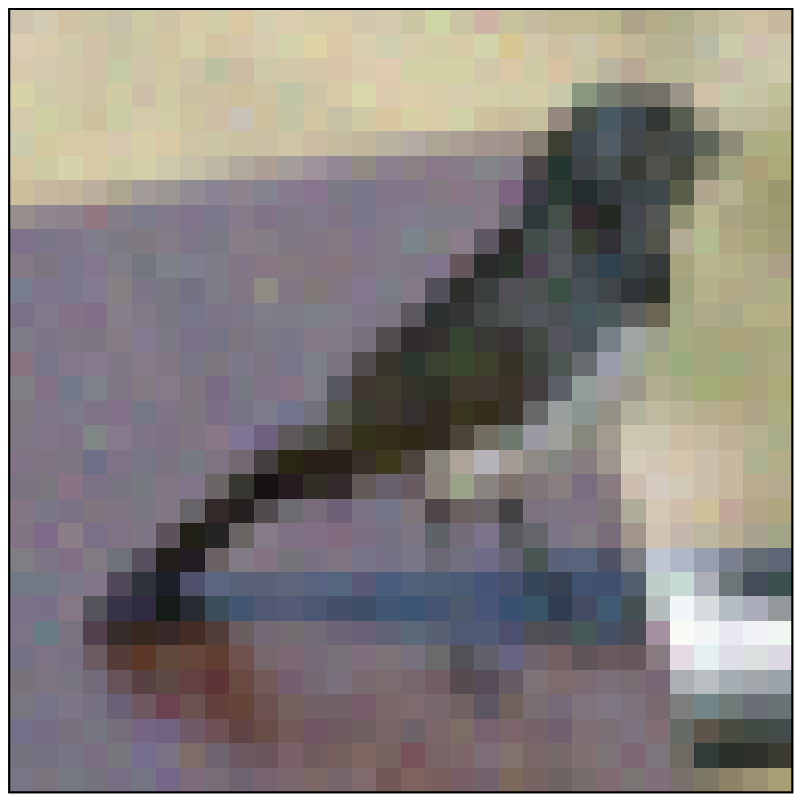}
      \hspace{0.005\textwidth}
      \includegraphics[width=0.08\textwidth]{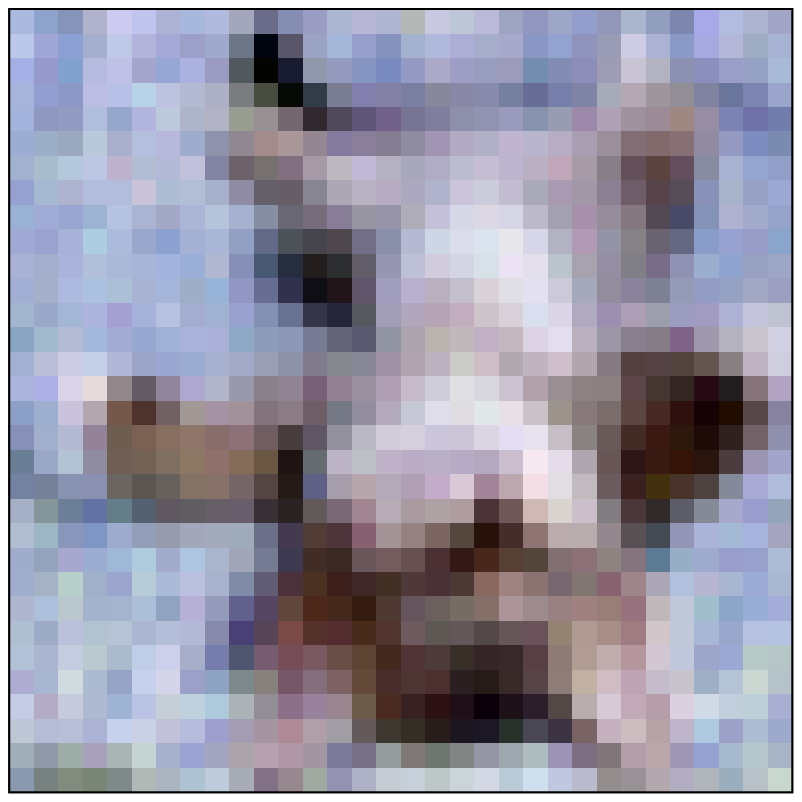}
      \hspace{0.005\textwidth}
      \includegraphics[width=0.08\textwidth]{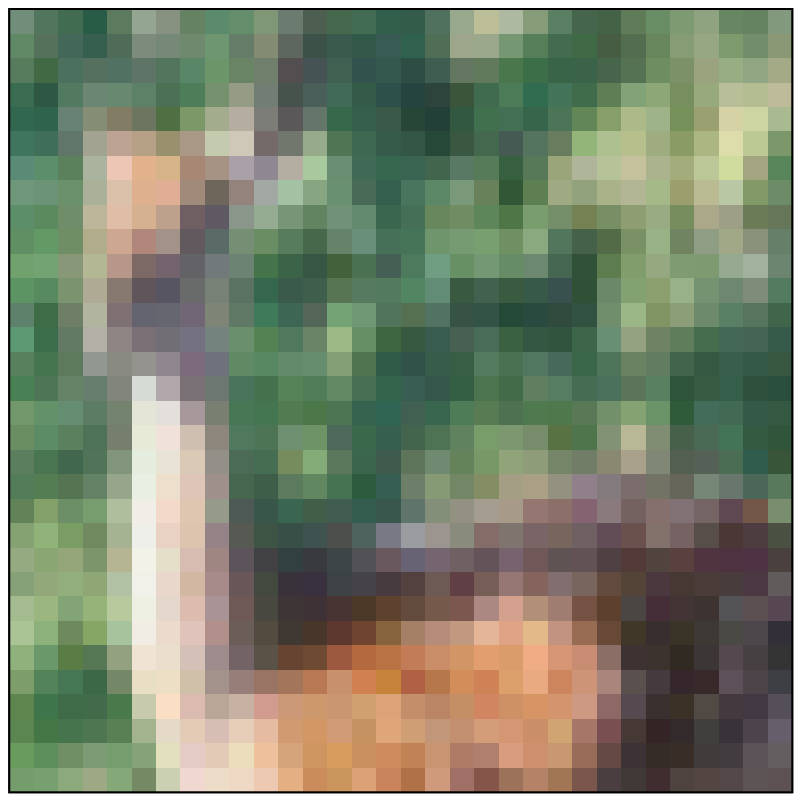}
      \hspace{0.005\textwidth}
      \includegraphics[width=0.08\textwidth]{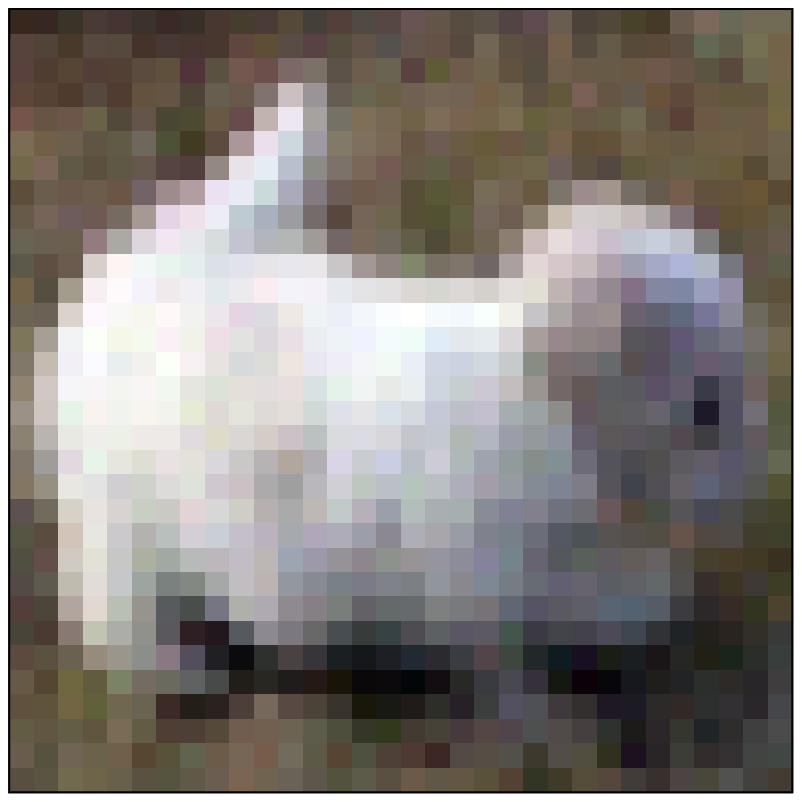}
      \hspace{0.005\textwidth}
      \includegraphics[width=0.08\textwidth]{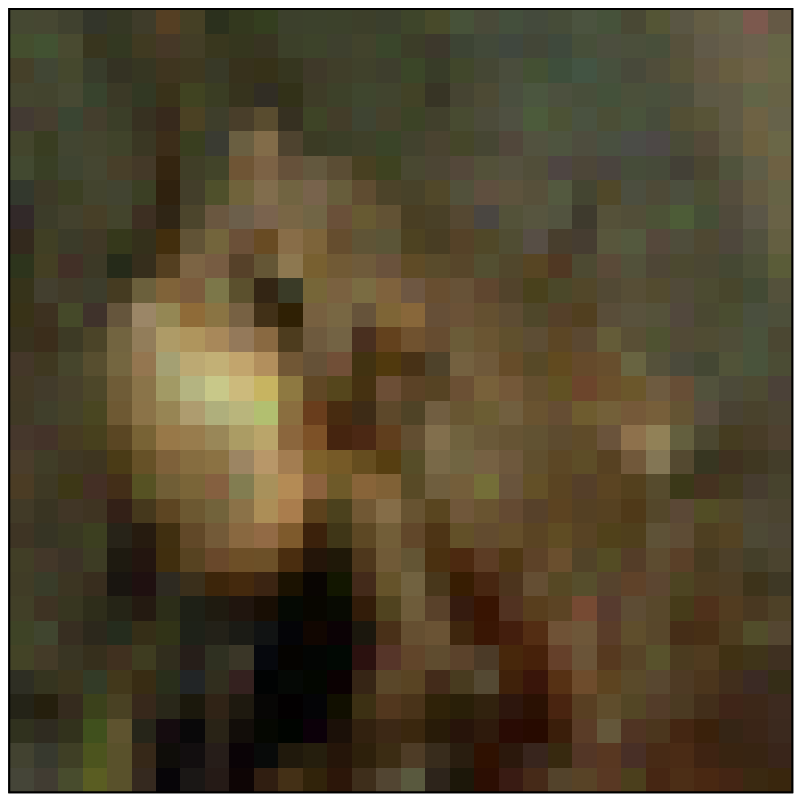}
      \hspace{0.005\textwidth}
      \includegraphics[width=0.08\textwidth]{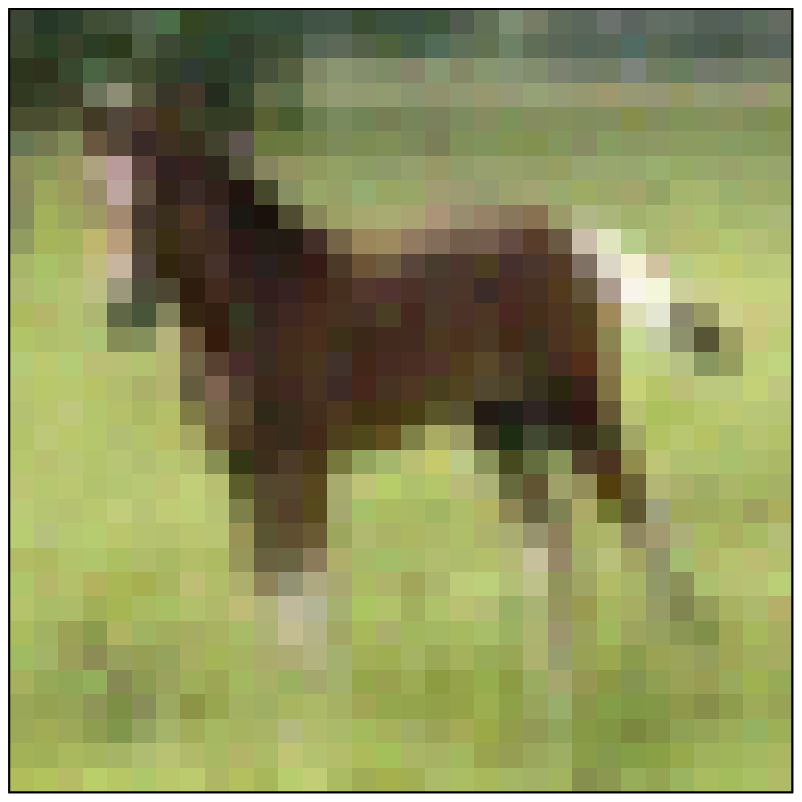}
      \hspace{0.005\textwidth}
      \includegraphics[width=0.08\textwidth]{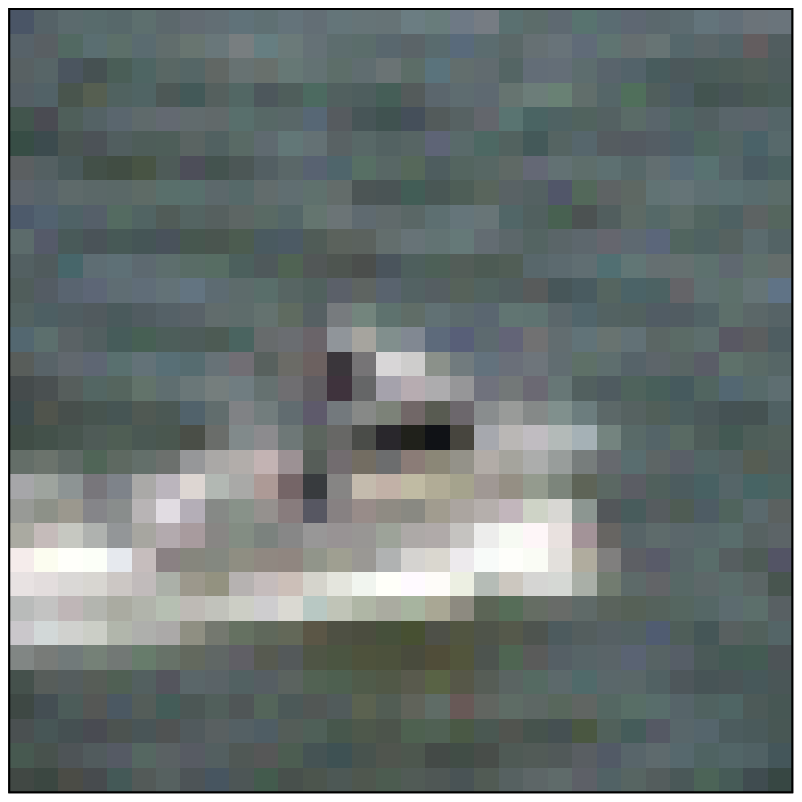}
      \hspace{0.005\textwidth}
      \includegraphics[width=0.08\textwidth]{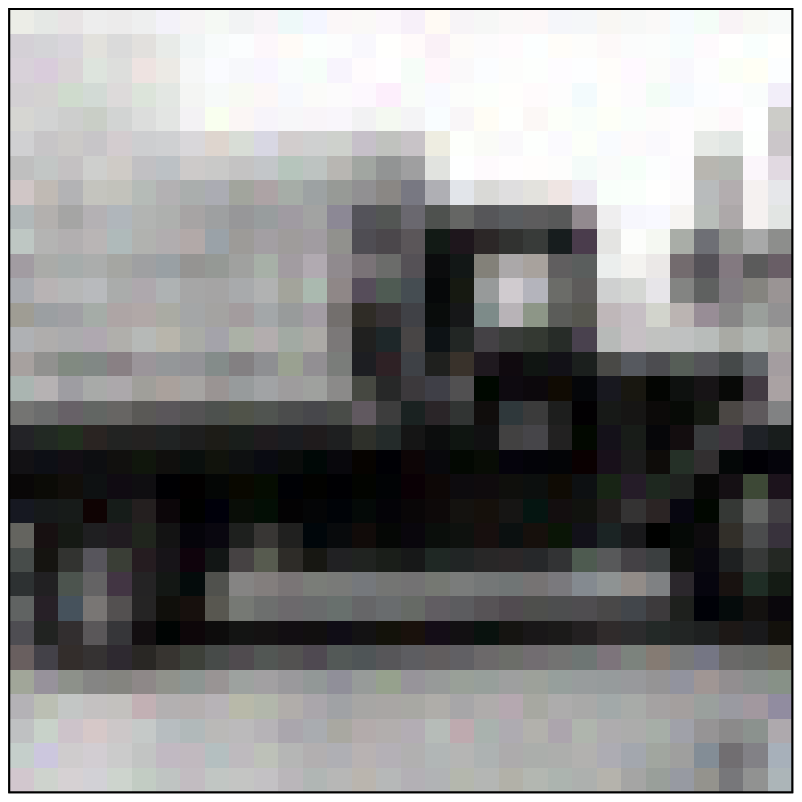}
    \end{minipage}
    \label{fig:cifar-dp100}
  }
  \subfigure[Noise-added images in $\epsilon$-DP-DNN ($\epsilon=10$)]
  {
    \begin{minipage}{\textwidth}
      \includegraphics[width=0.08\textwidth]{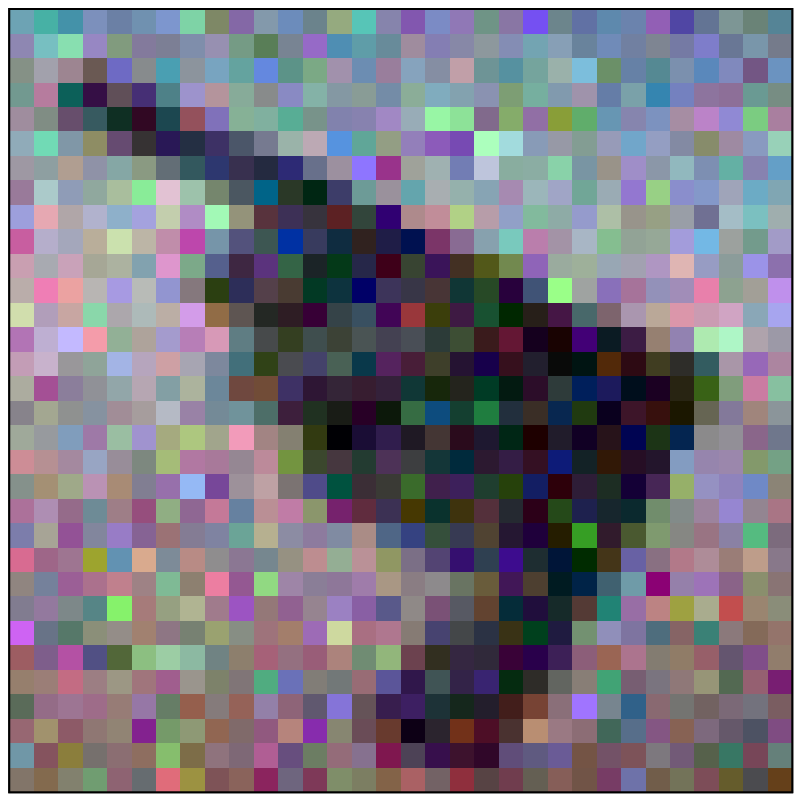}
      \hspace{0.005\textwidth}
      \includegraphics[width=0.08\textwidth]{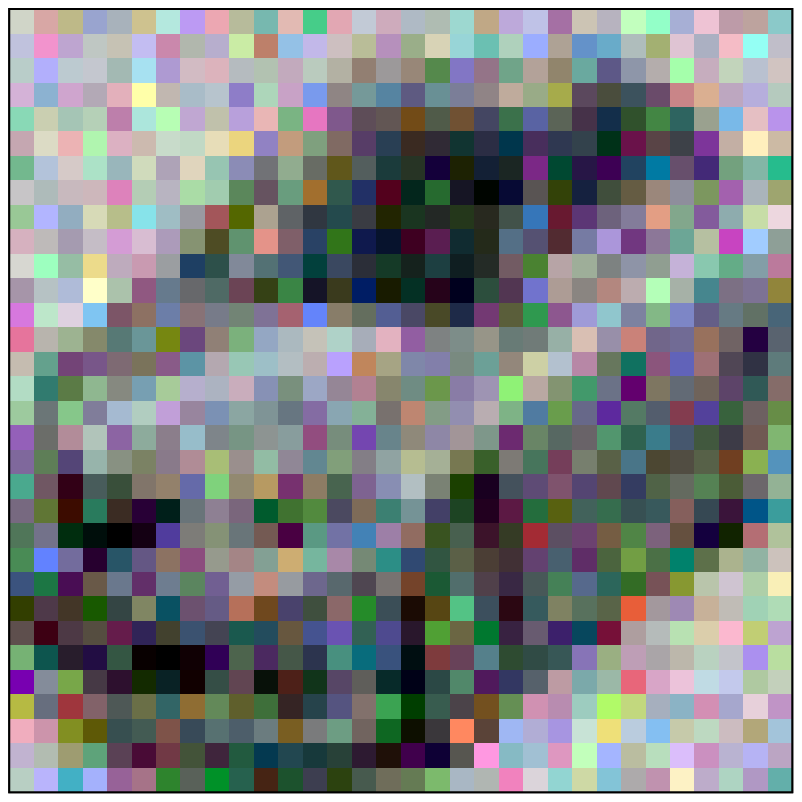}
      \hspace{0.005\textwidth}
      \includegraphics[width=0.08\textwidth]{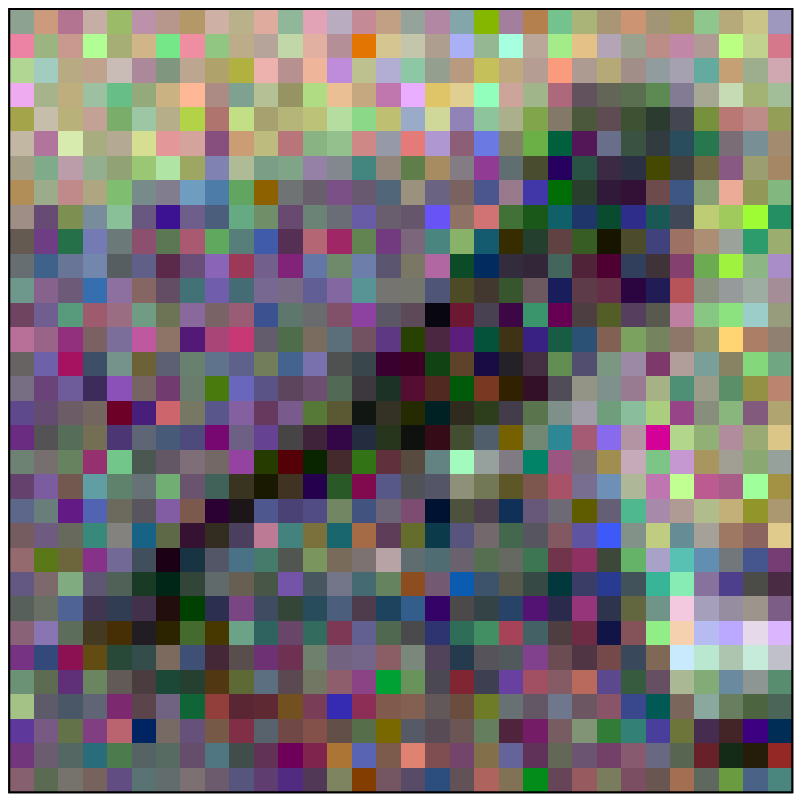}
      \hspace{0.005\textwidth}
      \includegraphics[width=0.08\textwidth]{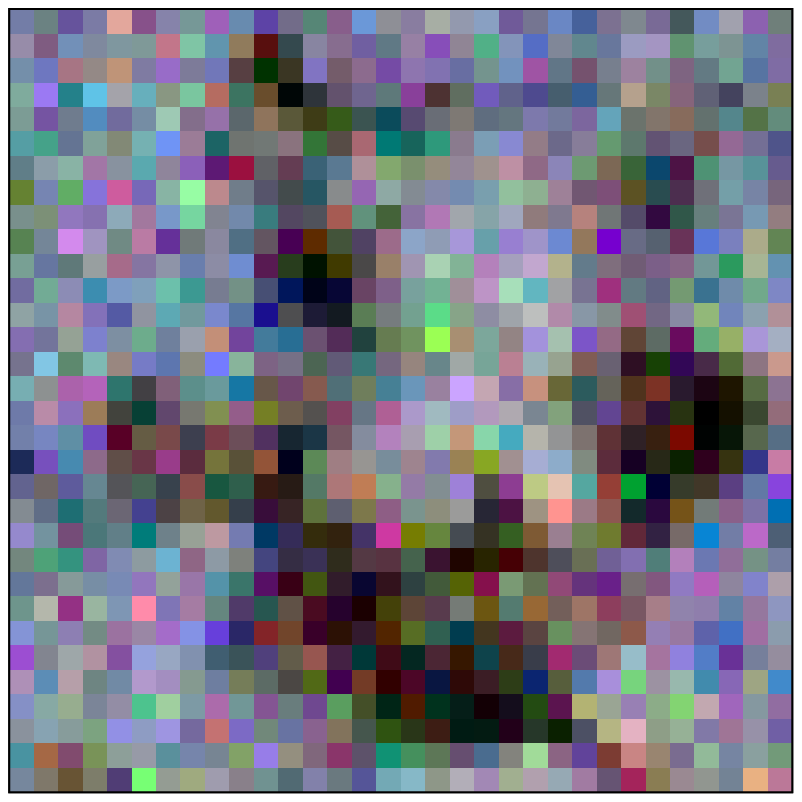}
      \hspace{0.005\textwidth}
      \includegraphics[width=0.08\textwidth]{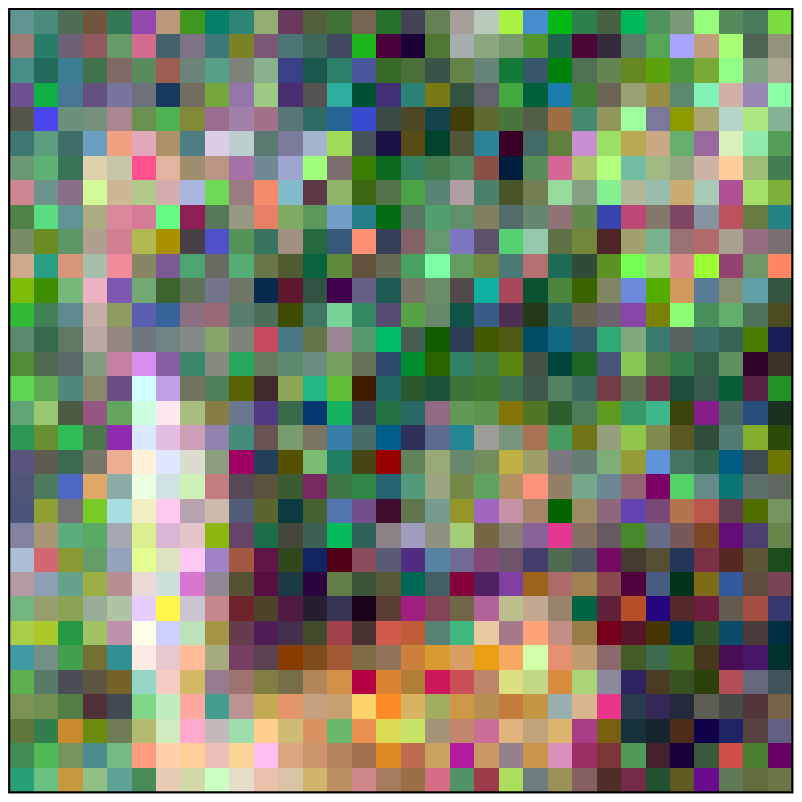}
      \hspace{0.005\textwidth}
      \includegraphics[width=0.08\textwidth]{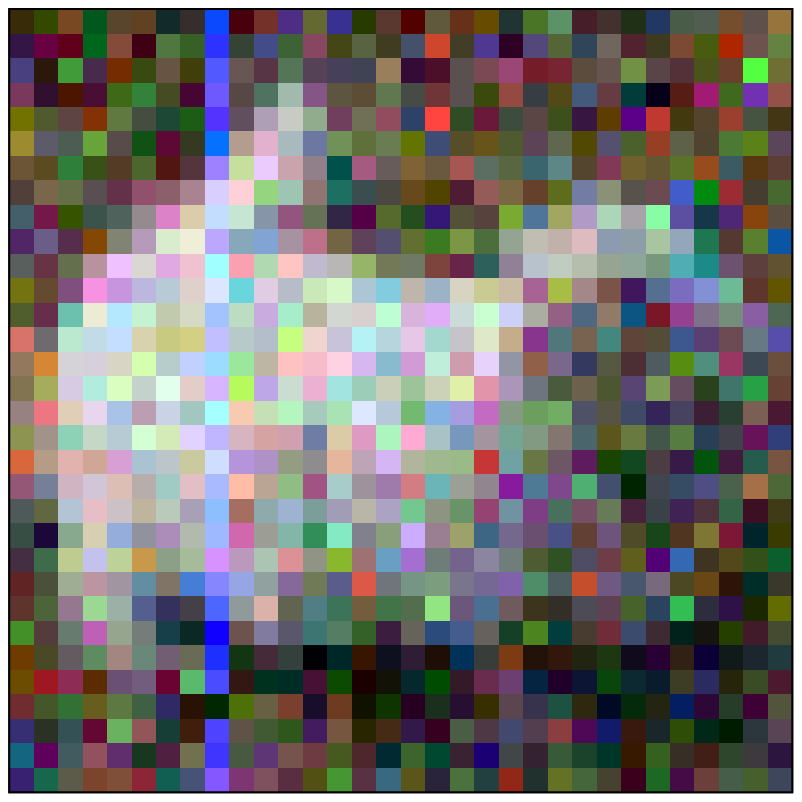}
      \hspace{0.005\textwidth}
      \includegraphics[width=0.08\textwidth]{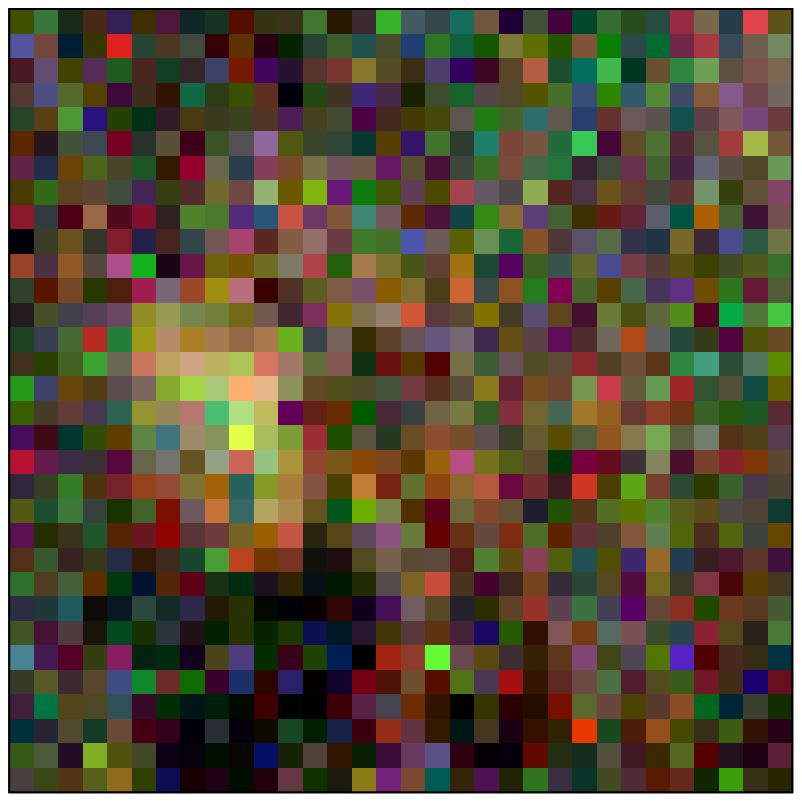}
      \hspace{0.005\textwidth}
      \includegraphics[width=0.08\textwidth]{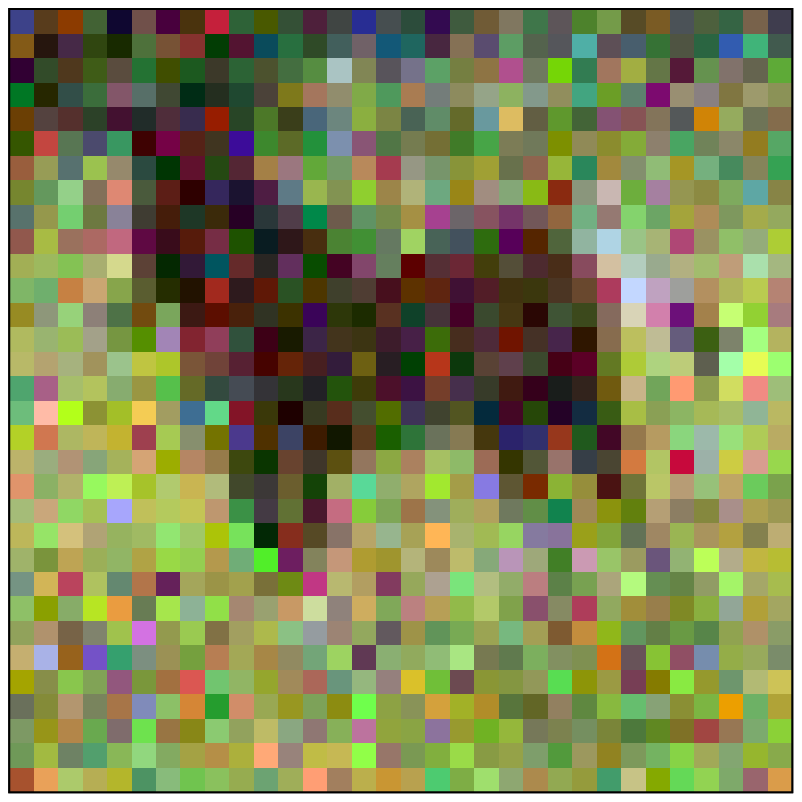}
      \hspace{0.005\textwidth}
      \includegraphics[width=0.08\textwidth]{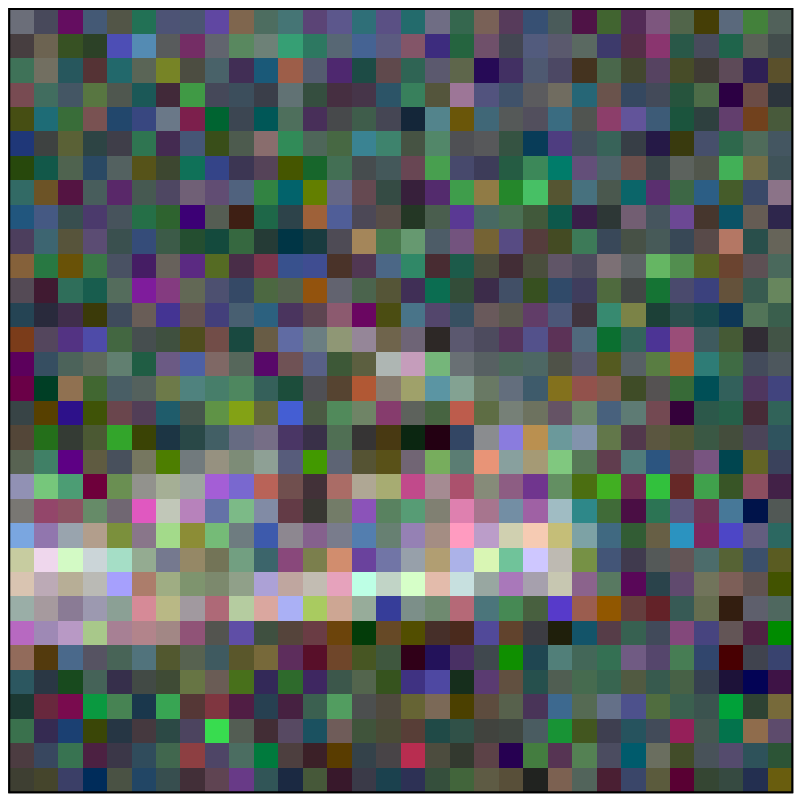}
      \hspace{0.005\textwidth}
      \includegraphics[width=0.08\textwidth]{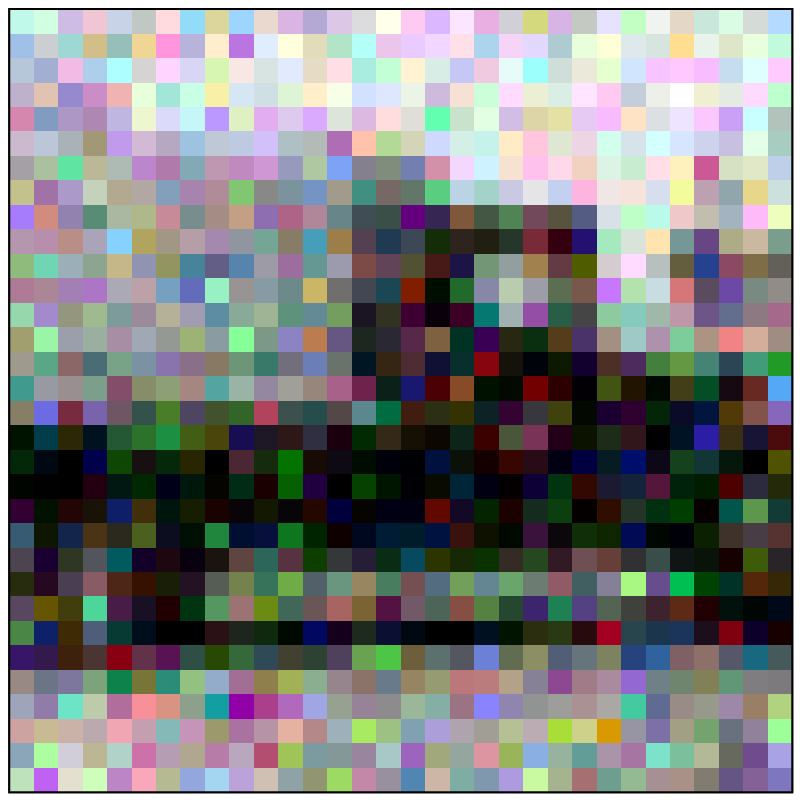}
    \end{minipage}
    \label{fig:cifar-dp10}
  }
  \caption{CIFAR-10 image samples. The classes are airplanes, cars, birds, cats, deers, dogs, frogs, horses, ships, and trucks.}
\end{figure*}

We design a 5-layer MLP classifier to detect spams. The numbers of ReLUs in the five layers are 57, 100, 50, 10, and 2, respectively. A softmax function is used lastly to make the final detection decision. Dropout is used during training to suppress overfitting. Without random projection, the MLP and the SVM with grid research for kernel parameters achieve test accuracy of 96.52\% and 96.25\%, respectively. This shows that the MLP and SVM {\revised{can capture the patterns of spambase well}}.

We evaluate the impact of the number of participants $N$ on the learning performance of GRP-DNN, GRP-NCL, and GRP-SVM. Fig.~\ref{fig:spam-n} shows the results.  The two horizontal lines in Fig.~\ref{fig:spam-n} represent the test accuracy of the plain MLP and SVM without any privacy protection. When $N$ increases from 1 to 200, the test accuracy of GRP-DNN decreases from 96\% to 83.25\%. If $N$ is no greater 100, GRP-DNN can maintain a test accuracy of about 90\%. The average test accuracy of GRP-NCL is about 5\% lower than that of the GRP-DNN, because GRP-NCL lacks the advantages of collaborative learning. The test accuracy of the GRP-SVM is about 1.25\% to 2.75\% lower than that of the GRP-DNN. Thus, the GRP-SVM performs satisfactorily for this spambase dataset. The reasons are two-fold. First, in this spambase dataset, the classifiers operate on the e-mail features, rather than the raw data. Second, the RBF kernel is effective in capturing the features. In fact, the nature of this spambase dataset is similar to that of the 2-dimensional and 10-dimensional generated feature datasets used in \sect\ref{subsec:example}, on which the GRP-DNN and GRP-SVM perform similarly.

\subsection{Evaluation Results with FSD Dataset}
{\blue  We adopt a modified version of the CNN used in \cite{xu2019lightweight} to recognize spoken digits. Fig.~\ref{fig:soundcnn} shows the structure of the CNN. The CNN consists of three convolutional layers, one max-pooling layer, and three dense layers. Zero padding is performed to the input image in the convolutional layers and the maxpooling layer. We apply ReLu activation function  to the output of every convolutional and dense layer except for the last layer. ReLU rectifies a negative input to zero. The last dense layer has 10 neurons with a softmax activation function corresponding
 to the 10 classes of FSD. Three dropout layers with dropout rate 0.25, 0.1 and 0.25 are applied after the max-pooling
 layer and in the first two dense layers. Specifically, 25$\%$, 10$\%$, and 25$\%$ of the neurons will be abandoned randomly
 from the neural network in the training process. Without random projection, the CNN  achieves test accuracy of 98.24$\%$.

We evaluate the impact of the number of participants $N$ on the learning performance of GRP-DNN in Fig.~\ref{fig:sound-n}. Without any projection, the CNN achieves test accuracy of 98.24$\%$. When $N$ increases from 10 to 50, the test accuracy decreases from 95.27$\%$ to 86.21$\%$. The results imply that our approach works well on the FSD Dataset.

}

 \begin{figure}
 	\centering
 	\normalsize
 	\resizebox{0.85\textwidth}{!}{
 		
 		\begin{tabular}[h]{ccc|c|c|c|c|c|c|c|c|c|c|c|c|cc}
 			\cline{4-8} \cline{10-15}
 			& \rotatebox[origin=c]{90}{MFCC representation}  &$\!\!\!\!\Rightarrow \!\!$
 			& \rotatebox[origin=c]{90}{$\;$ 32 $2\!\!\times \!\!2$ conv filters $\;$}
 			&\rotatebox[origin=c]{90}{$\;$ 48 $3\!\!\times \!\!3$ conv filters  $\;$}
 			&\rotatebox[origin=c]{90}{$\;$ 64 $3\!\!\times \!\!6$ conv filters  $\;$}
 			&\rotatebox[origin=c]{90}{max-pooling ($2\!\!\times \!\!2$)}
 			&\rotatebox[origin=c]{90}{dropout ($0.25$)}
 			&$\!\!\Rightarrow \!\!$	
 			& \rotatebox[origin=c]{90}{128 neurons}
 			&\rotatebox[origin=c]{90}{dropout ($0.1$)}
 			& \rotatebox[origin=c]{90}{64 neurons}
 			&\rotatebox[origin=c]{90}{dropout ($0.25$)}
 			& \rotatebox[origin=c]{90}{10 neurons}
 			& \rotatebox[origin=c]{90}{softmax}
 			& $\!\!\Rightarrow \!\!$
 			& \rotatebox[origin=c]{90}{classification result}
 			\\
 			
 			\cline{4-8} \cline{10-15}
 			&\multicolumn{2}{c}{}
 			& \multicolumn{5}{c}{conv layers}
 			& \multicolumn{1}{c}{}
 			& \multicolumn{6}{c}{dense layers}
 			& \multicolumn{1}{c}{}
 			\\
 		\end{tabular}
 	}
 	\caption{{\blue Structure of CNN for FSD recognition.}}
 	\label{fig:soundcnn}
 	
 \end{figure}

 \begin{figure}
 	\includegraphics[width=0.85\textwidth]{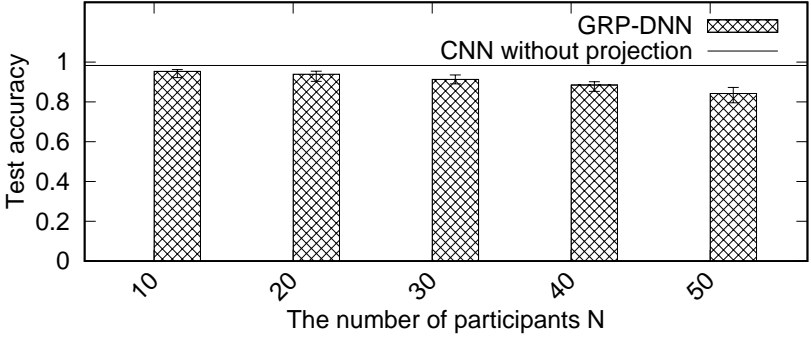}
 	\caption{{\blue Impact of the number of participants on the learning performance (FSD). The error bars represent min and max.}}
 	\label{fig:sound-n}
 \end{figure}

\subsection{Evaluation Results with CIFAR-10 Dataset}
\label{subsec:cifar-10}

To classify the more complex CIFAR-10 images, we adopt the residual neural network (ResNet) \cite{He16}.
In general, to capture more complex patterns, deeper neural networks will be needed, which often face degraded learning performance, however. ResNet is designed to address this challenge for very deep neural networks.
In our experiments, we use the ResNet-152, which contains 152 layers. Specifically, it consists of {\em blocks}, each of which consists of convolutional layers and ReLU-based dense layers.
After the blocks, ResNet-152 has a fully-connected neural network to make the final classification decision.
Without random projection, the ResNet-152 achieves a test accuracy of 95\%. This shows that the ResNet-152 {\revised{can capture the patterns of CIFRA-10 well}}. In contrast, without random projection, the SVM with grid search for kernel parameters achieves a test accuracy of 33\% only. This shows that, due to the high complexity of the patterns in CIFAR-10, no efficient RBF kernels can be found to create proper hyperplanes for classification.


\begin{figure}
  \centering
  \includegraphics[width=0.85\textwidth]{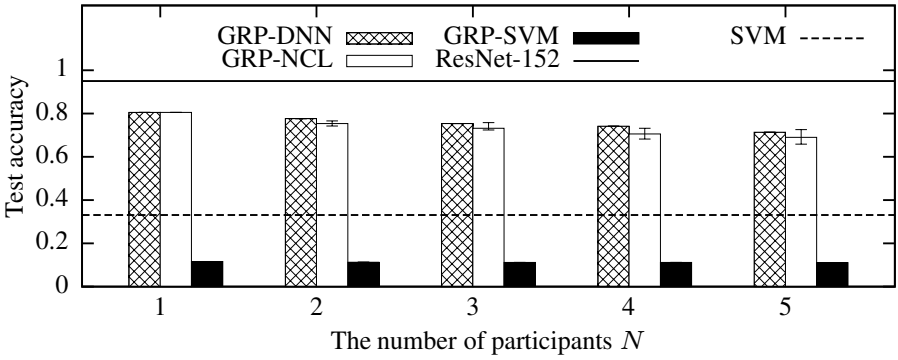}
  \caption{Impact of the number of participants on the learning performance (CIFAR-10).}
  \label{fig:Cifradata1}
\end{figure}

First, we evaluate the impact of the number of the participants $N$ on the learning performance of different approaches. Fig.~\ref{fig:Cifradata1} shows the results.  The two horizontal lines in Fig.~\ref{fig:Cifradata1} represent the test accuracy of the plain ResNet-152 and SVM without any privacy protection.
When $N=1$, the test accuracy of GRP-DNN is 80.6\%. Thus, compared with the test accuracy of ResNet-152 without  privacy protection, the random projection results in a test accuracy drop of 14.4\%. The test accuracy of GRP-DNN decreases with the number of participants. We think the performance drops are caused by the much more complicated data patterns after the projection, {\revised{that exceed the complexity that ResNet-152 can handle well.}}  Note that CIFAR-10 had been a challenging dataset until the high accuracy achieved by deep models in recent years. To address the substantially additional pattern complexity introduced by GRP, deeper ResNets may help. But they will require more training data to avoid overfitting. The average test accuracy of the GRP-NCL is slightly lower than the test accuracy of the GRP-DNN. This result is similar to that based on the MNIST and spambase datasets. The test accuracy of GRP-SVM is around 11\%, close to that of random guessing.


\begin{figure}
   \centering
   \begin{minipage}[t]{0.48\textwidth}
   \centering
   \includegraphics{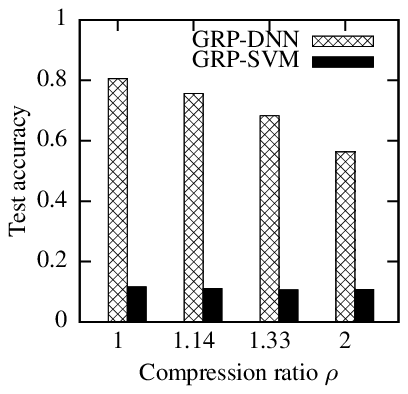}
   \caption{Impact of data compression (CIFAR-10, $N=1$).}
   \label{fig:Cifradata2}
  \end{minipage}
  \hspace{0.01\textwidth}
  \begin{minipage}[t]{0.48\textwidth}
   \centering
   \includegraphics{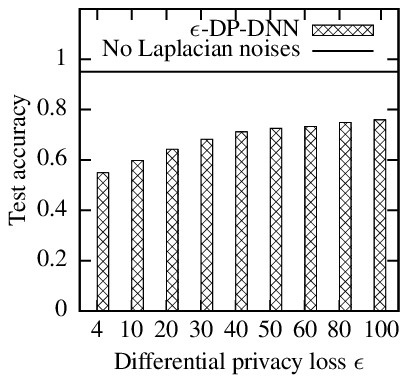}
   \caption{Impact of differential privacy loss (CIFAR-10).}
   \label{fig:cifar-dp-acc}
  \end{minipage}
\end{figure}


Fig.~\ref{fig:Cifradata2} shows the impact of the compression ratio of the projection on the learning performance. The test accuracy of the GRP-DNN decreases with the compression ratio. Compared with the results in Fig.~\ref{fig:mnistdata1} for MNIST, the GRP-DNN on the CIFAR-10 is more sensitive to the compression ratio because that CIFAR-10 images are less sparse, and thus less compressible, than the MNIST images. In Fig.~\ref{fig:Cifradata2}, under all settings for compression ratio, the GRP-SVM's performance is consistently close to random guessing.


Fig.~\ref{fig:cifar-dp-acc} shows the test accuracy of $\epsilon$-DP-DNN versus the DP loss level $\epsilon$. When $\epsilon=100$ (small Laplacian noises and large differential privacy loss), the $\epsilon$-DP-DNN achieves a test accuracy of 75.9\%, almost 20\% lower than the test accuracy achieved without Laplacian noises. Fig.~\ref{fig:cifar-dp100} shows the noise-added CIFAR-10 images under the setting $\epsilon=100$. We can see that it is almost identical to the original CIFAR-10 images in Fig.~\ref{fig:cifar10}. This result echos the understanding that deep learning is not robust to small perturbations \cite{goodfellow15,bose2018adversarial,zheng2016improving}. When $\epsilon=10$, the content of the noise-added images as shown in Fig.~\ref{fig:cifar-dp10} can still be interpreted. However, from Fig.~\ref{fig:cifar-dp-acc}, the test accuracy further reduces to 59.8\% only. For comparison, Fig.~\ref{fig:cifar-rp} shows the projected images. The content of the projected images cannot be interpreted.


\subsection{Summary and Discussion}

We have several observations from the results in \sect\ref{subsec:MNIST}, \sect\ref{subsec:spambase}, and \sect\ref{subsec:cifar-10}.
\begin{itemize}
\item Compared with SVM, deep learning can better adapt to the complexity introduced by the multiplicative projections.
\item Although the GRP-NCL approach additionally uses the identities of the participants, it gives inferior performance compared with the collaborative GRP-DNN. This shows the advantage of collaborative learning even with the privacy preservation requirement.
\item Compared with RRP-DNN and BRP-DNN, GRP-DNN gives higher test accuracy. However, there exists a trade-off between computation overhead and test accuracy  in choosing the type of random projection.
\item Compared with GRP-DNN, the additive noisification for $\epsilon$-DP achieves inferior trade-off between learning performance and protecting confidentiality of raw forms of training data.
\item GRP-DNN shows promising scalability with the number of participants sensing modalities including image, text and voice with low-complexity patterns to be recognized. For the MNIST and spambase datasets, the GRP-DNN can well support 100 participants with a few percents test accuracy drop. {\blue For the FSD dataset, the GRP-DNN can support at least {\revised{40 participants at a cost of a few percentage points in test accuracy}}.} {\blue  Besides, as our approach is based on the deep learning in IoT, sufficient amount of labeled training data from each participant is needed.} For large-scale PPCL systems involving more participants, we envision a two-tier system architecture as follows. The participants are divided into groups. At the first tier, our GRP-DNN is applied within each group; at the second tier, the DML approach is applied among the group coordinators.
\end{itemize}



\section{implementation and Benchmark}
\label{sec:implementation}

In this section, we measure the overhead of two PPCL approaches (i.e., our GRP-DNN and Crowd-ML \cite{Hamm15}) and a privacy-preserving classification outsourcing approach (i.e., CryptoNets \cite{gilad2016cryptonets}) on a testbed of 14 Raspberry Pi 2 Model B nodes \cite{pi} and a powerful workstation computer. The Raspberry Pi nodes act as PPCL participants and the workstation acts as the coordinator. They are interconnected using a 24-port network switch. We benchmark these approaches using the MNIST dataset. The training and testing samples are evenly allocated to the participants, resulting in 4,285 training samples and 714 testing samples on each participant. The implementations of the three approaches (GRP-DNN, Crowd-ML, CryptoNets) on the same platform, i.e., Raspberry Pi, allow fair comparisons. The participant part of our GRP-DNN can be implemented on mote-class platforms. Our previous work \cite{tan2017joint} has implemented Gaussian matrix generation and GRP on the MSP430-based Kmote platform. However, it is difficult/impossible to implement Crowd-ML and CryptoNets on mote-class platforms.


We implement our GRP-DNN approach on the testbed. The compression ratio $\rho=1$ (i.e., no compression). Table~\ref{tab:performance} shows the benchmark results. During the training phase, each GRP-DNN participant needs to transmit a total of $33.6\,\text{MB}$ projected data.
A participant can complete projecting all the 4,285 training images within $0.96\,\text{s}$. The coordinator needs $928.34\,\text{s}$ to train the CNN. In our GRP-DNN implementation, the testing phase is performed on the coordinator. During the testing phase, each participant completes projecting all the 714 testing images within $0.16\,\text{s}$ and transmits a total of $5.6\,\text{MB}$ data to the coordinator. The coordinator needs $40.88\,\text{s}$ to classify all projected testing images from the participants. Note that GPU acceleration is not used in this benchmark for GRP-DNN during both the training and testing phases.

\begin{table}
  \caption{The overhead of various approaches.}
  \label{tab:performance}
  \begin{tabular}{c|c|c|c|c}
    \hline
    & Overhead & GRP-DNN & Crowd-ML & $\!\!$CryptoNets$\!\!$ \\
    \hline
    \multirow{3}{*}{\rotatebox[origin=c]{90}{Training}} & Participant comm. vol. & $33.6\,\text{MB}$ & $117.2\,\text{MB}$ & n/a \\
    & Participant compute time & $0.96\,\text{s}$ & $367.24\,\text{s}$  & n/a \\
    & $\!\!$Coordinator compute time$\!\!$ & $928.34\,\text{s}$ & $1.04\,\text{s}$ & n/a \\
    \hline
    \multirow{3}{*}{\rotatebox[origin=c]{90}{Testing}} & Participant comm. vol. & $5.6\,\text{MB}$ & n/a & $15.0\,\text{MB}$ \\
    & Participant compute time & $0.16\,\text{s}$ & $4.67\,\text{s}$ & 116 hours \\
    & $\!\!$Coordinator compute time$\!\!$ & $40.88\,\text{s}$ & n/a & \\
    \hline
    \multicolumn{5}{l}{n/a represents ``not applicable.''}
  \end{tabular}
\end{table}


The Crowd-ML \cite{Hamm15} is a DML approach. In Crowd-ML, a participant checks out the global classifier parameters from the coordinator and computes the gradients using its own training data. Then, the participants transmit the gradients to the coordinator that will update the global classifier parameters. Thus, during the training phase, the participants and the coordinator repeatedly exchange parameters. We apply an existing implementation of Crowd-ML \cite{crowd-ml-impl} on our testbed.
Our measurement shows that, during the training phase, each participant needs to upload and download a total of $117.2\,\text{MB}$ data, which is 3.5x of our GRP-DNN.  The participant compute time is more than 350x of that under GRP-DNN. Despite the larger volume of data exchanges, Crowd-ML achieves 91.28\% test accuracy only, which is lower than the 95.58\% test accuracy achieved by GRP-DNN. This is because Crowd-ML uses a simple multiclass logistic classifier, which is inferior compared with the CNN used by GRP-DNN in terms of learning performance. Note that during the testing phase of Crowd-ML, the participants execute their local classifiers. Thus, they do not need to transmit the testing samples to the coordinator for classification.


CryptoNets \cite{gilad2016cryptonets} uses homomorphic encryption to encrypt a testing sample during the classification phase and transmits the encrypted sample to the coordinator. Then, the coordinator uses a neural network trained with plaintext data to classify the encrypted testing sample. {\blue Within the homomorphic encryption implementation provided by Microsoft SEAL \cite{Seal}, we have implemented the homomorphic encryption part of CrytoNets that runs on the Raspberry Pis.}
The volume of the 714 encrypted testing images is $15\,\text{MB}$, almost 3x of the data volume generated by random projection. In particular, a Raspberry Pi node takes about 10 minutes and a total of 116 hours to encrypt an image and all the testing images, respectively. This is 2.6 million times slower than the random projection computation. This result clearly shows that the high computation complexity of the homomorphic encryption makes CryptoNets ill-suited for resource-constrained devices.



\section{Conclusion}
\label{sec:conclude}

This paper proposes a practical privacy-preserving collaborative learning approach, in which the resource-constrained learning participants apply independent random projections on their training data vectors and the coordinator applies deep learning to train a classifier based on the projected data vectors. Our approach protects the confidentiality of the raw forms of the training data against the honest-but-curious coordinator. Evaluation using four datasets shows that our approach outperforms various baselines and exhibits promising scalability with respect to the number of participants observing low- to moderate-complexity data patterns. Benchmark on a testbed shows the practicality and efficiency of our approach.


\bibliographystyle{ACM-Reference-Format}
\bibliography{ref}


\begin{thebibliography}{65}


\ifx \showCODEN    \undefined \def \showCODEN     #1{\unskip}     \fi
\ifx \showDOI      \undefined \def \showDOI       #1{#1}\fi
\ifx \showISBNx    \undefined \def \showISBNx     #1{\unskip}     \fi
\ifx \showISBNxiii \undefined \def \showISBNxiii  #1{\unskip}     \fi
\ifx \showISSN     \undefined \def \showISSN      #1{\unskip}     \fi
\ifx \showLCCN     \undefined \def \showLCCN      #1{\unskip}     \fi
\ifx \shownote     \undefined \def \shownote      #1{#1}          \fi
\ifx \showarticletitle \undefined \def \showarticletitle #1{#1}   \fi
\ifx \showURL      \undefined \def \showURL       {\relax}        \fi
\providecommand\bibfield[2]{#2}
\providecommand\bibinfo[2]{#2}
\providecommand\natexlab[1]{#1}
\providecommand\showeprint[2][]{arXiv:#2}

\bibitem[\protect\citeauthoryear{??}{cro}{2018}]%
        {crowd-ml-impl}
 \bibinfo{year}{2018}\natexlab{}.
\newblock \bibinfo{title}{Crowd-ML}.
\newblock
\newblock
\newblock
\shownote{\url{https://github.com/jihunhamm/Crowd-ML}.}


\bibitem[\protect\citeauthoryear{??}{pyt}{2018}]%
        {pytorch}
 \bibinfo{year}{2018}\natexlab{}.
\newblock \bibinfo{title}{{PyTorch}}.
\newblock
\newblock
\newblock
\shownote{\url{https://pytorch.org/}.}


\bibitem[\protect\citeauthoryear{??}{pi}{2018}]%
        {pi}
 \bibinfo{year}{2018}\natexlab{}.
\newblock \bibinfo{title}{{Raspberry Pi 2 Model B}}.
\newblock
\newblock
\newblock
\shownote{\url{https://bit.ly/1b75SRj}.}


\bibitem[\protect\citeauthoryear{??}{spa}{2018}]%
        {spambase}
 \bibinfo{year}{2018}\natexlab{}.
\newblock \bibinfo{title}{Spambase data set}.
\newblock
\newblock
\newblock
\shownote{\url{https://archive.ics.uci.edu/ml/datasets/spambase}.}


\bibitem[\protect\citeauthoryear{??}{sou}{2019}]%
        {sound-mnist}
 \bibinfo{year}{2019}\natexlab{}.
\newblock \bibinfo{title}{free-spoken-digit-dataset}.
\newblock
\newblock
\newblock
\shownote{\url{https://github.com/Jakobovski/free-spoken-digit-dataset}.}


\bibitem[\protect\citeauthoryear{??}{Sea}{2020}]%
        {Seal}
 \bibinfo{year}{2020}\natexlab{}.
\newblock \bibinfo{title}{Microsoft SEAL}.
\newblock
\newblock
\newblock
\shownote{\url{https://www.microsoft.com/en-us/research/project/microsoft-seal/}.}


\bibitem[\protect\citeauthoryear{??}{cod}{2020}]%
        {codes}
 \bibinfo{year}{2020}\natexlab{}.
\newblock \bibinfo{title}{source codes of the evaluation}.
\newblock
\newblock
\newblock
\shownote{\url{https://github.com/jls2007/TIOT_code}.}


\bibitem[\protect\citeauthoryear{Abadi, Chu, Goodfellow, McMahan, Mironov,
  Talwar, and Zhang}{Abadi et~al\mbox{.}}{2016}]%
        {Abadi16}
\bibfield{author}{\bibinfo{person}{M. Abadi}, \bibinfo{person}{A. Chu},
  \bibinfo{person}{I. Goodfellow}, \bibinfo{person}{H. McMahan},
  \bibinfo{person}{I. Mironov}, \bibinfo{person}{K. Talwar}, {and}
  \bibinfo{person}{L. Zhang}.} \bibinfo{year}{2016}\natexlab{}.
\newblock \showarticletitle{Deep learning with differential privacy}. In
  \bibinfo{booktitle}{\emph{Proc. CCS}}. ACM, \bibinfo{pages}{308--318}.
\newblock


\bibitem[\protect\citeauthoryear{Ailon and Chazelle}{Ailon and
  Chazelle}{2009}]%
        {Ailon09}
\bibfield{author}{\bibinfo{person}{Nir Ailon} {and} \bibinfo{person}{Bernard
  Chazelle}.} \bibinfo{year}{2009}\natexlab{}.
\newblock \showarticletitle{The fast Johnson--Lindenstrauss transform and
  approximate nearest neighbors}.
\newblock \bibinfo{journal}{\emph{SIAM Journal on computing}}
  \bibinfo{volume}{39}, \bibinfo{number}{1} (\bibinfo{year}{2009}),
  \bibinfo{pages}{302--322}.
\newblock


\bibitem[\protect\citeauthoryear{Aslett, Esperan{\c{c}}a, and Holmes}{Aslett
  et~al\mbox{.}}{2015}]%
        {aslett2015review}
\bibfield{author}{\bibinfo{person}{Louis~JM Aslett}, \bibinfo{person}{Pedro~M
  Esperan{\c{c}}a}, {and} \bibinfo{person}{Chris~C Holmes}.}
  \bibinfo{year}{2015}\natexlab{}.
\newblock \showarticletitle{A review of homomorphic encryption and software
  tools for encrypted statistical machine learning}.
\newblock \bibinfo{journal}{\emph{arXiv preprint arXiv:1508.06574}}
  (\bibinfo{year}{2015}).
\newblock


\bibitem[\protect\citeauthoryear{Bebensee}{Bebensee}{2019}]%
        {bebensee2019local}
\bibfield{author}{\bibinfo{person}{Bj{\"o}rn Bebensee}.}
  \bibinfo{year}{2019}\natexlab{}.
\newblock \showarticletitle{Local Differential Privacy: a tutorial}.
\newblock \bibinfo{journal}{\emph{arXiv preprint arXiv:1907.11908}}
  (\bibinfo{year}{2019}).
\newblock


\bibitem[\protect\citeauthoryear{Ben-Israel and Greville}{Ben-Israel and
  Greville}{2003}]%
        {ben2003generalized}
\bibfield{author}{\bibinfo{person}{Adi Ben-Israel} {and}
  \bibinfo{person}{Thomas~NE Greville}.} \bibinfo{year}{2003}\natexlab{}.
\newblock \bibinfo{booktitle}{\emph{Generalized inverses: theory and
  applications}}. Vol.~\bibinfo{volume}{15}.
\newblock \bibinfo{publisher}{Springer Science \& Business Media}.
\newblock


\bibitem[\protect\citeauthoryear{Berinde, Gilbert, Indyk, Karloff, and
  Strauss}{Berinde et~al\mbox{.}}{2008}]%
        {berinde2008combining}
\bibfield{author}{\bibinfo{person}{Radu Berinde}, \bibinfo{person}{Anna~C
  Gilbert}, \bibinfo{person}{Piotr Indyk}, \bibinfo{person}{Howard Karloff},
  {and} \bibinfo{person}{Martin~J Strauss}.} \bibinfo{year}{2008}\natexlab{}.
\newblock \showarticletitle{Combining geometry and combinatorics: A unified
  approach to sparse signal recovery}. In \bibinfo{booktitle}{\emph{2008 46th
  Annual Allerton Conference on Communication, Control, and Computing}}. IEEE,
  \bibinfo{pages}{798--805}.
\newblock


\bibitem[\protect\citeauthoryear{Berr}{Berr}{2018}]%
        {Equifax}
\bibfield{author}{\bibinfo{person}{Jonathan Berr}.}
  \bibinfo{year}{2018}\natexlab{}.
\newblock \bibinfo{title}{Equifax breach exposed data for 143 million
  consumers}.
\newblock
\newblock
\newblock
\shownote{\url{https://cbsn.ws/2Qc8VOg}.}


\bibitem[\protect\citeauthoryear{Bierlaire, Toint, and Tuyttens}{Bierlaire
  et~al\mbox{.}}{1991}]%
        {bierlaire1991iterative}
\bibfield{author}{\bibinfo{person}{Michel Bierlaire}, \bibinfo{person}{Ph~L
  Toint}, {and} \bibinfo{person}{Daniel Tuyttens}.}
  \bibinfo{year}{1991}\natexlab{}.
\newblock \showarticletitle{On iterative algorithms for linear least squares
  problems with bound constraints}.
\newblock \bibinfo{journal}{\emph{Linear Algebra Appl.}}  \bibinfo{volume}{143}
  (\bibinfo{year}{1991}), \bibinfo{pages}{111--143}.
\newblock


\bibitem[\protect\citeauthoryear{Bonawitz, Ivanov, Kreuter, Marcedone, McMahan,
  Patel, Ramage, Segal, and Seth}{Bonawitz et~al\mbox{.}}{2017}]%
        {Bonawitz17}
\bibfield{author}{\bibinfo{person}{Keith Bonawitz}, \bibinfo{person}{Vladimir
  Ivanov}, \bibinfo{person}{Ben Kreuter}, \bibinfo{person}{Antonio Marcedone},
  \bibinfo{person}{H.~Brendan McMahan}, \bibinfo{person}{Sarvar Patel},
  \bibinfo{person}{Daniel Ramage}, \bibinfo{person}{Aaron Segal}, {and}
  \bibinfo{person}{Karn Seth}.} \bibinfo{year}{2017}\natexlab{}.
\newblock \showarticletitle{Practical secure aggregation for privacy preserving
  machine learning}. In \bibinfo{booktitle}{\emph{Proc. CCS}}. ACM,
  \bibinfo{pages}{1175--1191}.
\newblock


\bibitem[\protect\citeauthoryear{Bose and Aarabi}{Bose and Aarabi}{2018}]%
        {bose2018adversarial}
\bibfield{author}{\bibinfo{person}{Avishek~Joey Bose} {and}
  \bibinfo{person}{Parham Aarabi}.} \bibinfo{year}{2018}\natexlab{}.
\newblock \showarticletitle{Adversarial Attacks on Face Detectors using Neural
  Net based Constrained Optimization}. In \bibinfo{booktitle}{\emph{Proc. Intl.
  Workshop Multimedia Signal Process.}}
\newblock


\bibitem[\protect\citeauthoryear{Cand{\`e}s et~al\mbox{.}}{Cand{\`e}s
  et~al\mbox{.}}{2006}]%
        {candes2006compressive}
\bibfield{author}{\bibinfo{person}{Emmanuel~J Cand{\`e}s} {et~al\mbox{.}}}
  \bibinfo{year}{2006}\natexlab{}.
\newblock \showarticletitle{Compressive sampling}. In
  \bibinfo{booktitle}{\emph{Proceedings of the international congress of
  mathematicians}}, Vol.~\bibinfo{volume}{3}. Madrid, Spain,
  \bibinfo{pages}{1433--1452}.
\newblock


\bibitem[\protect\citeauthoryear{Cand{\`e}s and Wakin}{Cand{\`e}s and
  Wakin}{2008}]%
        {candes2008introduction}
\bibfield{author}{\bibinfo{person}{Emmanuel~J Cand{\`e}s} {and}
  \bibinfo{person}{Michael~B Wakin}.} \bibinfo{year}{2008}\natexlab{}.
\newblock \showarticletitle{An introduction to compressive sampling}.
\newblock \bibinfo{journal}{\emph{IEEE Signal Process. Mag.}}
  \bibinfo{volume}{25}, \bibinfo{number}{2} (\bibinfo{year}{2008}),
  \bibinfo{pages}{21--30}.
\newblock


\bibitem[\protect\citeauthoryear{Chabanne, de~Wargny, Milgram, Morel, and
  Prouff}{Chabanne et~al\mbox{.}}{2017}]%
        {chabanne2017privacy}
\bibfield{author}{\bibinfo{person}{Herv{\'e} Chabanne}, \bibinfo{person}{Amaury
  de Wargny}, \bibinfo{person}{Jonathan Milgram}, \bibinfo{person}{Constance
  Morel}, {and} \bibinfo{person}{Emmanuel Prouff}.}
  \bibinfo{year}{2017}\natexlab{}.
\newblock \showarticletitle{Privacy-Preserving Classification on Deep Neural
  Network}.
\newblock \bibinfo{journal}{\emph{IACR Cryptology ePrint Archive}}
  \bibinfo{volume}{2017} (\bibinfo{year}{2017}), \bibinfo{pages}{35}.
\newblock


\bibitem[\protect\citeauthoryear{Chang and Lin}{Chang and Lin}{2018}]%
        {libSVM}
\bibfield{author}{\bibinfo{person}{Chih-Chung Chang} {and}
  \bibinfo{person}{Chih-Jen Lin}.} \bibinfo{year}{2018}\natexlab{}.
\newblock \bibinfo{title}{LIBSVM -- a library for support vector machines}.
\newblock
\newblock
\newblock
\shownote{\url{https://www.csie.ntu.edu.tw/~cjlin/libsvm/}.}


\bibitem[\protect\citeauthoryear{Chaudhuri and Monteleoni}{Chaudhuri and
  Monteleoni}{2009}]%
        {chaudhuri2009privacy}
\bibfield{author}{\bibinfo{person}{Kamalika Chaudhuri} {and}
  \bibinfo{person}{Claire Monteleoni}.} \bibinfo{year}{2009}\natexlab{}.
\newblock \showarticletitle{Privacy-preserving logistic regression}. In
  \bibinfo{booktitle}{\emph{Proc. NIPS}}. \bibinfo{pages}{289--296}.
\newblock


\bibitem[\protect\citeauthoryear{Chen and Dongarra}{Chen and Dongarra}{2005}]%
        {chen2005condition}
\bibfield{author}{\bibinfo{person}{Zizhong Chen} {and} \bibinfo{person}{Jack~J
  Dongarra}.} \bibinfo{year}{2005}\natexlab{}.
\newblock \showarticletitle{Condition numbers of Gaussian random matrices}.
\newblock \bibinfo{journal}{\emph{SIAM J. Matrix Anal. Appl.}}
  \bibinfo{volume}{27}, \bibinfo{number}{3} (\bibinfo{year}{2005}),
  \bibinfo{pages}{603--620}.
\newblock


\bibitem[\protect\citeauthoryear{Cormode, Jha, Kulkarni, Li, Srivastava, and
  Wang}{Cormode et~al\mbox{.}}{2018}]%
        {cormode2018privacy}
\bibfield{author}{\bibinfo{person}{Graham Cormode}, \bibinfo{person}{Somesh
  Jha}, \bibinfo{person}{Tejas Kulkarni}, \bibinfo{person}{Ninghui Li},
  \bibinfo{person}{Divesh Srivastava}, {and} \bibinfo{person}{Tianhao Wang}.}
  \bibinfo{year}{2018}\natexlab{}.
\newblock \showarticletitle{Privacy at scale: Local differential privacy in
  practice}. In \bibinfo{booktitle}{\emph{Proceedings of the 2018 International
  Conference on Management of Data}}. \bibinfo{pages}{1655--1658}.
\newblock


\bibitem[\protect\citeauthoryear{Danezis and Diaz}{Danezis and Diaz}{2008}]%
        {danezis2008survey}
\bibfield{author}{\bibinfo{person}{George Danezis} {and}
  \bibinfo{person}{Claudia Diaz}.} \bibinfo{year}{2008}\natexlab{}.
\newblock \bibinfo{booktitle}{\emph{A survey of anonymous communication
  channels}}.
\newblock \bibinfo{type}{{T}echnical {R}eport}. \bibinfo{institution}{Microsoft
  Research}.
\newblock
\newblock
\shownote{MSR-TR-2008-35.}


\bibitem[\protect\citeauthoryear{Dwork}{Dwork}{2006}]%
        {Dwork06}
\bibfield{author}{\bibinfo{person}{C. Dwork}.} \bibinfo{year}{2006}\natexlab{}.
\newblock \showarticletitle{Differential privacy}. In
  \bibinfo{booktitle}{\emph{Proc. ICALP}}.
\newblock


\bibitem[\protect\citeauthoryear{Dwork, McSherry, Nissim, and Smith}{Dwork
  et~al\mbox{.}}{2006}]%
        {Dwork061}
\bibfield{author}{\bibinfo{person}{C. Dwork}, \bibinfo{person}{F. McSherry},
  \bibinfo{person}{K. Nissim}, {and} \bibinfo{person}{A. Smith}.}
  \bibinfo{year}{2006}\natexlab{}.
\newblock \showarticletitle{Calibrating noise to sensitivity in private data
  analysis}.
\newblock \bibinfo{journal}{\emph{Conf. Theory of Cryptography}}
  (\bibinfo{year}{2006}), \bibinfo{pages}{265--284}.
\newblock


\bibitem[\protect\citeauthoryear{Erlingsson, Pihur, and Korolova}{Erlingsson
  et~al\mbox{.}}{2014}]%
        {erlingsson2014rappor}
\bibfield{author}{\bibinfo{person}{{\'U}lfar Erlingsson},
  \bibinfo{person}{Vasyl Pihur}, {and} \bibinfo{person}{Aleksandra Korolova}.}
  \bibinfo{year}{2014}\natexlab{}.
\newblock \showarticletitle{Rappor: Randomized aggregatable privacy-preserving
  ordinal response}. In \bibinfo{booktitle}{\emph{Proceedings of the 2014 ACM
  SIGSAC conference on computer and communications security}}.
  \bibinfo{pages}{1054--1067}.
\newblock


\bibitem[\protect\citeauthoryear{Gilad-Bachrach, Dowlin, Laine, Lauter,
  Naehrig, and Wernsing}{Gilad-Bachrach et~al\mbox{.}}{2016}]%
        {gilad2016cryptonets}
\bibfield{author}{\bibinfo{person}{Ran Gilad-Bachrach}, \bibinfo{person}{Nathan
  Dowlin}, \bibinfo{person}{Kim Laine}, \bibinfo{person}{Kristin Lauter},
  \bibinfo{person}{Michael Naehrig}, {and} \bibinfo{person}{John Wernsing}.}
  \bibinfo{year}{2016}\natexlab{}.
\newblock \showarticletitle{Cryptonets: Applying neural networks to encrypted
  data with high throughput and accuracy}. In \bibinfo{booktitle}{\emph{Proc.
  ICML}}. \bibinfo{pages}{201--210}.
\newblock


\bibitem[\protect\citeauthoryear{Goodfellow, Shlens, and Szegedy}{Goodfellow
  et~al\mbox{.}}{2015}]%
        {goodfellow15}
\bibfield{author}{\bibinfo{person}{Ian~J. Goodfellow},
  \bibinfo{person}{Jonathon Shlens}, {and} \bibinfo{person}{Christian
  Szegedy}.} \bibinfo{year}{2015}\natexlab{}.
\newblock \showarticletitle{Explaining and Harnessing Adversarial Examples}. In
  \bibinfo{booktitle}{\emph{Proc. ICLR}}.
\newblock


\bibitem[\protect\citeauthoryear{{Google Cloud}}{{Google Cloud}}{2018}]%
        {edge-tpu}
\bibfield{author}{\bibinfo{person}{{Google Cloud}}.}
  \bibinfo{year}{2018}\natexlab{}.
\newblock \bibinfo{title}{Edge {TPU}}.
\newblock
\newblock
\newblock
\shownote{\url{https://cloud.google.com/edge-tpu/}.}


\bibitem[\protect\citeauthoryear{Graepel, Lauter, and Naehrig}{Graepel
  et~al\mbox{.}}{2012}]%
        {graepel2012ml}
\bibfield{author}{\bibinfo{person}{Thore Graepel}, \bibinfo{person}{Kristin
  Lauter}, {and} \bibinfo{person}{Michael Naehrig}.}
  \bibinfo{year}{2012}\natexlab{}.
\newblock \showarticletitle{ML confidential: Machine learning on encrypted
  data}. In \bibinfo{booktitle}{\emph{Proc. Intl. Conf. Inf. Security \&
  Cryptology}}. Springer, \bibinfo{pages}{1--21}.
\newblock


\bibitem[\protect\citeauthoryear{Hamm, Champion, Chen, Belkin, and Xuan}{Hamm
  et~al\mbox{.}}{2015}]%
        {Hamm15}
\bibfield{author}{\bibinfo{person}{J. Hamm}, \bibinfo{person}{A. Champion},
  \bibinfo{person}{G. Chen}, \bibinfo{person}{M. Belkin}, {and}
  \bibinfo{person}{D. Xuan}.} \bibinfo{year}{2015}\natexlab{}.
\newblock \showarticletitle{Crowd-ML: A Privacy-Preserving Learning Framework
  for a Crowd of Smart Devices}. In \bibinfo{booktitle}{\emph{Proc. ICDCS}}.
  IEEE, \bibinfo{pages}{11--20}.
\newblock


\bibitem[\protect\citeauthoryear{He, Zhang, Ren, and Sun}{He
  et~al\mbox{.}}{2016}]%
        {He16}
\bibfield{author}{\bibinfo{person}{K. He}, \bibinfo{person}{X. Zhang},
  \bibinfo{person}{S. Ren}, {and} \bibinfo{person}{J. Sun}.}
  \bibinfo{year}{2016}\natexlab{}.
\newblock \showarticletitle{Deep Residual Learning for Image Recognition}. In
  \bibinfo{booktitle}{\emph{CVPR}}.
\newblock


\bibitem[\protect\citeauthoryear{Hitaj, Ateniese, and Perez-Cruz}{Hitaj
  et~al\mbox{.}}{2017}]%
        {Hitaj17}
\bibfield{author}{\bibinfo{person}{B. Hitaj}, \bibinfo{person}{G. Ateniese},
  {and} \bibinfo{person}{F. Perez-Cruz}.} \bibinfo{year}{2017}\natexlab{}.
\newblock \showarticletitle{Deep Models Under the GAN: Information Leakage from
  Collaborative Deep Learning}. In \bibinfo{booktitle}{\emph{Proc. CCS}}. ACM,
  \bibinfo{pages}{603--618}.
\newblock


\bibitem[\protect\citeauthoryear{Huynh, Lee, and Balan}{Huynh
  et~al\mbox{.}}{2017}]%
        {huynh2017deepmon}
\bibfield{author}{\bibinfo{person}{Loc~N Huynh}, \bibinfo{person}{Youngki Lee},
  {and} \bibinfo{person}{Rajesh~Krishna Balan}.}
  \bibinfo{year}{2017}\natexlab{}.
\newblock \showarticletitle{Deepmon: Mobile gpu-based deep learning framework
  for continuous vision applications}. In \bibinfo{booktitle}{\emph{Proc.
  MobiSys}}. ACM, \bibinfo{pages}{82--95}.
\newblock


\bibitem[\protect\citeauthoryear{Krizhevsky and Hinton}{Krizhevsky and
  Hinton}{2009}]%
        {krizhevsky2009learning}
\bibfield{author}{\bibinfo{person}{Alex Krizhevsky} {and}
  \bibinfo{person}{Geoffrey Hinton}.} \bibinfo{year}{2009}\natexlab{}.
\newblock \bibinfo{booktitle}{\emph{Learning multiple layers of features from
  tiny images}}.
\newblock \bibinfo{type}{{T}echnical {R}eport}.
\newblock


\bibitem[\protect\citeauthoryear{LeCun, Bengio, and Hinton}{LeCun
  et~al\mbox{.}}{2015}]%
        {lecun2015deep}
\bibfield{author}{\bibinfo{person}{Yann LeCun}, \bibinfo{person}{Yoshua
  Bengio}, {and} \bibinfo{person}{Geoffrey Hinton}.}
  \bibinfo{year}{2015}\natexlab{}.
\newblock \showarticletitle{Deep learning}.
\newblock \bibinfo{journal}{\emph{Nature}} \bibinfo{volume}{521},
  \bibinfo{number}{7553} (\bibinfo{year}{2015}), \bibinfo{pages}{436--444}.
\newblock


\bibitem[\protect\citeauthoryear{LeCun, Corts, and Burges}{LeCun
  et~al\mbox{.}}{2018}]%
        {mnist}
\bibfield{author}{\bibinfo{person}{Yann LeCun}, \bibinfo{person}{Corinna
  Corts}, {and} \bibinfo{person}{Christopher~J.C. Burges}.}
  \bibinfo{year}{2018}\natexlab{}.
\newblock \bibinfo{title}{The MNIST Database of Handwritten Digits}.
\newblock
\newblock
\newblock
\shownote{\url{http://yann.lecun.com/exdb/mnist/}.}


\bibitem[\protect\citeauthoryear{Li, Da~Xu, and Wang}{Li et~al\mbox{.}}{2013}]%
        {li2013compressed}
\bibfield{author}{\bibinfo{person}{Shancang Li}, \bibinfo{person}{Li Da~Xu},
  {and} \bibinfo{person}{Xinheng Wang}.} \bibinfo{year}{2013}\natexlab{}.
\newblock \showarticletitle{Compressed sensing signal and data acquisition in
  wireless sensor networks and internet of things}.
\newblock \bibinfo{journal}{\emph{IEEE Trans. Ind. Informat.}}
  \bibinfo{volume}{9}, \bibinfo{number}{4} (\bibinfo{year}{2013}),
  \bibinfo{pages}{2177--2186}.
\newblock


\bibitem[\protect\citeauthoryear{Liu, Jiang, Sha, and Govindan}{Liu
  et~al\mbox{.}}{2012}]%
        {liu2012cloud}
\bibfield{author}{\bibinfo{person}{Bin Liu}, \bibinfo{person}{Yurong Jiang},
  \bibinfo{person}{Fei Sha}, {and} \bibinfo{person}{Ramesh Govindan}.}
  \bibinfo{year}{2012}\natexlab{}.
\newblock \showarticletitle{Cloud-enabled privacy-preserving collaborative
  learning for mobile sensing}. In \bibinfo{booktitle}{\emph{Proc. SenSys}}.
  ACM, \bibinfo{pages}{57--70}.
\newblock


\bibitem[\protect\citeauthoryear{Liu, Kargupta, and Ryan}{Liu
  et~al\mbox{.}}{2006}]%
        {liu2006random}
\bibfield{author}{\bibinfo{person}{Kun Liu}, \bibinfo{person}{Hillol Kargupta},
  {and} \bibinfo{person}{Jessica Ryan}.} \bibinfo{year}{2006}\natexlab{}.
\newblock \showarticletitle{Random projection-based multiplicative data
  perturbation for privacy preserving distributed data mining}.
\newblock \bibinfo{journal}{\emph{IEEE Trans. knowl. Data Eng.}}
  \bibinfo{volume}{18}, \bibinfo{number}{1} (\bibinfo{year}{2006}),
  \bibinfo{pages}{92--106}.
\newblock


\bibitem[\protect\citeauthoryear{Logan et~al\mbox{.}}{Logan
  et~al\mbox{.}}{2000}]%
        {logan2000mel}
\bibfield{author}{\bibinfo{person}{Beth Logan} {et~al\mbox{.}}}
  \bibinfo{year}{2000}\natexlab{}.
\newblock \showarticletitle{Mel frequency cepstral coefficients for music
  modeling.}. In \bibinfo{booktitle}{\emph{Ismir}}, Vol.~\bibinfo{volume}{270}.
  \bibinfo{pages}{1--11}.
\newblock


\bibitem[\protect\citeauthoryear{McMahan, Moore, Ramage, Hampson, and
  y~Arcas}{McMahan et~al\mbox{.}}{2017}]%
        {mcmahan2016communication}
\bibfield{author}{\bibinfo{person}{H~Brendan McMahan}, \bibinfo{person}{Eider
  Moore}, \bibinfo{person}{Daniel Ramage}, \bibinfo{person}{Seth Hampson},
  {and} \bibinfo{person}{Blaise~Ag\"{u}era y Arcas}.}
  \bibinfo{year}{2017}\natexlab{}.
\newblock \showarticletitle{Communication-efficient learning of deep networks
  from decentralized data}. In \bibinfo{booktitle}{\emph{AISTATS}}.
\newblock


\bibitem[\protect\citeauthoryear{McMahan, Ramage, Talwar, and Zhang}{McMahan
  et~al\mbox{.}}{2018}]%
        {mcmahan2018learning}
\bibfield{author}{\bibinfo{person}{H~Brendan McMahan}, \bibinfo{person}{Daniel
  Ramage}, \bibinfo{person}{Kunal Talwar}, {and} \bibinfo{person}{Li Zhang}.}
  \bibinfo{year}{2018}\natexlab{}.
\newblock \showarticletitle{Learning Differentially Private Recurrent Language
  Models}. In \bibinfo{booktitle}{\emph{Proc. ICLR}}.
\newblock


\bibitem[\protect\citeauthoryear{Narayanan and Shmatikov}{Narayanan and
  Shmatikov}{2006}]%
        {narayanan2006break}
\bibfield{author}{\bibinfo{person}{Arvind Narayanan} {and}
  \bibinfo{person}{Vitaly Shmatikov}.} \bibinfo{year}{2006}\natexlab{}.
\newblock \showarticletitle{How to break anonymity of the netflix prize
  dataset}.
\newblock \bibinfo{journal}{\emph{arXiv preprint cs/0610105}}
  (\bibinfo{year}{2006}).
\newblock


\bibitem[\protect\citeauthoryear{O'Donnell}{O'Donnell}{2018}]%
        {Lindsey2018}
\bibfield{author}{\bibinfo{person}{Lindsey O'Donnell}.}
  \bibinfo{year}{2018}\natexlab{}.
\newblock \bibinfo{title}{Zero-Day Flash Exploit Targeting Middle East}.
\newblock
\newblock
\newblock
\shownote{\url{https://threatpost.com/zero-day-flash-exploit-targeting-middle-east/132659/}.}


\bibitem[\protect\citeauthoryear{Paige and Saunders}{Paige and
  Saunders}{1982}]%
        {paige1982lsqr}
\bibfield{author}{\bibinfo{person}{Christopher~C Paige} {and}
  \bibinfo{person}{Michael~A Saunders}.} \bibinfo{year}{1982}\natexlab{}.
\newblock \showarticletitle{LSQR: An algorithm for sparse linear equations and
  sparse least squares}.
\newblock \bibinfo{journal}{\emph{ACM Trans. Math. Software}}
  \bibinfo{volume}{8}, \bibinfo{number}{1} (\bibinfo{year}{1982}),
  \bibinfo{pages}{43--71}.
\newblock


\bibitem[\protect\citeauthoryear{Phong, Aono, Hayashi, Wang, and Moriai}{Phong
  et~al\mbox{.}}{2018}]%
        {Phong17}
\bibfield{author}{\bibinfo{person}{L. Phong}, \bibinfo{person}{Y. Aono},
  \bibinfo{person}{T. Hayashi}, \bibinfo{person}{L. Wang}, {and}
  \bibinfo{person}{S. Moriai}.} \bibinfo{year}{2018}\natexlab{}.
\newblock \showarticletitle{Privacy-Preserving Deep Learning via Additively
  Homomorphic Encryption}.
\newblock \bibinfo{journal}{\emph{IEEE Trans. Inf. Forensics Security}}
  \bibinfo{volume}{13}, \bibinfo{number}{5} (\bibinfo{year}{2018}).
\newblock


\bibitem[\protect\citeauthoryear{Rachlin and Baron}{Rachlin and Baron}{2008}]%
        {rachlin2008secrecy}
\bibfield{author}{\bibinfo{person}{Yaron Rachlin} {and} \bibinfo{person}{Dror
  Baron}.} \bibinfo{year}{2008}\natexlab{}.
\newblock \showarticletitle{The secrecy of compressed sensing measurements}. In
  \bibinfo{booktitle}{\emph{Proc. Allerton}}. IEEE, \bibinfo{pages}{813--817}.
\newblock


\bibitem[\protect\citeauthoryear{{Reuters}}{{Reuters}}{2018}]%
        {FB2018}
\bibfield{author}{\bibinfo{person}{{Reuters}}.}
  \bibinfo{year}{2018}\natexlab{}.
\newblock \bibinfo{title}{Facebook critics want regulation, investigation after
  data misuse}.
\newblock
\newblock
\newblock
\shownote{\url{https://reut.rs/2GwKF8p}.}


\bibitem[\protect\citeauthoryear{Shen, Luo, Yin, Wen, Daniela, and Hu}{Shen
  et~al\mbox{.}}{2018}]%
        {shen2018privacy}
\bibfield{author}{\bibinfo{person}{Yiran Shen}, \bibinfo{person}{Chengwen Luo},
  \bibinfo{person}{Dan Yin}, \bibinfo{person}{Hongkai Wen},
  \bibinfo{person}{Rus Daniela}, {and} \bibinfo{person}{Wen Hu}.}
  \bibinfo{year}{2018}\natexlab{}.
\newblock \showarticletitle{Privacy-preserving sparse representation
  classification in cloud-enabled mobile applications}.
\newblock \bibinfo{journal}{\emph{Comput. Netw.}}  \bibinfo{volume}{133}
  (\bibinfo{year}{2018}), \bibinfo{pages}{59--72}.
\newblock


\bibitem[\protect\citeauthoryear{Shokri and Shmatikov}{Shokri and
  Shmatikov}{2015}]%
        {Shokri15}
\bibfield{author}{\bibinfo{person}{R. Shokri} {and} \bibinfo{person}{V.
  Shmatikov}.} \bibinfo{year}{2015}\natexlab{}.
\newblock \showarticletitle{Privacy-preserving deep learning}. In
  \bibinfo{booktitle}{\emph{Proc. CCS}}. ACM, \bibinfo{pages}{1310--1321}.
\newblock


\bibitem[\protect\citeauthoryear{Song, Chaudhuri, and Sarwate}{Song
  et~al\mbox{.}}{2013}]%
        {song2013stochastic}
\bibfield{author}{\bibinfo{person}{Shuang Song}, \bibinfo{person}{Kamalika
  Chaudhuri}, {and} \bibinfo{person}{Anand~D Sarwate}.}
  \bibinfo{year}{2013}\natexlab{}.
\newblock \showarticletitle{Stochastic gradient descent with differentially
  private updates}. In \bibinfo{booktitle}{\emph{Proc. GlobalSIP}}. IEEE,
  \bibinfo{pages}{245--248}.
\newblock


\bibitem[\protect\citeauthoryear{Suykens}{Suykens}{2003}]%
        {suykens2003}
\bibfield{author}{\bibinfo{person}{Johan~AK Suykens}.}
  \bibinfo{year}{2003}\natexlab{}.
\newblock \bibinfo{booktitle}{\emph{Advances in learning theory: methods,
  models, and applications}}. Vol.~\bibinfo{volume}{190}.
\newblock \bibinfo{publisher}{IOS Press}.
\newblock


\bibitem[\protect\citeauthoryear{Tan, Chiu, Nguyen, Yau, and Jung}{Tan
  et~al\mbox{.}}{2017}]%
        {tan2017joint}
\bibfield{author}{\bibinfo{person}{Rui Tan}, \bibinfo{person}{Sheng-Yuan Chiu},
  \bibinfo{person}{Hoang~Hai Nguyen}, \bibinfo{person}{David~KY Yau}, {and}
  \bibinfo{person}{Deokwoo Jung}.} \bibinfo{year}{2017}\natexlab{}.
\newblock \showarticletitle{A Joint Data Compression and Encryption Approach
  for Wireless Energy Auditing Networks}.
\newblock \bibinfo{journal}{\emph{ACM Trans. Sensor Networks}}
  \bibinfo{volume}{13}, \bibinfo{number}{2} (\bibinfo{year}{2017}),
  \bibinfo{pages}{9}.
\newblock


\bibitem[\protect\citeauthoryear{Tram{\`e}r, Zhang, Juels, Reiter, and
  Ristenpart}{Tram{\`e}r et~al\mbox{.}}{2016}]%
        {tramer2016stealing}
\bibfield{author}{\bibinfo{person}{Florian Tram{\`e}r}, \bibinfo{person}{Fan
  Zhang}, \bibinfo{person}{Ari Juels}, \bibinfo{person}{Michael~K Reiter},
  {and} \bibinfo{person}{Thomas Ristenpart}.} \bibinfo{year}{2016}\natexlab{}.
\newblock \showarticletitle{Stealing machine learning models via prediction
  apis}. In \bibinfo{booktitle}{\emph{25th $\{$USENIX$\}$ Security Symposium
  ($\{$USENIX$\}$ Security 16)}}. \bibinfo{pages}{601--618}.
\newblock


\bibitem[\protect\citeauthoryear{Truex, Baracaldo, Anwar, Steinke, Ludwig,
  Zhang, and Zhou}{Truex et~al\mbox{.}}{2019}]%
        {truex2019hybrid}
\bibfield{author}{\bibinfo{person}{Stacey Truex}, \bibinfo{person}{Nathalie
  Baracaldo}, \bibinfo{person}{Ali Anwar}, \bibinfo{person}{Thomas Steinke},
  \bibinfo{person}{Heiko Ludwig}, \bibinfo{person}{Rui Zhang}, {and}
  \bibinfo{person}{Yi Zhou}.} \bibinfo{year}{2019}\natexlab{}.
\newblock \showarticletitle{A hybrid approach to privacy-preserving federated
  learning}. In \bibinfo{booktitle}{\emph{Proceedings of the 12th ACM Workshop
  on Artificial Intelligence and Security}}. \bibinfo{pages}{1--11}.
\newblock


\bibitem[\protect\citeauthoryear{Wang, Zhang, Ren, and Roveda}{Wang
  et~al\mbox{.}}{2013}]%
        {wang2013privacy}
\bibfield{author}{\bibinfo{person}{Cong Wang}, \bibinfo{person}{Bingsheng
  Zhang}, \bibinfo{person}{Kui Ren}, {and} \bibinfo{person}{Janet~M Roveda}.}
  \bibinfo{year}{2013}\natexlab{}.
\newblock \showarticletitle{Privacy-assured outsourcing of image reconstruction
  service in cloud}.
\newblock \bibinfo{journal}{\emph{IEEE Trans. Emerg. Topics Comput.}}
  \bibinfo{volume}{1}, \bibinfo{number}{1} (\bibinfo{year}{2013}),
  \bibinfo{pages}{166--177}.
\newblock


\bibitem[\protect\citeauthoryear{Warner}{Warner}{1965}]%
        {warner1965randomized}
\bibfield{author}{\bibinfo{person}{Stanley~L Warner}.}
  \bibinfo{year}{1965}\natexlab{}.
\newblock \showarticletitle{Randomized response: A survey technique for
  eliminating evasive answer bias}.
\newblock \bibinfo{journal}{\emph{J. Amer. Statist. Assoc.}}
  \bibinfo{volume}{60}, \bibinfo{number}{309} (\bibinfo{year}{1965}),
  \bibinfo{pages}{63--69}.
\newblock


\bibitem[\protect\citeauthoryear{W{\'o}jcik and Kurdziel}{W{\'o}jcik and
  Kurdziel}{2018}]%
        {wojcik2018training}
\bibfield{author}{\bibinfo{person}{Piotr~Iwo W{\'o}jcik} {and}
  \bibinfo{person}{Marcin Kurdziel}.} \bibinfo{year}{2018}\natexlab{}.
\newblock \showarticletitle{Training neural networks on high-dimensional data
  using random projection}.
\newblock \bibinfo{journal}{\emph{Pattern Anal. Appl.}} (\bibinfo{year}{2018}),
  \bibinfo{pages}{1--11}.
\newblock


\bibitem[\protect\citeauthoryear{Xu, Zheng, Jiang, Gu, Tan, and Cheng}{Xu
  et~al\mbox{.}}{2019}]%
        {xu2019lightweight}
\bibfield{author}{\bibinfo{person}{Dixing Xu}, \bibinfo{person}{Mengyao Zheng},
  \bibinfo{person}{Linshan Jiang}, \bibinfo{person}{Chaojie Gu},
  \bibinfo{person}{Rui Tan}, {and} \bibinfo{person}{Peng Cheng}.}
  \bibinfo{year}{2019}\natexlab{}.
\newblock \showarticletitle{Lightweight and Unobtrusive Privacy Preservation
  for Remote Inference via Edge Data Obfuscation}.
\newblock \bibinfo{journal}{\emph{arXiv preprint arXiv:1912.09859}}
  (\bibinfo{year}{2019}).
\newblock


\bibitem[\protect\citeauthoryear{Xue, Luo, Lan, Rana, Hu, and Seneviratne}{Xue
  et~al\mbox{.}}{2017}]%
        {xue2017kryptein}
\bibfield{author}{\bibinfo{person}{Wanli Xue}, \bibinfo{person}{Chenwen Luo},
  \bibinfo{person}{Guohao Lan}, \bibinfo{person}{Rajib Rana},
  \bibinfo{person}{Wen Hu}, {and} \bibinfo{person}{Aruna Seneviratne}.}
  \bibinfo{year}{2017}\natexlab{}.
\newblock \showarticletitle{Kryptein: a compressive-sensing-based encryption
  scheme for the internet of things}. In \bibinfo{booktitle}{\emph{Proc.
  IPSN}}. IEEE, \bibinfo{pages}{169--180}.
\newblock


\bibitem[\protect\citeauthoryear{Yao, Zhao, Zhang, Su, and Abdelzaher}{Yao
  et~al\mbox{.}}{2017}]%
        {yao2017deepiot}
\bibfield{author}{\bibinfo{person}{Shuochao Yao}, \bibinfo{person}{Yiran Zhao},
  \bibinfo{person}{Aston Zhang}, \bibinfo{person}{Lu Su}, {and}
  \bibinfo{person}{Tarek Abdelzaher}.} \bibinfo{year}{2017}\natexlab{}.
\newblock \showarticletitle{{DeepIoT}: Compressing deep neural network
  structures for sensing systems with a compressor-critic framework}. In
  \bibinfo{booktitle}{\emph{Proc. SenSys}}. ACM, \bibinfo{pages}{4:1--4:14}.
\newblock


\bibitem[\protect\citeauthoryear{Zheng, Song, Leung, and Goodfellow}{Zheng
  et~al\mbox{.}}{2016}]%
        {zheng2016improving}
\bibfield{author}{\bibinfo{person}{Stephan Zheng}, \bibinfo{person}{Yang Song},
  \bibinfo{person}{Thomas Leung}, {and} \bibinfo{person}{Ian Goodfellow}.}
  \bibinfo{year}{2016}\natexlab{}.
\newblock \showarticletitle{Improving the robustness of deep neural networks
  via stability training}. In \bibinfo{booktitle}{\emph{Proc. CVPR}}. IEEE,
  \bibinfo{pages}{4480--4488}.
\newblock


\end{thebibliography}

\end{document}